\documentclass{aa}  
\usepackage{graphicx}
\usepackage{natbib}
\usepackage{longtable}
\usepackage{rotating}
\usepackage{lscape}
\usepackage{amsmath}
\usepackage{txfonts}
\begin{document}
   \title{X-ray source population study of the starburst galaxy M\,83 with XMM-\emph{Newton}
   \thanks{Based on 
   observations obtained with XMM-Newton, an ESA science mission with 
   instruments and contributions directly funded by ESA Member States 
   and NASA.}\fnmsep
   \thanks{Appendix \ref{sect. Discussion of classification and identification of the XMM-Newton sources}
   is available in electronic form at http://www.aanda.org}\fnmsep
   \thanks{Tables~\ref{Tab. source list} and~\ref{Tab. source list classification} 
    are only available in electronic form
    at the CDS via anonymous ftp to cdsarc.u-strasbg.fr (130.79.128.5)
    or via http://cdsweb.u-strasbg.fr/cgi-bin/qcat?J/A+A/ }
}
\titlerunning{X-ray source population study of the galaxy M\,83 with XMM-\emph{Newton}}

   \subtitle{}

   \author{L. Ducci
          \inst{1}
          \and
          M. Sasaki\inst{1}
          \and
          F. Haberl\inst{2}
          \and
          W. Pietsch\inst{2}
          }
   \institute{Institut f\"ur Astronomie und Astrophysik, Eberhard Karls Universit\"at, 
              Sand 1, 72076 T\"ubingen, Germany\\
              \email{ducci@astro.uni-tuebingen.de}
              \and
              Max-Planck-Institut f\"ur Extraterrestrische Physik, 
              Giessenbachstra{\ss}e, 85741 Garching, Germany
             }
   \date{}

 
  \abstract
{}
{We present the results obtained from the analysis of three XMM-\emph{Newton} observations
of M\,83.
The aims of the paper are studying the X-ray source populations in M\,83
and calculating the X-ray luminosity functions of X-ray binaries for different
regions of the galaxy.}
{We detected 189 sources in the XMM-\emph{Newton} field of view
in the energy range of $0.2-12$ keV. We constrained their nature
by means of spectral analysis, hardness ratios, studies of the X-ray variability,
and cross-correlations with catalogues in X-ray, optical, infrared, and radio wavelengths.}
{We identified and classified 12 background objects, five foreground stars,
two X-ray binaries, one supernova remnant candidate, one super-soft source candidate and one ultra-luminous X-ray source.
Among these sources, we classified for the first time three active galactic nuclei (AGN) candidates.
We derived X-ray luminosity functions of the X-ray sources in M\,83
in the $2-10$ keV energy range, within and outside the $D_{25}$ ellipse,
correcting the total X-ray luminosity function for incompleteness and subtracting the AGN contribution.
The X-ray luminosity function inside the $D_{25}$ ellipse is consistent with
that previously observed by \emph{Chandra}.
The Kolmogorov-Smirnov test shows that the X-ray luminosity function
of the outer disc and the AGN luminosity distribution 
are uncorrelated with a probability of $\sim 99.3$\%.
We also found that the X-ray sources detected outside the $D_{25}$ ellipse
and the uniform spatial distribution of AGNs
are spatially uncorrelated with a significance
of 99.5\%. We interpret these results as an indication that part of the
observed X-ray sources are X-ray binaries in the outer disc of M\,83.}
{}
%
%



   \maketitle
%

\section{Introduction}
\label{section introduction}

\begin{figure*}
\begin{center}
\includegraphics[width=16cm]{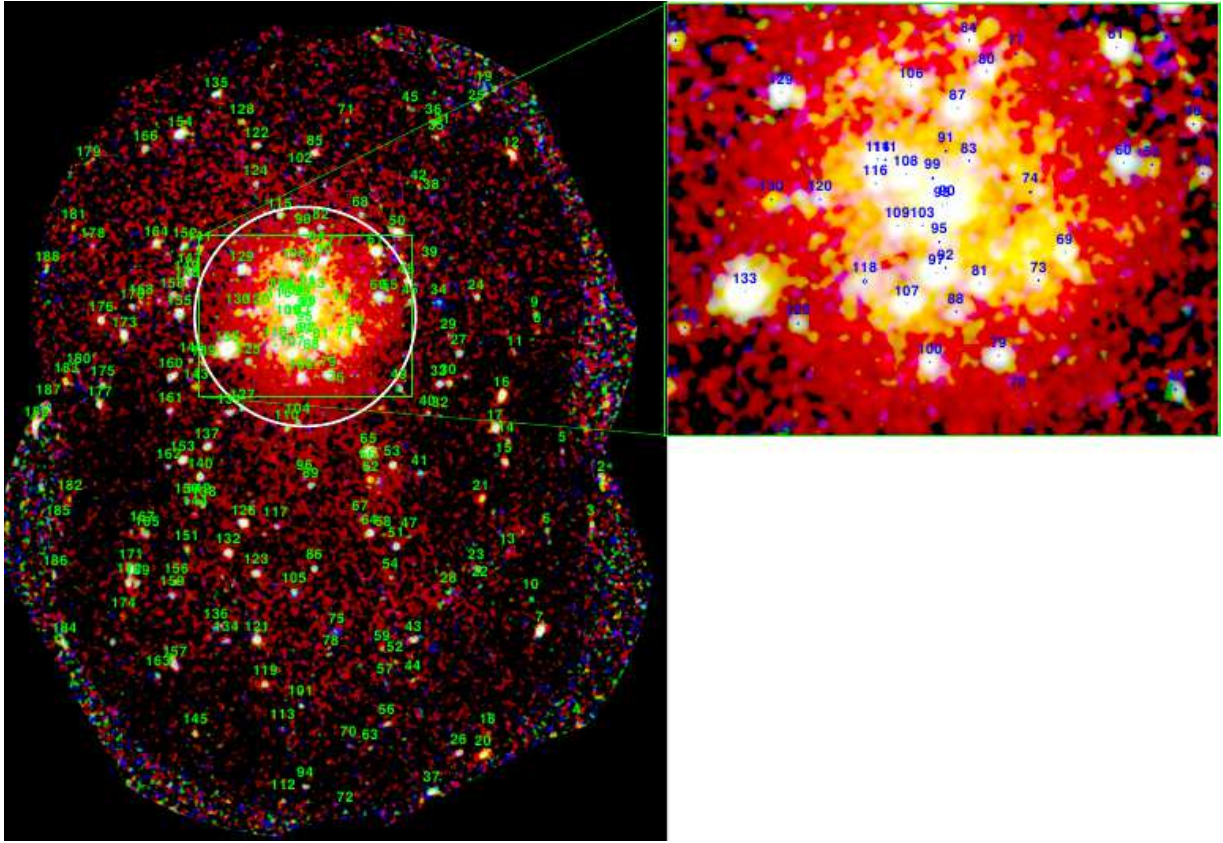} 
\end{center}
\caption{Combined PN, MOS1, and MOS2 three-colour mosaic image of M\,83. The crowded central region is shown in higher resolution.
The white circle is the $D_{25}$ ellipse (diameter$=11.5^\prime$; \citealt{Tully88}).}
\label{figure RGB}
\end{figure*}

M\,83 (NGC\,5236) is a grand-design barred spiral galaxy (SAB(s)c; \citealt{deVaucouleurs92})
located at $4.5 \pm 0.3$ Mpc from the Milky Way \citep{Thim03}.
M\,83 is oriented nearly face-on ($i=24^\circ$; \citealt{Rogstad74})
and shows a galactic disc spanning $12.9^\prime \times 11.5^\prime$ 
($17$ kpc $\times$ $15.2$ kpc; \citealt{Wofford11}).

M83 is experiencing a starburst activity 
with a present-day star formation rate (SFR) of $3-4$ M$_\odot$ yr$^{-1}$ \citep{Boissier05}
in three regions:
the nuclear region (galactocentric distance $d\lesssim 300$ pc; \citealt{Harris01}),
the inner disc ($300$ pc $\lesssim d \lesssim 7.5$ kpc),
and the outer disc ($7.5$ kpc$\lesssim d \lesssim 20$ kpc; \citealt{Dong08}).
Ultraviolet (UV) images of M\,83 obtained with the 
\emph{Galaxy Evolution Explorer} (GALEX) satellite revealed 
a population of young stars ($\lesssim 400$ Myr)
in the outer disc of M\,83 \citep{Thilker05}.
Although this would indicate recent star-forming activity, 
using \emph{Spitzer} and GALEX data,
\citet{Dong08} discovered that the star formation in the outer disc started
at least 1 Gyr ago. These results are confirmed by the study of AGB stars of \citet{Davidge10}.
\citet{Bigiel10} compared the HI data from
the National Radio Astronomy Observatory (NRAO) Very Large Array (VLA)
and far-ultraviolet (FUV) data from GALEX in the outer disc of M\,83,
and discovered that the star formation traced by the FUV emission
and HI are spatially correlated out to almost four optical radii.
\citet{Bigiel10} also found that the star formation rate 
in the outer disc ($\sim 0.01$ M$_\odot$ yr$^{-1}$; \citealt{Bresolin09})
implies that the star formation activity is not 
completely consuming the HI reservoir,
which will be available as fuel for star formation in the inner disc.

%
M\,83 was observed in the X-ray bands by \emph{Einstein} in
1979-1981 \citep{Trinchieri85},  Ginga in 1987 \citep{Ohashi90},
ROSAT in 1992-1994 (\citealt{Ehle98}; \citealt{Immler99}), 
ASCA in 1994 \citep{Okada97}, and 
\emph{Chandra} in 2000 (\citealt{SoriaWu02}; \citealt{SoriaWu03}, SW03 hereafter). 
SW03 identified 127 discrete sources near the centre of M\,83 ($8.3^\prime \times 8.3^\prime$) 
and resolved for the first time the nuclear region in X-rays.
The diffuse X-ray emission of M\,83 has been studied by \citet{Owen09} with 
an XMM-\emph{Newton} observation performed in January 27, 2003 (obsid 0110910201).
They obtained a good fit to the spectrum assuming a two-temperature thermal model, which
is typical of the diffuse emission in normal
and starburst galaxies.
They also found that the soft X-ray emission mainly overlaps  with the
inner spiral arm, and shows a strong correlation with the distribution of UV emission,
indicative of a correlation between X-ray emission and recent star formation.

%
The recent high star formation activity experienced by the nucleus
and the spiral arms of M\,83 provided an unusually large number 
of supernova remnants (SNRs). 
In fact, the optical survey performed
at the Cerro Tololo Inter-American Observatory in Chile by \citet{Blair04}
identified 71 sources as SNR candidates,
the \emph{Hubble} Space Telescope (HST) observations of the nuclear region of M\,83 \citep{Dopita10}
provided the identification of 60 SNR candidates,
and the Magellan I survey 271 SNR candidates \citep{Blair12}.

%
In a normal galaxy such as M\,83, X-ray binaries (XRBs) are
the most prominent class of X-ray sources.
XRBs show X-ray luminosities ranging from $\sim 10^{32}$ erg s$^{-1}$ to 
the Eddington luminosity, 
and sometimes they can exceed this limit (see e.g. \citealt{White78}).
They are composed of a compact object (a neutron star or a black hole)
and a companion star, which can be a main-sequence, giant, or supergiant star, 
and in some cases a white dwarf (e.g. \citealt{vanParadijs98}).
The strong X-ray emission is produced by the
accretion of matter from the companion star onto the compact object.
XRBs are usually divided into two classes: low mass X-ray binaries (LMXBs),
and high mass X-ray binaries (HMXBs).
The companion stars of LMXBs have masses lower than $\sim 1$ M$_\odot$.
The lifetime of an LMXB is determined by the nuclear evolution time-scale of the companion star
to $10^8 - 10^9$ yr (e.g. \citealt{Tauris06}), 
and their number is correlated to the total stellar mass of a galaxy \citep{Gilfanov04}.
The companion star of LMXBs usually tranfers mass by Roche-lobe overflow,
and the compact object accretes from an accretion disc (e.g. \citealt{vanParadijs98}).
The donors in HMXBs have masses $\gtrsim 8$ M$_\odot$,
and their typical lifetime does not exceed $10^6 -10^7$ yr. 
Therefore, the presence of HMXBs in a particular region of a galaxy is associated with 
a relatively recent star formation event (e.g. \citealt{Fabbiano06}).
The X-ray emission from HMXBs is usually explained with 
the accretion of a fraction of the stellar wind ejected by the donor star
onto the compact object,
or through mass transfer via Roche-lobe overflow
(see e.g. \citealt{Treves88} and references therein).
%
As a first approximation, two standard models are commonly used to describe the X-ray spectra of XRBs
in nearby galaxies: an absorbed disc-blackbody model, 
with temperatures ranging from $\sim 0.5$ to $\sim 1$ keV (e.g. \citealt{Makishima86}),
or an absorbed powerlaw model.
X-ray spectra of LMXBs below 10 keV are described by absorbed powerlaw with photon indices $1-3$.
HMXBs usually show harder X-ray spectra in the energy range $1-10$ keV, 
with photon indices $1-2$ and a high intrinsic absorption \citep{White95}.
Within each of these classes, the properties of the X-ray spectra 
can also depend on the type of the accreting compact object.
Accreting black holes can show states of high luminosity (e.g. \citealt{Jones77}),
with very soft spectra, with slopes steeper than those shown by accreting neutron stars
(see e.g. \citealt{White84}).
Given the wide variety of spectral shapes shown by XRBs,
they can be confused with background AGNs,
whose X-ray spectra have roughly a powerlaw shape,
with indices ranging from 1.6 to 2.5 
(see e.g. \citealt{Walter93}; \citealt{Vignali99}; \citealt{Turner91}).

In this paper we report the results obtained from a
study of the X-ray source populations of M\,83,
using three XMM-\emph{Newton} observations covering both
the inner and outer disc regions.
The higher spatial resolution and sensitivity of XMM-\emph{Newton}
compared to the previous observations of ROSAT and \emph{Einstein}
allowed an increase of the number of detected sources in M\,83.
While the spatial coverage of the \emph{Chandra} observation was limited to a region
located at the centre of M\,83 with a size of  
$8.3^\prime \times 8.3^\prime$ (the ACIS S3 field of view), 
the XMM-\emph{Newton} observations allowed us to
obtain a complete coverage of M\,83,
and to study also the outer parts of the galaxy,
which in total provided us with a more representative sample of X-ray sources in M\,83.

The paper is organised as follows:
in Sect. \ref{section Reduction and Data Analysis} we describe the data reduction
and analysis of XMM-\emph{Newton} observations.
In Sect. \ref{section Astrometrical corrections} we show the astrometrical corrections
that have been applied.
In Sect. \ref{sect. Analysis}
we present the techniques adopted to classify the X-ray sources 
(X-ray variability, spectral analysis, and hardness ratios).
In Sect. \ref{sect. Source classification} we describe the properties
and classification of the detected sources.
In Sect. \ref{sect. X-ray Luminosity Functions} we derive the X-ray luminosity functions (XLFs)
of X-ray binaries within and outside the $D_{25}$ ellipse, 
after correcting them for incompleteness
and subtracting the AGN contribution, and we discuss our results.
We examine in detail the properties of the sources that have been identified and classified
in this work in Appendix \ref{sect. Discussion of classification and identification of the XMM-Newton sources}.

\begin{table*}
\begin{center}
\caption{XMM-\emph{Newton} observations of M\,83. 
The exposure times after the screening for high background are given in units of ks.
Mode: EFF=extended full frame imaging mode; FF=full frame imaging mode.}
\label{Tab. OBS ID XMM}
\resizebox{\columnwidth+\columnwidth}{!}{
\begin{tabular}{lcccccccccccc}
\hline
\hline
  & Obs. ID.   &     Date     &  \multicolumn{2}{c}{Pointing direction} & \multicolumn{2}{c}{EPIC PN} & \multicolumn{2}{c}{EPIC MOS1} & \multicolumn{2}{c}{EPIC MOS2} &\multicolumn{2}{c}{Mode} \\
  &            &              &        RA         &          Dec        &    filter    &   $T_{\rm exp}$   &    filter    &    $T_{\rm exp}$    &      filter     &  $T_{\rm exp}$   &     PN     &    MOS     \\  
\hline
1 & 0110910201 &  2003-01-27  &    13:37:05.16    &     -29:51:46.1     &     thin     &     21.2     &    medium    &      24.6      &      medium     &   24.6      &     EFF    &     FF     \\
2 & 0503230101 &  2008-01-16  &    13:37:01.09    &     -30:03:49.9     &    medium    &     15.4     &    medium    &      19.0      &      medium     &   19.0      &     EFF    &     FF     \\
3 & 0552080101 &  2008-08-16  &    13:36:50.87    &     -30:03:55.2     &    medium    &     25.0     &    medium    &      28.8      &      medium     &   28.8      &     EFF    &     FF     \\
\hline
\end{tabular}
}
\end{center}
\end{table*}

\section{Reduction and data analysis}
\label{section Reduction and Data Analysis}

We analysed the public archival XMM-\emph{Newton} data of M\,83
(PIs: M. Watson, K.D. Kuntz). 
Table \ref{Tab. OBS ID XMM} lists the three observations that we analysed,
one pointing at the centre of the galaxy (obs.\,1) and 
two in the south, which covered the outer arms with a young population of stars
discovered with GALEX.
The data analysis was performed using the XMM-\emph{Newton} Science Analysis System (SAS) 12.0.
For each pointing we produced PN, MOS1, and MOS2 event files.
We excluded times of high background due to soft proton flares
as follows. For each observation and instrument, 
we created background lightcurves (with sources removed) 
in  the 7$-$15 keV energy band. Good time intervals (GTIs) 
were determined by selecting count rates lower than 
8 cts ks$^{-1}$ arcmin$^{-2}$ and 2.5 cts ks$^{-1}$ arcmin$^{-2}$ 
for PN and MOS, respectively.

For each observation, data were divided into five energy bands:
\begin{itemize}
\item \emph{R1:} 0.2--0.5 keV;
\item \emph{R2:} 0.5--1 keV;
\item \emph{R3:} 1--2 keV;
\item \emph{R4:} 2--4.5 keV;
\item \emph{R5:} 4.5--12 keV.
\end{itemize}
For the PN data we used single-pixel events (PATTERN=0)
in the first energy band 
and for the other energy bands single- and double-pixel events 
(PATTERN$\leq$4) were selected.
For the MOS data, single-pixel to quadruple-pixel events 
(PATTERN$\leq$12) were used for all five bands.

We ran the source detection procedure separately for images of each
observation, and simultaneously for five energy bands and 
three instruments with the SAS task {\tt edetect\_chain}. 
The source detection consists of three steps.
The first step provides a list of source positions used to
create the background maps. 
We adopted a minimum-detection likelihood\footnote{The detection likelihood $L$ is defined by the relationship
$L = - \ln (p)$, where $p$ is the probability that a Poissonian fluctuation in the background is 
detected as a spurious source.} 
 of 7 to obtain this list of sources.
After removing the sources, a two-dimensional spline with 20 nodes
was fitted to the exposure-corrected image.
In the second step the background maps are used to improve the detection sensitivity
and hence to create a new source list, assuming a minimum-detection likelihood of 4.
In the last step, a maximum-likelihood point-spread function (PSF) fit to the source count distribution is performed
simultaneously in all energy bands and each EPIC instrument, from the input list of 
source positions obtained in the previous step (a description of this algorithm is given by 
\citealt{Cruddace88}).
For each observation we generated the final source list adopting a 
lower threshold for the
maximum-detection likelihood
of 6, which corresponds to a detection probability of $\sim 99.75$\%. 
The source detection gives several parameters for each source, such as the
coordinates, count rates, and likelihood of detection (see Table \ref{Tab. source list} 
in the appendix \ref{sect. catalogue-table}). 
As mentioned above, 20 nodes (more than the default 16) 
for the background spline map were used to better follow the central
diffuse emission and to minimise false detections.
We removed the remaining false detections due to diffuse emission
structures by visual inspection.

Fig. \ref{figure RGB} shows the combined PN, MOS1, and MOS2 three-colour mosaic image
obtained from the three observations.
The numbers of the detected sources are overplotted on the image.
The red, green, and blue colours represent the 
$0.2-1$ keV, $1-2$ keV, and $2-4.5$ keV energy bands.

\begin{figure*}
\begin{center}
\includegraphics[bb= 96 375 555 716, clip, width=7cm]{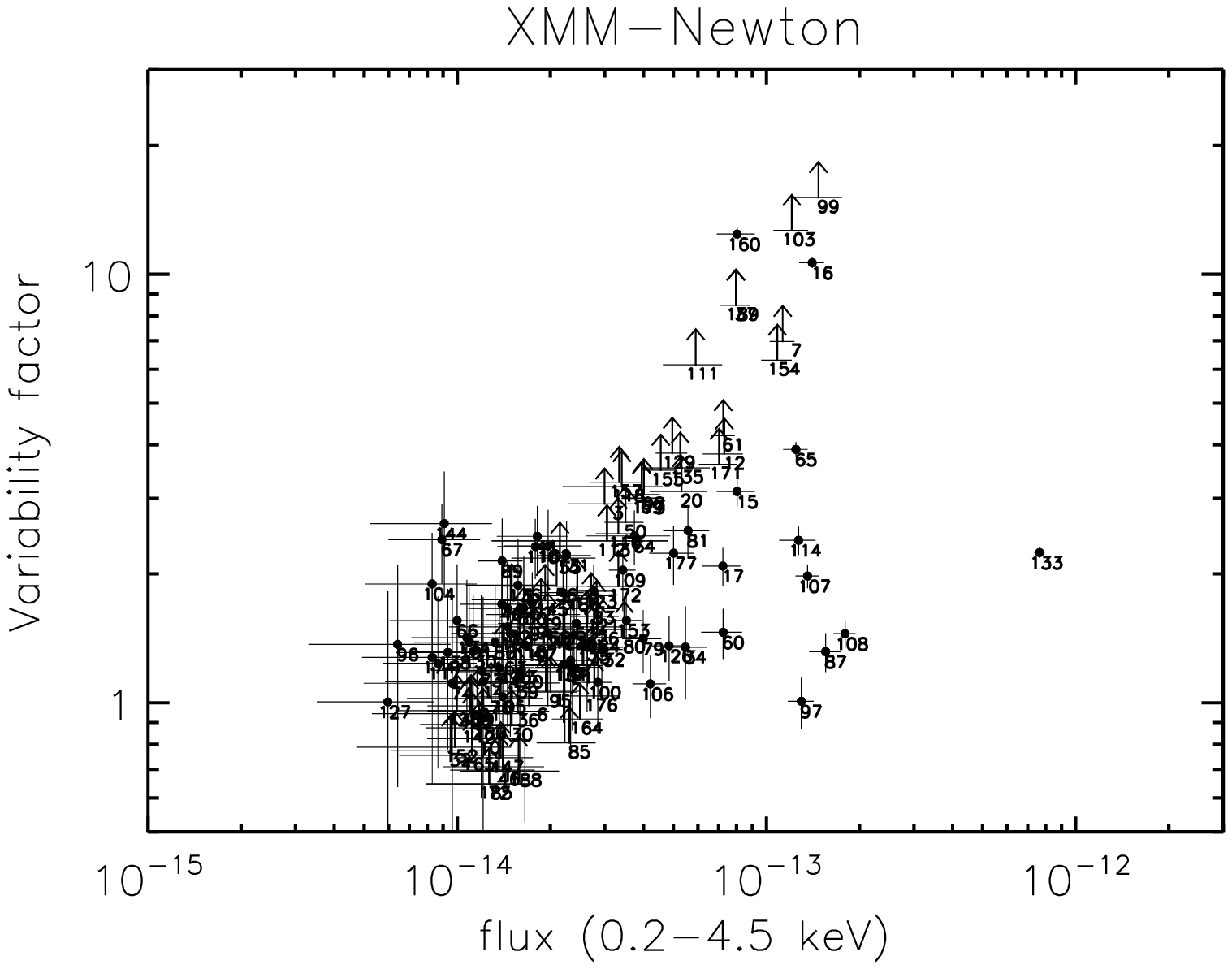}
\includegraphics[bb= 96 375 555 716, clip, width=7cm]{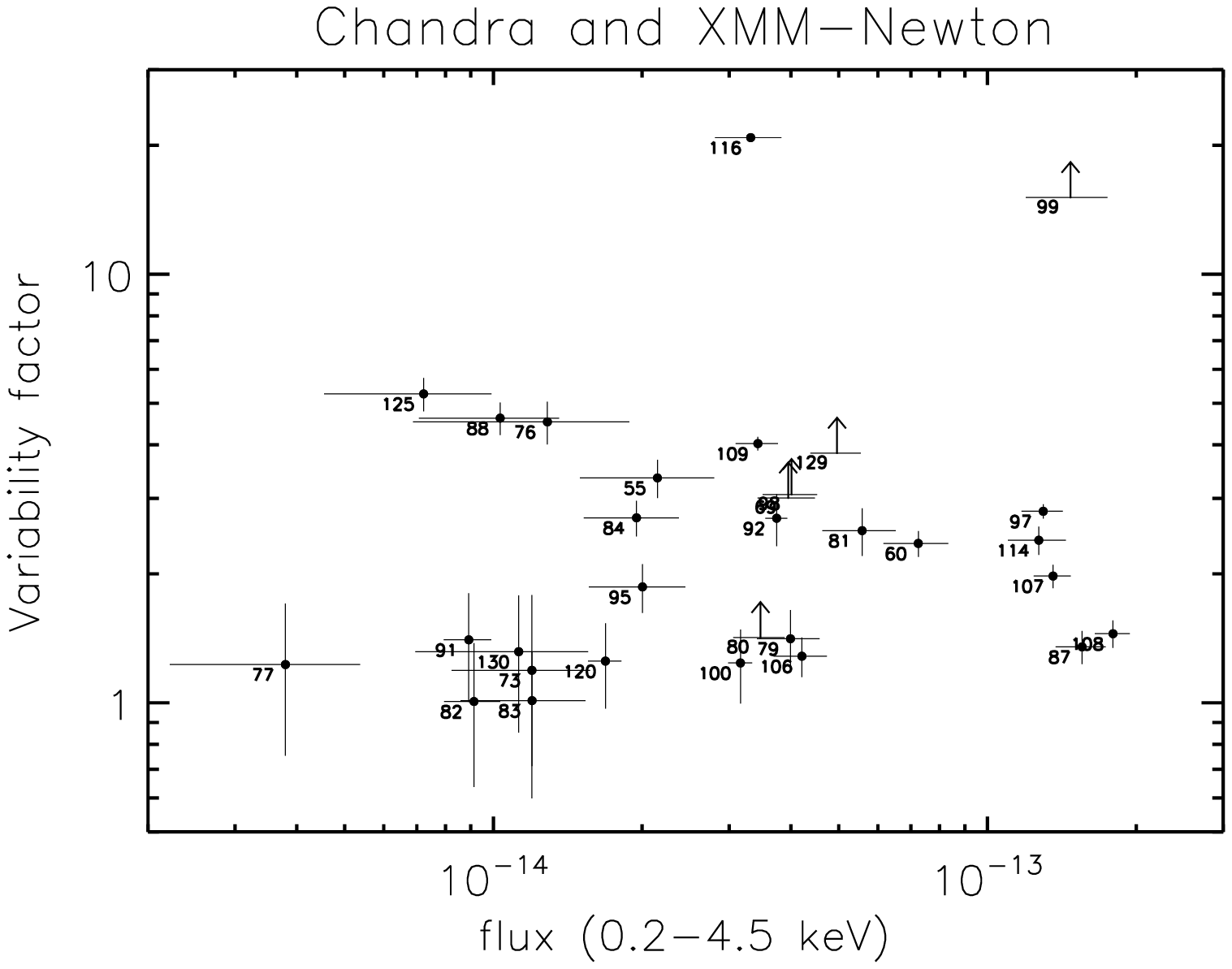}
\includegraphics[bb= 96 375 555 700, clip, width=7cm]{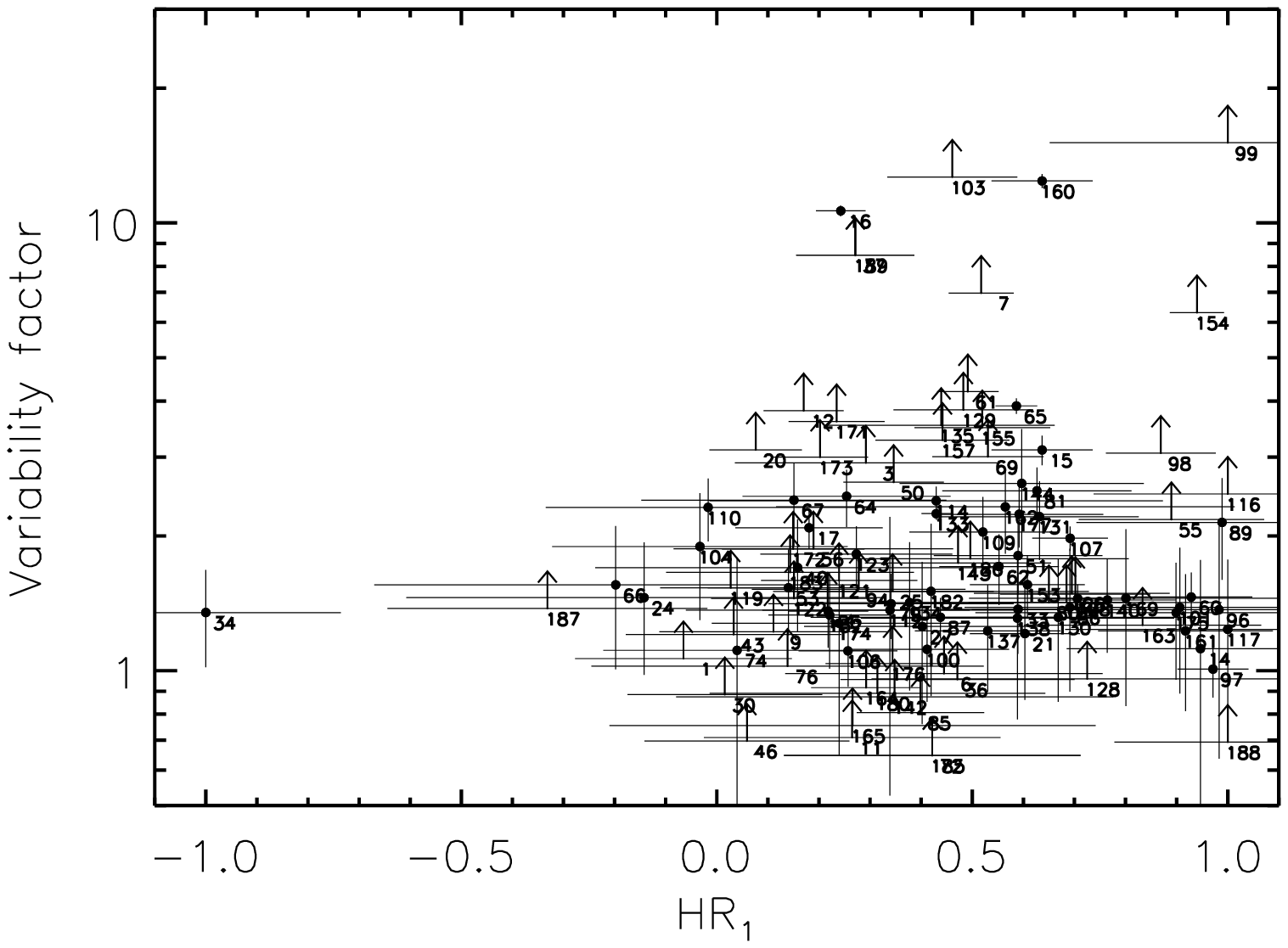}
\includegraphics[bb= 96 375 555 700, clip, width=7cm]{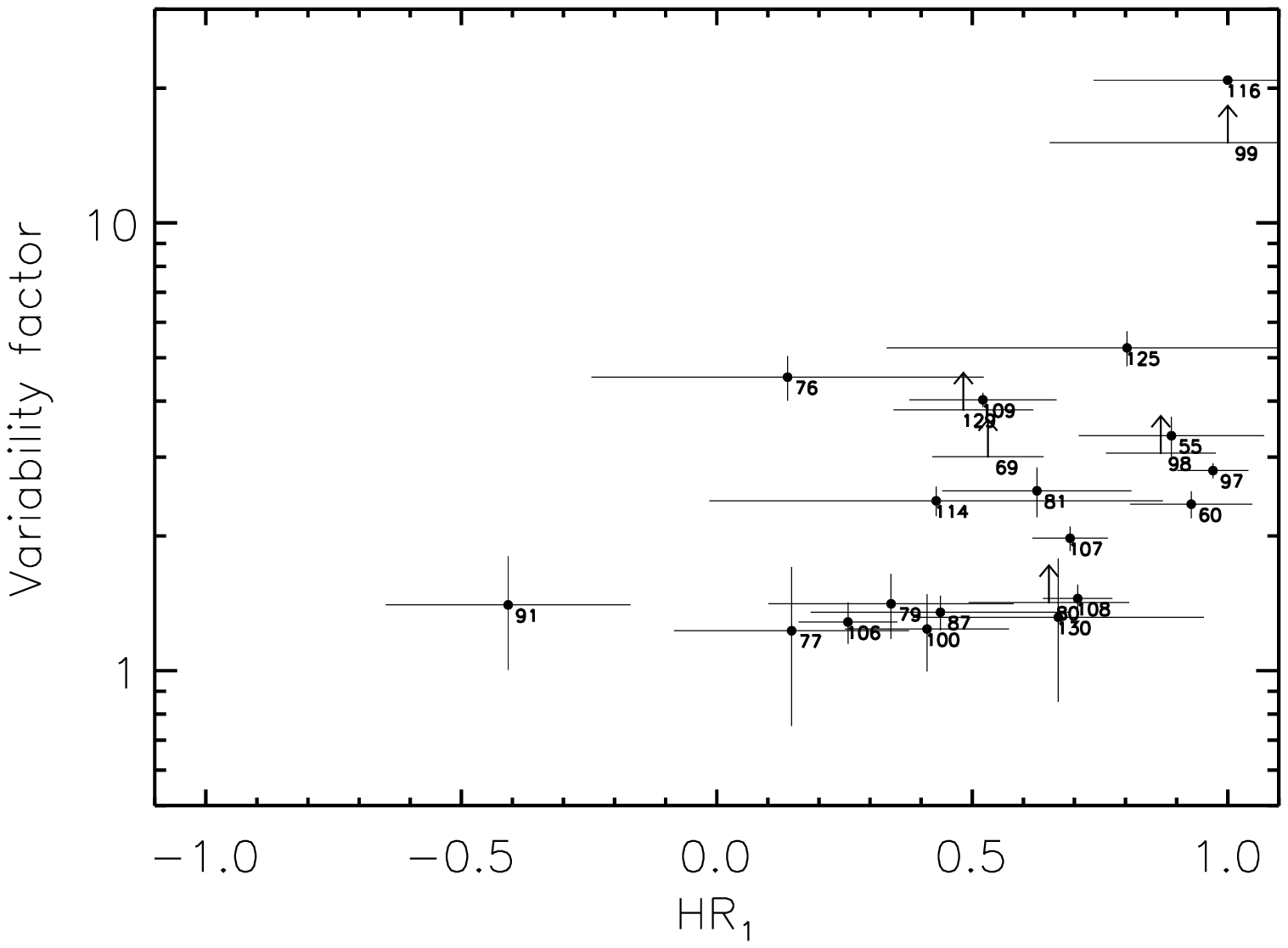}
\end{center}
\caption{\emph{Left panels:} variability factor as a function of the maximum flux 
(\emph{upper panel}) and hardness ratio HR$_1$ (\emph{bottom panel})
based on XMM-\emph{Newton} observations.  
\emph{Right panels:} variability factor as a function of the maximum flux (\emph{upper panel})
and hardness ratio HR$_1$ (\emph{bottom panel})
based on XMM-\emph{Newton} and \emph{Chandra} observations. 
The lower limits of the variability factors are marked as arrows.}
\label{fig. variab}
\end{figure*}

\begin{table}
\begin{center}
\caption{Count rate to energy conversion factors for thin and medium filters of the
EPIC instruments in the energy ranges R1-R5, assuming an absorbed powerlaw
with a photon index of 1.7 and the Galactic foreground absorption $3.69 \times 10^{20}$ cm$^{-2}$
in the direction of M\,83.}
\label{Tab. ecfs}
\resizebox{\columnwidth}{!}{
\begin{tabular}{l|cccccc}
\hline
\hline
Detector  & Filter &      R1      &      R2     &      R3    &       R4       &       R5      \\
EPIC      &        & \multicolumn{5}{c}{$(10^{-12})$ erg cm$^{-2}$ ct$^{-1}$}                 \\
\hline
PN        & Thin   &    0.8850    &     1.091   &    1.731   &      5.020     &     17.97     \\
          & Medium &    1.055     &     1.110   &    1.730   &      4.977     &     17.47     \\
MOS       & Medium &    6.796     &     5.582   &    5.189   &     13.71      &     69.63     \\
\hline
\end{tabular}
}
\end{center}
\end{table}

\section{Astrometrical corrections} 
\label{section Astrometrical corrections}

\subsection{Corrections between XMM-Newton observations}

We calculated the RA and Dec offsets of the three XMM-\emph{Newton} observations
using position of the sources
detected in at least two observations.
Sources were considered as detections in at least two different observations if their position
was closer than $3\times$ the combined statistical positional errors.
We calculated the offsets of observations 2 and 3 with respect 
to the reference observation 1 as the 
weighted mean of RA and Dec of all sources,
then recalculated all X-ray positions
correcting for the shifts relative to the observation 1.

\subsection{Correcting the position of the detected sources using X-ray and optical observations}

We also applied the cross-correlation procedure described above to
determine the systematic errors in the X-ray positions of the XMM-\emph{Newton} observations
by calculating the offsets in the X-ray positions of the XMM-\emph{Newton}
sources with respect to the X-ray sources observed by SW03 
with \emph{Chandra}. 
The offset between the XMM-\emph{Newton} and \emph{Chandra} positions 
(given as the weighted mean of RA and Dec in arcsec) is
$\Delta$RA$ = -1.22 \pm 0.16$, $\Delta$Dec$ = -0.72 \pm  0.16$.
We point out that SW03 corrected the \emph{Chandra} positions
using only the position of the infrared nucleus of M\,83
deduced from HST/WFPC2 observations.
Therefore, to obtain more accurate positions from possible optical counterparts,
we cross-correlated the XMM-\emph{Newton} list of sources
with the optical catalogue of the United States Naval Observatory 
USNO-B1 \citep{Monet03}.
For this calculation we were interested in associations
between X-ray sources and foreground stars. 
%
%
As discussed in Sect. \ref{sect. Foreground stars},
we classified five sources as foreground star candidates.
The offset between the X-ray positions and optical positions
corrected for proper motion
(given as the weighted mean of RA and Dec in arcsec) is
$\Delta$RA$ = -2.02 \pm 0.43$, $\Delta$Dec$ = -0.44 \pm  0.43$.
The measured offset in RA agrees with the expected
precision of the XMM-\emph{Newton} Attitude Measurement System \citep{Guainazzi12}.
We used these systematic offsets to correct the position of all detected sources.

\section{Analysis}
\label{sect. Analysis}

\begin{table*}[!]
\begin{center}
\caption{Variability factors ($V_{\rm f}$) with errors of sources 
observed in at least two XMM-\emph{Newton} observations
and in XMM-\emph{Newton} and \emph{Chandra} observations.
Table also includes maximum fluxes and errors 
in the energy range $0.2-4.5$ keV in units erg cm$^{-2}$ s$^{-1}$,
and the significance of the difference $S$.}
\label{Tab. variability}
\begin{tabular}{lcccc|lcccc}
\hline
\hline
\multicolumn{10}{c}{XMM-\emph{Newton}:}\\
\hline
Source &  flux max. &   $V_{\rm f}$    &    error $V_{\rm f}$  &      $S$    & Source &  flux max. &   $V_{\rm f}$    &    error $V_{\rm f}$  &      $S$    \\
\hline
{ } { } 2 &$(3.05 \pm 1.76)\times 10^{-14}$ &  3.0  &  0.6  &  1.2 &     104 &$(8.29 \pm 3.25)\times 10^{-15}$ &  1.9  &  0.6  &  1.0 \\
{ } { } 3 &$(3.00 \pm 0.71)\times 10^{-14}$ &  3.1  &  0.2  &  2.9 &     107 &$(1.36 \pm 0.12)\times 10^{-13}$ &  1.97 &  0.12 &  5.1 \\
{ } { } 4 &$(1.65 \pm 0.60)\times 10^{-14}$ &  1.7  &  0.4  &  1.2 &     108 &$(1.79 \pm 0.15)\times 10^{-13}$ &  1.44 &  0.10 &  3.3 \\
{ } { } 7 &$(1.13 \pm 0.10)\times 10^{-13}$ &  7.16 &  0.09 &  9.4 &     109 &$(3.43 \pm 0.34)\times 10^{-14}$ &  2.0  &  0.4  &  2.4 \\
{ } { } 9 &$(2.22 \pm 0.71)\times 10^{-14}$ &  1.5  &  0.3  &  1.1 &     110 &$(1.79 \pm 0.44)\times 10^{-14}$ &  2.3  &  0.4  &  2.1 \\
{ } 12 &$(7.31 \pm 1.08)\times 10^{-14}$ &  3.95 &  0.14 &  5.0 &     111 &$(5.91 \pm 1.28)\times 10^{-14}$ &  6.4  &  0.2  &  3.9 \\
{ } 15 &$(8.04 \pm 1.13)\times 10^{-14}$ &  3.1  &  0.2  &  4.4 &     113 &$(3.05 \pm 1.76)\times 10^{-14}$ &  3.0  &  0.6  &  1.2 \\
{ } 16 &$(1.41 \pm 0.13)\times 10^{-13}$ & 10.6  &  0.3  &  9.5 &     114 &$(1.27 \pm 0.17)\times 10^{-13}$ &  2.39 &  0.18 &  4.1 \\
{ } 17 &$(7.23 \pm 1.01)\times 10^{-14}$ &  2.1  &  0.2  &  3.3 &     116 &$(3.32 \pm 0.51)\times 10^{-14}$ &  2.63 &  0.15 &  4.0 \\
{ } 19 &$(1.86 \pm 0.63)\times 10^{-14}$ &  1.9  &  0.3  &  1.4 &     118 &$(3.41 \pm 1.21)\times 10^{-14}$ &  3.6  &  0.4  &  2.0 \\
{ } 20 &$(5.31 \pm 1.12)\times 10^{-14}$ &  3.3  &  0.2  &  3.3 &     119 &$(1.64 \pm 0.29)\times 10^{-14}$ &  1.70 &  0.17 &  2.4 \\
{ } 25 &$(2.62 \pm 0.82)\times 10^{-14}$ &  1.8  &  0.3  &  1.4 &     121 &$(2.82 \pm 0.50)\times 10^{-14}$ &  1.76 &  0.17 &  2.4 \\
{ } 26 &$(2.85 \pm 0.87)\times 10^{-14}$ &  1.8  &  0.3  &  1.4 &     122 &$(2.21 \pm 0.38)\times 10^{-14}$ &  1.62 &  0.17 &  2.2 \\
{ } 33 &$(2.64 \pm 0.55)\times 10^{-14}$ &  1.4  &  0.3  &  1.1 &     123 &$(2.76 \pm 0.49)\times 10^{-14}$ &  1.8  &  0.3  &  2.1 \\
{ } 37 &$(7.96 \pm 0.89)\times 10^{-14}$ &  8.58 &  0.11 &  7.9 &     126 &$(4.84 \pm 0.62)\times 10^{-14}$ &  1.4  &  0.2  &  1.4 \\
{ } 40 &$(1.39 \pm 0.37)\times 10^{-14}$ &  1.7  &  0.5  &  1.2 &     129 &$(4.96 \pm 0.58)\times 10^{-14}$ &  3.94 &  0.11 &  6.4 \\
{ } 41 &$(1.82 \pm 0.47)\times 10^{-14}$ &  2.4  &  0.4  &  2.0 &     131 &$(2.26 \pm 0.45)\times 10^{-14}$ &  2.2  &  0.4  &  2.1 \\
{ } 50 &$(3.50 \pm 0.51)\times 10^{-14}$ &  2.77 &  0.14 &  4.4 &     133 &$(7.64 \pm 0.29)\times 10^{-13}$ &  2.24 &  0.06 & 12.4 \\
{ } 51 &$(2.20 \pm 0.36)\times 10^{-14}$ &  1.8  &  0.3  &  2.0 &     135 &$(5.26 \pm 1.29)\times 10^{-14}$ &  3.8  &  0.2  &  3.0 \\
{ } 53 &$(2.43 \pm 0.51)\times 10^{-14}$ &  1.5  &  0.3  &  1.3 &     136 &$(1.57 \pm 0.51)\times 10^{-14}$ &  1.9  &  0.5  &  1.2 \\
{ } 55 &$(2.15 \pm 0.65)\times 10^{-14}$ &  2.5  &  0.3  &  2.0 &     140 &$(2.39 \pm 0.48)\times 10^{-14}$ &  1.4  &  0.3  &  1.1 \\
{ } 56 &$(2.09 \pm 0.69)\times 10^{-14}$ &  2.2  &  0.3  &  1.7 &     143 &$(1.50 \pm 0.39)\times 10^{-14}$ &  2.0  &  0.3  &  1.9 \\
{ } 60 &$(7.24 \pm 1.09)\times 10^{-14}$ &  1.46 &  0.19 &  1.8 &     144 &$(9.08 \pm 3.86)\times 10^{-15}$ &  2.6  &  0.8  &  1.2 \\
{ } 61 &$(7.25 \pm 0.64)\times 10^{-14}$ &  4.28 &  0.08 &  8.6 &     145 &$(1.92 \pm 0.56)\times 10^{-14}$ &  2.0  &  0.3  &  1.7 \\
{ } 62 &$(1.75 \pm 0.40)\times 10^{-14}$ &  1.7  &  0.3  &  1.6 &     153 &$(3.52 \pm 0.43)\times 10^{-14}$ &  1.6  &  0.2  &  1.9 \\
{ } 64 &$(3.74 \pm 1.14)\times 10^{-14}$ &  2.5  &  0.4  &  1.9 &     154 &$(1.08 \pm 0.12)\times 10^{-13}$ &  6.41 &  0.11 &  7.5 \\
{ } 65 &$(1.25 \pm 0.11)\times 10^{-13}$ &  3.90 &  0.15 &  7.8 &     155 &$(4.55 \pm 0.58)\times 10^{-14}$ &  3.61 &  0.12 &  5.7 \\
{ } 67 &$(8.92 \pm 2.93)\times 10^{-15}$ &  2.4  &  0.5  &  1.6 &     157 &$(3.34 \pm 0.66)\times 10^{-14}$ &  3.46 &  0.19 &  3.6 \\
{ } 69 &$(3.95 \pm 0.52)\times 10^{-14}$ &  3.13 &  0.13 &  5.1 &     160 &$(8.04 \pm 1.13)\times 10^{-14}$ & 12.4  &  0.4  &  6.4 \\
{ } 75 &$(1.46 \pm 0.36)\times 10^{-14}$ &  1.7  &  0.5  &  1.1 &     162 &$(1.97 \pm 0.56)\times 10^{-14}$ &  2.3  &  0.5  &  1.7 \\
{ } 79 &$(3.99 \pm 0.58)\times 10^{-14}$ &  1.4  &  0.2  &  1.5 &     163 &$(1.40 \pm 0.50)\times 10^{-14}$ &  1.6  &  0.4  &  1.1 \\
{ } 80 &$(3.47 \pm 0.41)\times 10^{-14}$ &  1.53 &  0.11 &  2.9 &     166 &$(1.96 \pm 0.39)\times 10^{-14}$ &  1.68 &  0.19 &  2.0 \\
{ } 81 &$(5.58 \pm 0.94)\times 10^{-14}$ &  2.5  &  0.3  &  3.0 &     171 &$(7.03 \pm 0.98)\times 10^{-14}$ &  3.74 &  0.13 &  5.3 \\
{ } 87 &$(1.55 \pm 0.18)\times 10^{-13}$ &  1.31 &  0.13 &  1.9 &     172 &$(3.32 \pm 0.68)\times 10^{-14}$ &  2.1  &  0.2  &  2.5 \\
{ } 89 &$(1.40 \pm 0.23)\times 10^{-14}$ &  2.1  &  0.5  &  1.8 &     173 &$(3.95 \pm 0.55)\times 10^{-14}$ &  3.13 &  0.13 &  4.9 \\
{ } 90 &$(1.34 \pm 0.04)\times 10^{-12}$ &  1.35 &  0.03 &  8.1 &     177 &$(5.01 \pm 0.83)\times 10^{-14}$ &  2.2  &  0.4  &  2.6 \\
{ } 93 &$(2.78 \pm 0.17)\times 10^{-13}$ & 16.42 &  0.06 & 15.6 &     183 &$(2.71 \pm 0.60)\times 10^{-14}$ &  1.9  &  0.2  &  2.1 \\
{ } 94 &$(1.73 \pm 0.59)\times 10^{-14}$ &  1.9  &  0.3  &  1.4 &     184 &$(2.83 \pm 0.60)\times 10^{-14}$ &  1.6  &  0.2  &  1.8 \\
{ } 98 &$(4.01 \pm 0.51)\times 10^{-14}$ &  3.18 &  0.12 &  5.4 &     186 &$(2.44 \pm 0.74)\times 10^{-14}$ &  2.1  &  0.3  &  1.7 \\
{ } 99 &$(1.47 \pm 0.28)\times 10^{-13}$ & 15.29 &  0.18 &  5.0 &     187 &$(1.78 \pm 0.63)\times 10^{-14}$ &  1.7  &  0.4  &  1.2 \\
103 &$(1.21 \pm 0.15)\times 10^{-13}$ & 12.78 &  0.12 &  7.2 &     189 &$(7.96 \pm 0.89)\times 10^{-14}$ &  8.58 &  0.11 &  7.9 \\
\hline
\multicolumn{10}{c}{\emph{Chandra} and XMM-\emph{Newton}:}\\
\hline
Source&  flux max.  &  $V_{\rm f}$    &    error $V_{\rm f}$  &      $S$    & Source &  flux max.  &  $V_{\rm f}$    &    error $V_{\rm f}$  &      $S$    \\
\hline
 55 &$(2.15 \pm 0.65)\times 10^{-14}$ & 3.3  & 0.3  & 2.3 &  { } 97 &$(1.30 \pm 0.12)\times 10^{-13}$ &  2.79 & 0.10 & 6.6 \\
 60 &$(7.24 \pm 1.09)\times 10^{-14}$ & 2.35 & 0.16 & 3.8 &  { } 98 &$(4.01 \pm 0.51)\times 10^{-14}$ &  3.18 & 0.12 & 5.4 \\
 69 &$(3.95 \pm 0.52)\times 10^{-14}$ & 3.13 & 0.13 & 5.1 &  { } 99 &$(1.47 \pm 0.28)\times 10^{-13}$ & 15.29 & 0.18 & 5.0 \\
 76 &$(1.29 \pm 0.60)\times 10^{-14}$ & 4.5  & 0.5  & 1.7 &     106 &$(4.21 \pm 0.52)\times 10^{-14}$ &  1.28 & 0.13 & 1.7 \\
 79 &$(3.99 \pm 0.58)\times 10^{-14}$ & 1.4  & 0.2  & 1.5 &     107 &$(1.36 \pm 0.17)\times 10^{-13}$ &  1.97 & 0.12 & 5.1 \\
 80 &$(3.47 \pm 0.41)\times 10^{-14}$ & 1.53 & 0.11 & 2.9 &     108 &$(1.79 \pm 0.15)\times 10^{-13}$ &  1.44 & 0.10 & 3.3 \\
 81 &$(5.58 \pm 0.94)\times 10^{-14}$ & 2.5  & 0.3  & 3.0 &     109 &$(3.43 \pm 0.34)\times 10^{-14}$ &  4.02 & 0.14 & 7.4 \\
 84 &$(1.95 \pm 0.42)\times 10^{-14}$ & 2.7  & 0.3  & 2.8 &     114 &$(1.27 \pm 0.17)\times 10^{-13}$ &  2.39 & 0.18 & 4.1 \\
 87 &$(1.55 \pm 0.18)\times 10^{-13}$ & 1.35 & 0.11 & 2.2 &     116 &$(3.32 \pm 0.51)\times 10^{-14}$ & 20.8  & 0.3  & 6.2 \\
 88 &$(1.03 \pm 0.33)\times 10^{-14}$ & 4.6  & 0.4  & 2.5 &     125 &$(7.23 \pm 0.27)\times 10^{-15}$ &  5.3  & 0.5  & 2.2 \\
 92 &$(3.74 \pm 0.19)\times 10^{-14}$ & 2.7  & 0.4  & 4.3 &     129 &$(4.96 \pm 0.58)\times 10^{-14}$ &  3.94 & 0.11 & 6.4 \\
 95 &$(2.00 \pm 0.44)\times 10^{-14}$ & 1.9  & 0.2  & 2.0 &         &                                &       &      &      \\
\hline
\end{tabular}
\end{center}
\end{table*}

\subsection{Variability of the sources}
\label{sect. Variability of the sources}

\subsubsection{Short-term variability}

For each XMM-\emph{Newton} observation, 
we searched for pulsations of the brightest sources (counts\,$\gtrsim 200$)
on time scales between $\sim 4$ s and the time duration of each observation.
After extracting the event files, we applied both a Fourier transform and a $Z^2_n$ analysis
\citep{Buccheri83}. No statistically significant variability 
from the analysed sources was detected.

\subsubsection{Long-term variability}

To study the long-term time variability of sources observable 
at least in two different observations,
we calculated the average flux (or the $3\sigma$ upper limit 
in case of non-detection) at the source position in each observation.
We considered fluxes in the $0.2-4.5$ keV energy band because, as \citet{Pietsch04} noted,
the band $4.5-12$ keV has a lower sensitivity and is contaminated by hard background.
We calculated the fluxes with the energy conversion factors (ECFs) reported in Table \ref{Tab. ecfs}.
Then, we searched for variable sources by comparing their fluxes
(or upper limits) in different observations.
We measured the X-ray variability of each source by its variability factor
$V_{\rm f} = F_{\rm max}/F_{\rm min}$, 
where $F_{\rm max}$ and $F_{\rm min}$ are the maximum and minimum (or upper-limit) fluxes.
To estimate the significance of the variability between different observations,
we calculated the significance parameter $S=(F_{\rm max} - F_{\rm min})/\sqrt{\sigma_{\rm max}^2 + \sigma_{\rm min}^2}$,
where $\sigma_{\rm max}$ and $\sigma_{\rm min}$ are the errors 
of the maximum and minimum flux \citep{Primini93}.

We also studied the X-ray variability considering the \emph{Chandra} observation of M\,83.
We converted the \emph{Chandra} counts ($0.3-8$ keV) of SW03 to $0.2-4.5$ keV fluxes
with the conversion factor calculated by SW03 and
the distance of M\,83 ($d=4.5$ Mpc) assumed in this work.
The conversion factor $CF=8\times 10^{37}/300$ erg s$^{-1}$ counts$^{-1}$
was calculated by SW03 assuming an absorbed powerlaw spectrum
with $\Gamma=1.7$, $N_{\rm H}=10^{21}$ cm$^{-2}$, and a distance of $3.7$ Mpc.
For each \emph{Chandra} source,
we obtained the flux in the energy range $0.2-4.5$ keV
correcting the luminosity $L_{\rm 0.3-8\ keV} =$ counts $\times CF$ 
by the absorption column density,
the galaxy distance, and the energy range. 
The results are reported in Table \ref{Tab. variability}.

Fig. \ref{fig. variab} shows the variability factor 
plotted versus the maximum detected flux
and the hardness ratio (R2-R1)/(R1+R2) 
(see section \ref{sect. Hardness ratios diagrams}) for each source.
The left column shows the variability factors calculated for sources
observed in at least two XMM-\emph{Newton} observations.
The right column shows the variability factors calculated for sources
observed with \emph{Chandra} and in at least one XMM-\emph{Newton} observation.

Applying a variability significance threshold of $S=3$,
we found 35 variable sources.
Like XRBs and AGNs, SSSs can show high variability, 
and because of their soft spectrum (see section \ref{sect. super-soft sources}),
they can be distinguished from the other sources: 
in Fig. \ref{fig. variab} (lower panels),
SSSs candidates should appear on the left-hand side, 
while XRBs (characterized by a much harder spectrum)
are expected to appear on the right-hand side.
%
%

\subsection{Spectral analysis}
\label{sect. spectral analysis}

We extracted the X-ray spectra of sources with $\gtrsim 300$ counts
in the energy range $0.2-12$ keV.
For each source, we fitted all three EPIC spectra simultaneously with different models:
powerlaw, disc-blackbody, thermal plasma model (APEC \citealt{Smith01}), and blackbody,
using XSPEC (ver. 12.7.0, \citealt{Arnaud96}).
For the absorption we used the PHABS model.

A good fit with one of the above-mentioned spectral models
can be used to classify the sources 
into one of the following classes of sources:
\begin{itemize}
\item X-ray binaries; 
\item supernova remnants; 
\item super-soft sources. 
\end{itemize}

In total, we fitted the spectra of  12 sources 
(see section \ref{sect. Discussion of classification and identification of the XMM-Newton sources}).  

For sources that are not bright enough for spectral modelling,
we only calculated their hardness ratios,
as described in section \ref{sect. Hardness ratios diagrams}.

\begin{figure}
\begin{center}
\includegraphics[bb=75 371 560 825,clip,width=8cm]{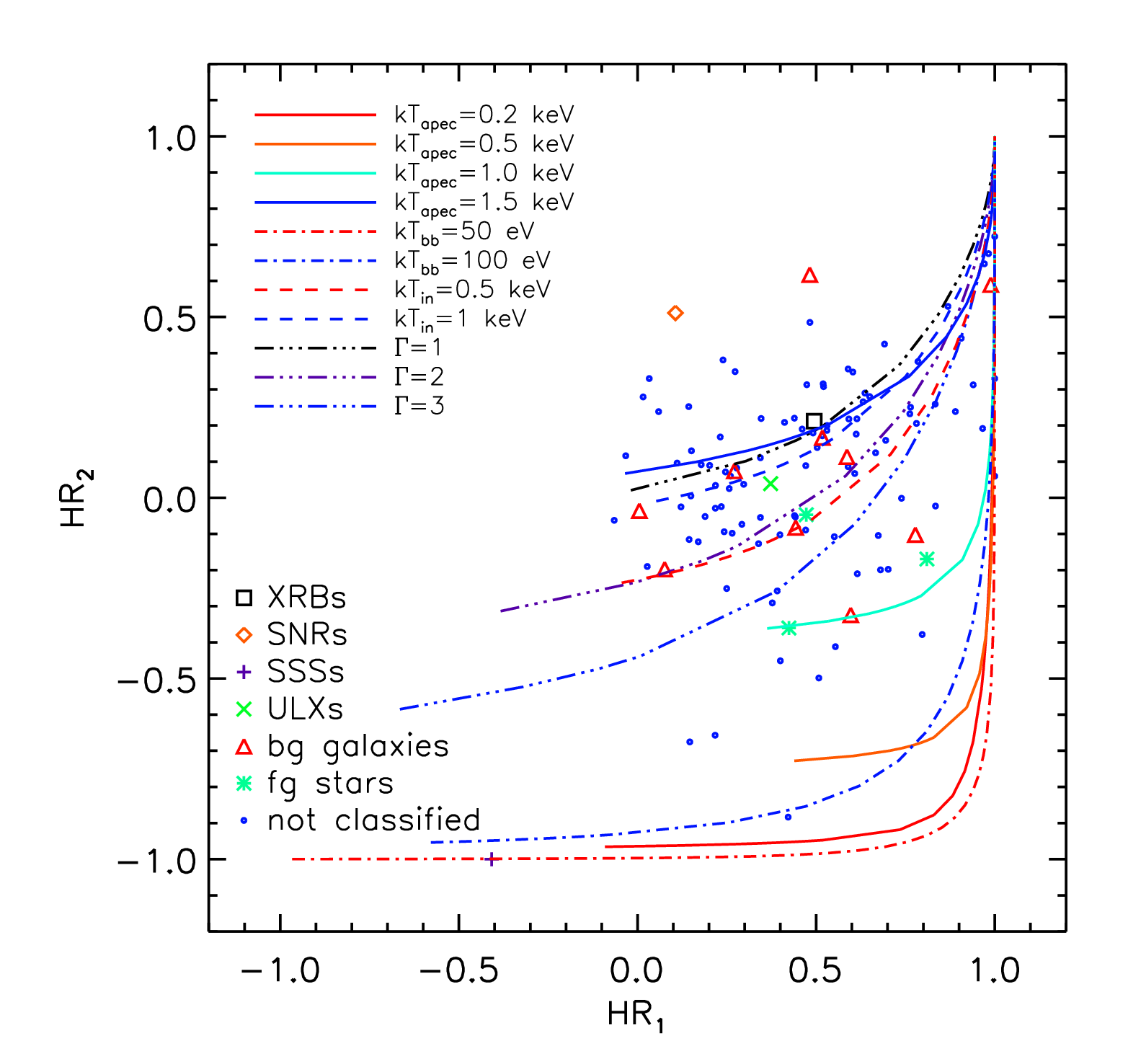}
\includegraphics[bb=75 371 560 825,clip,width=8cm]{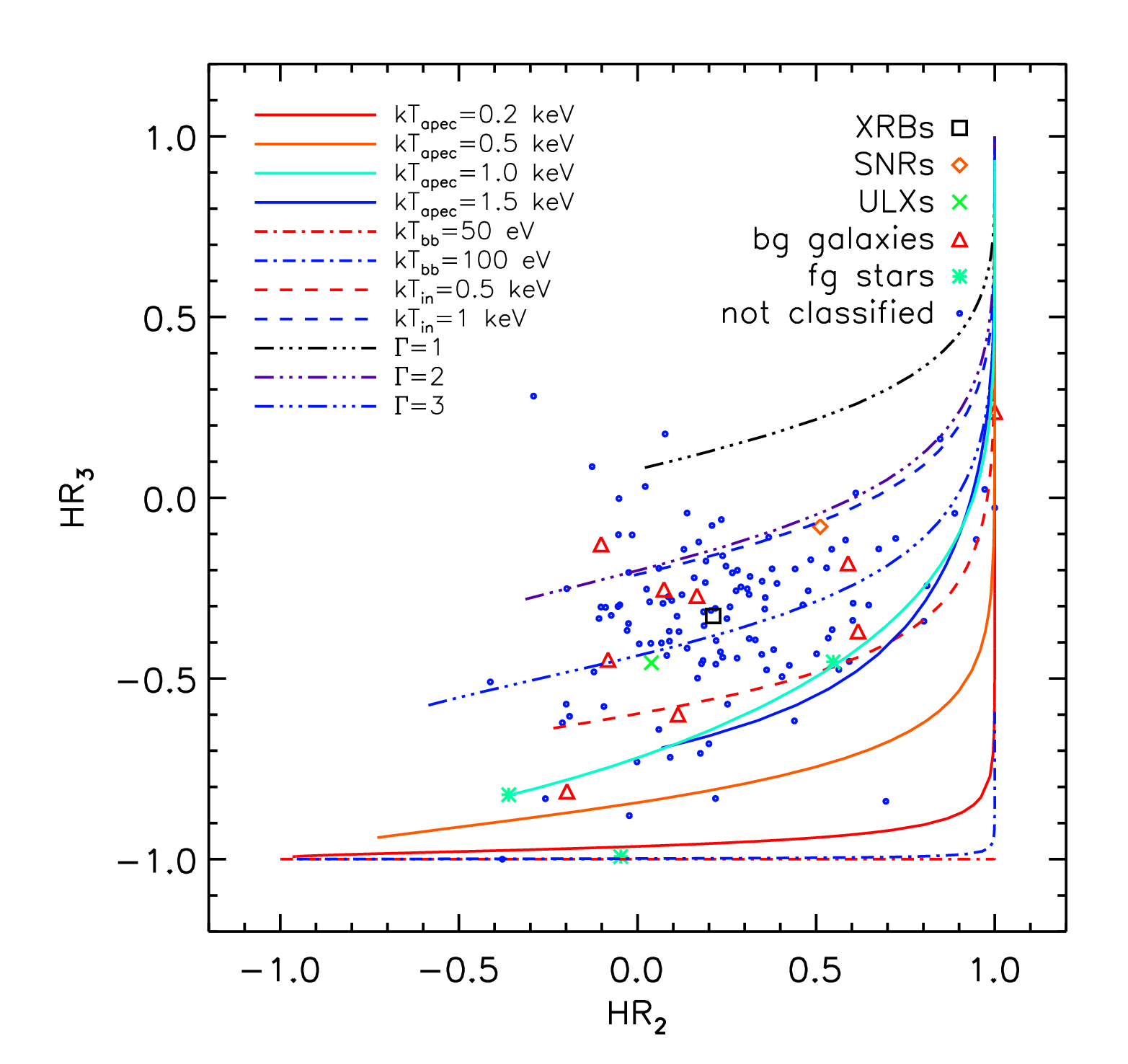}
\includegraphics[bb=75 371 560 825,clip,width=8cm]{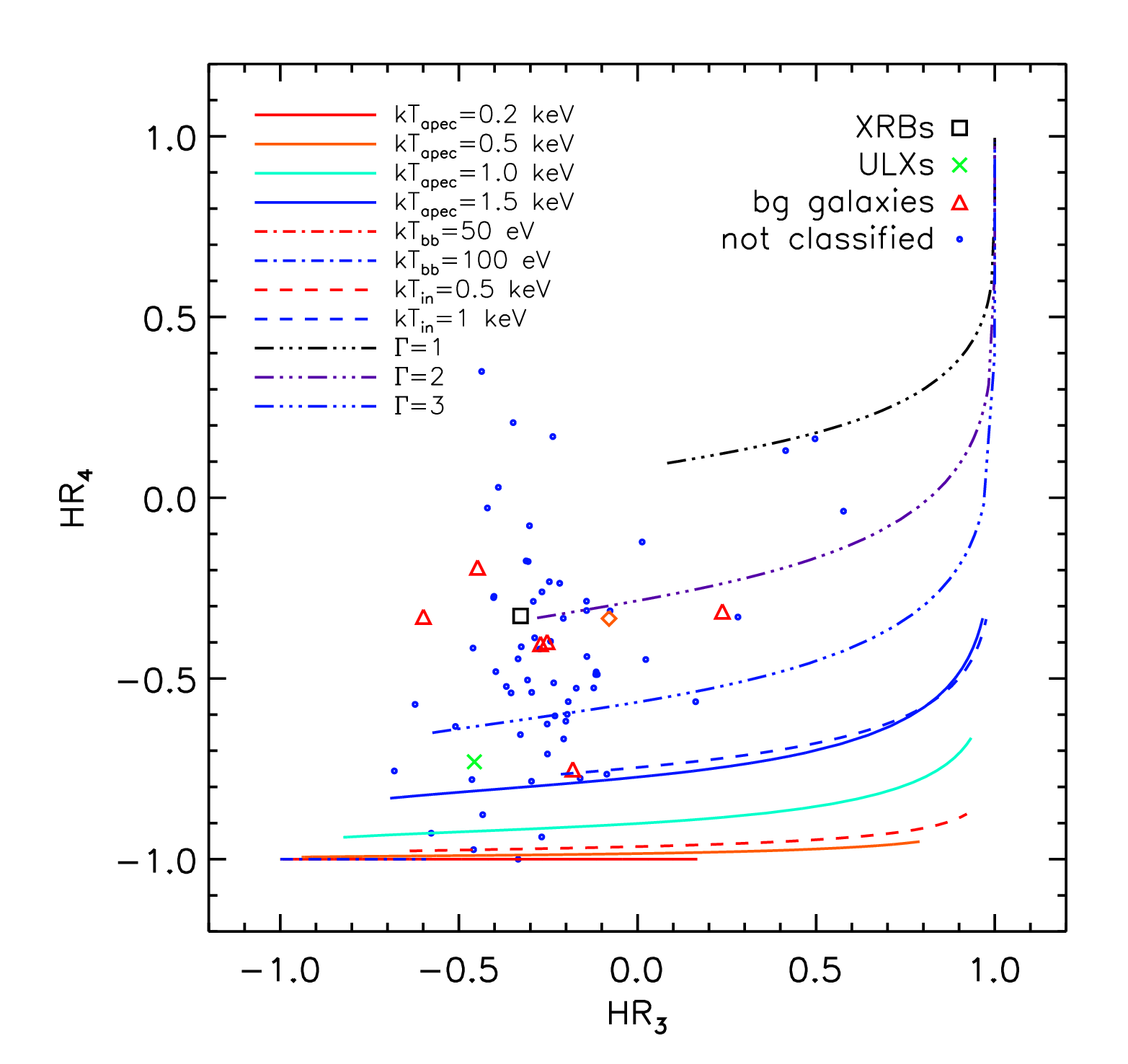}
\end{center}
\caption{Hardness-ratio diagrams of sources with error-bars smaller than 0.3.
Black squares are sources classified as XRBs (section \ref{sect. X-ray Binaries}), 
orange diamonds are SNRs (section \ref{sect. supernova remnants}),
violet plus signs are SSSs (section \ref{sect. super-soft sources}),
green crosses are ULXs (section \ref{sect. Observation of an ULX}),
cyan stars are foreground stars (section \ref{sect. Foreground stars}),
red triangles are background sources (section \ref{sect. Background objects}), 
and blue circles are sources not classified.
The lines are the hardness ratios calculated for different spectral models and column densities,
as described in section \ref{sect. Hardness ratios diagrams}.
}
\label{fig. col-col diagrams}
\end{figure}

\subsection{Hardness-ratio diagrams}
\label{sect. Hardness ratios diagrams}

We used the hardness-ratio diagrams to separate different classes of
sources according to their X-ray properties.
They are especially helpful for sources that are too faint, 
for which spectral fitting is not possible.
For each source, we computed four hardness ratios, defined as
\begin{equation} \label{eq. hrs}
HR_i = \frac{R_{i+1} - R_i}{R_{i+1} + R_i} \mbox{ \ \ for \emph{i} = 1, ... , 4,}
\end{equation}
where $R_i$ are the net source counts in five energy bands.
To obtain the best statistics we combined the hardness-ratios 
of all three instruments.

When a source was detected in more than one observation,
we considered the observation with the highest number of counts.
Some sources can exhibit different spectral states
(which can be correlated with the X-ray flux), 
resulting in hardness-ratio changes between different observations (see e.g. \citealt{Done07}).
Therefore, for some of these sources we only considered a state
by adopting the highest number of counts when determining the hardness ratio.
This approach allowed us to obtain the hardness ratios with small uncertainties
for bright sources in their bright states.
However, one has to be aware that if a source changes its state,
the hardness-ratio may change as well.
For fainter sources (with hardness ratio uncertainties $\gtrsim 0.2$), 
the hardness ratios are not sensitive to changes
of the state of the source within uncertainties.
The hardness ratios calculated for each source are reported in Table \ref{Tab. source list}.

Fig. \ref{fig. col-col diagrams} shows the hardness ratios 
of sources with errors smaller than $0.3$, 
detected in the field of view of M\,83.
We plotted sources 
classified as XRBs, SNRs, SSSs, ultra-luminous X-ray sources (ULXs), 
foreground stars, and background objects (see section \ref{sect. Source classification})
with different symbols.
On the same plot we also overlaid grids of hardness ratios
calculated for different spectral models:
three absorbed powerlaws with photon-index $\Gamma=1$, $2$, $3$
(XRBs in hard state),
two absorbed disc-blackbody models with temperatures at the inner disc radius 
of $kT_{\rm in}=0.5$ and $1$ keV (XRBs in soft state),
four thermal plasma models APEC with temperatures
$kT_{\rm apec}=0.2$, $0.5$, $1$, $1.5$ keV (SNRs),
and two blackbody models with temperatures $kT_{\rm bb}=50$ and $100$ eV 
(SSSs, see section \ref{sect. super-soft sources}).
The column densities range from $N_{\rm H}=10^{20}$ cm$^{-2}$ to $N_{\rm H}=10^{24}$ cm$^{-2}$.

\section{Source classification}
\label{sect. Source classification}

We cross-correlated the list of sources 
observed with XMM-\emph{Newton}
with existing catalogues. 
For this purpose we used X-ray
(\citealt{Trinchieri85};  \citealt{Ehle98}; \citealt{Immler99}; 
SW03; \citealt{DiStefano03}),
optical (\citealt{Blair04}; \citealt{Dopita10}; \citealt{Jones04}; 
\citealt{Rumstay83}; USNO-B1, \citealt{Monet03}),
radio (\citealt{Maddox06}; \citealt{Cowan94}; \citealt{Condon98}),
and infrared (2MASS, \citealt{Skrutskie06}) catalogues.

We considered two sources as associated to each other 
if their positions were closer than the $3\times$
combined statistical errors.
The optical counterparts of several X-ray sources cannot be determined
uniquely. In such cases we assumed as counterpart the brightest optical
object within the error circle.
The cross-correlations are reported in Table \ref{Tab. source list classification}
in appendix \ref{sect. catalogue-table}.

We used the previous classifications in X-rays and other wavelengths
and the methods of classification described in sections 
\ref{sect. Variability of the sources} (X-ray variability), 
\ref{sect. spectral analysis} (spectral analysis), 
and \ref{sect. Hardness ratios diagrams} (hardness ratios),
to identify and classify sources as background objects,
foreground stars, XRBs, SNRs, SSSs, and ULXs.

In this section we describe the observational properties
for each class of sources and define the classification criteria.

\subsection{Foreground stars}
\label{sect. Foreground stars}

X-ray observations of nearby galaxies
are contaminated by foreground stars,
which have X-ray luminosities
ranging from $\sim 10^{26}$ to $\sim 10^{30}$ erg s$^{-1}$
for stars of spectral type F to M,
and $\sim 10^{29}$ to $\sim 10^{34}$ erg s$^{-1}$ for stars
of spectral types O and B (\citealt{Vaiana81};
\citealt{Rosner85}).
Stars of spectral classes F to M emit X-rays  
because of the intense magnetic fields that form a corona,
in which the plasma is heated to temperatures
of about $\sim 10^6-10^8$ K (e.g. \citealt{Guedel02}).
A mechanism proposed to explain the X-ray emission
from stars of spectral types O-B
is the formation of shocks in the coronal regions
due to the instability of the wind-driven mechanism 
(see \citealt{Puls08} and references therein).
In A-type stars, none of the above mechanisms for X-ray emission
can operate efficiently. Therefore, A-type stars are expected to be weak
X-ray sources \citep{Schroeder07} and only very few have
been observed in X-rays (see e.g. \citealt{Robrade10}; \citealt{Schroeder08}).

\begin{figure}
\begin{center}
\includegraphics[bb= 92 360 564 841, clip, width=9cm]{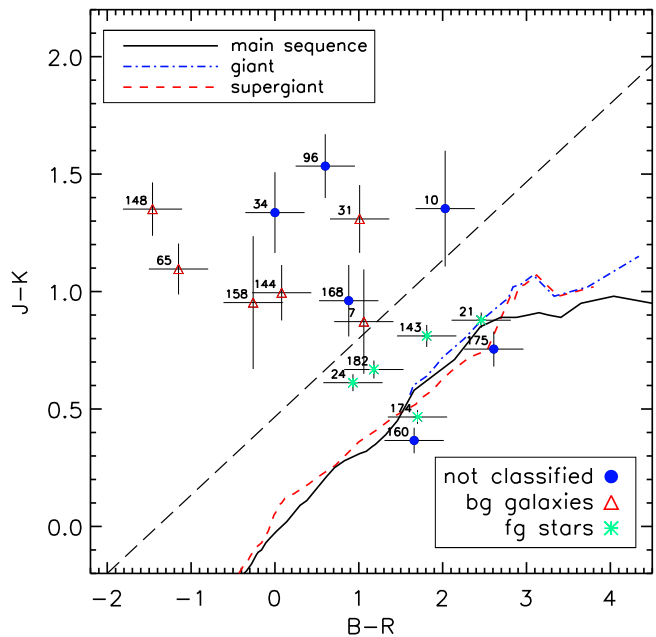}
\end{center}
\caption{Colour-colour diagram of XMM-\emph{Newton} sources with optical (USNO-B1) and infrared (2MASS) counterparts.
Sources located below the black dashed line are very likely foreground stars.}
\label{fig. b-r_j-k.ps}
\end{figure}

\begin{figure*}
\begin{center}
\includegraphics[bb=92 360 564 841, clip, width=9.1cm]{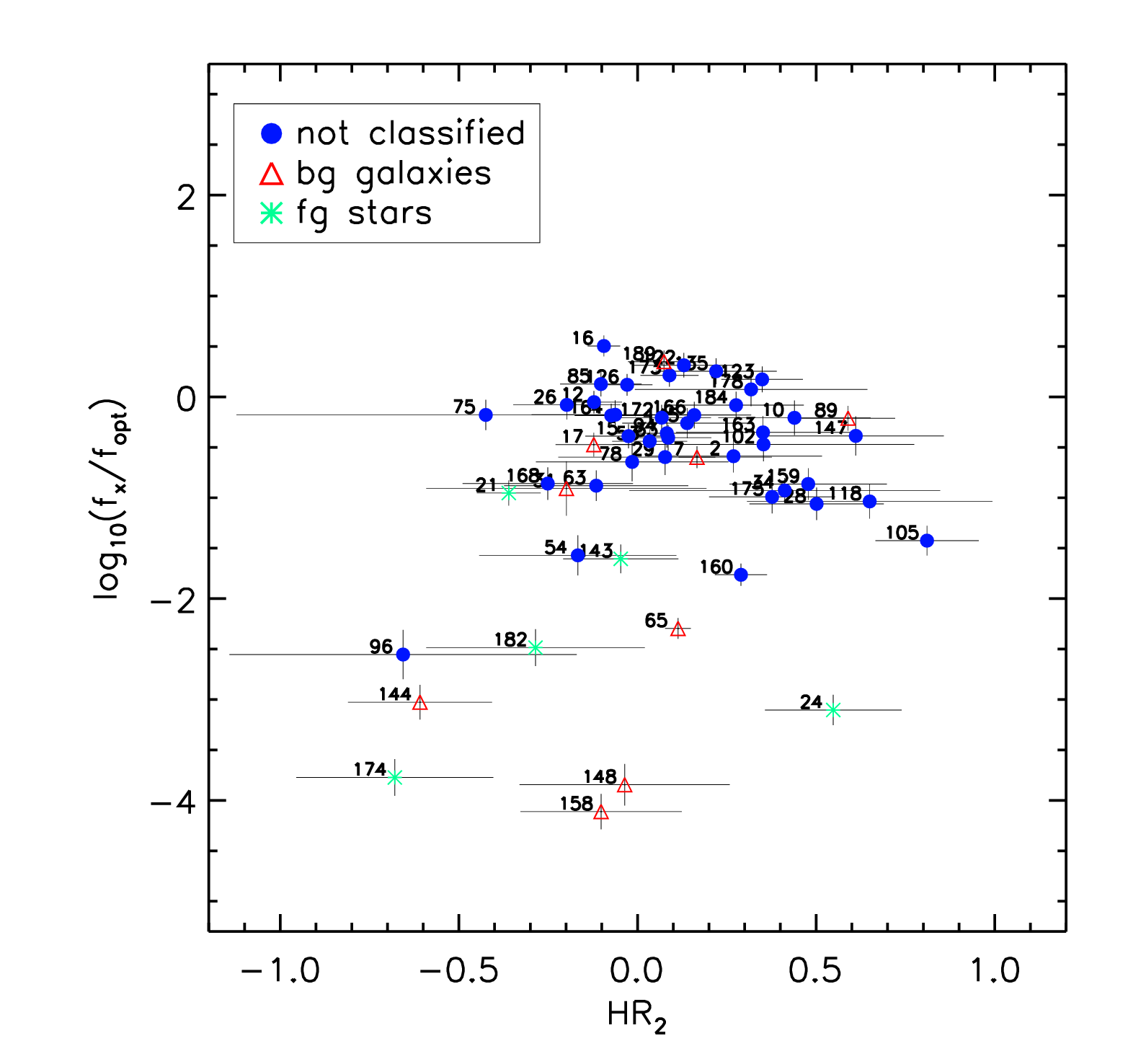}
\includegraphics[bb=92 360 564 841, clip, width=9.1cm]{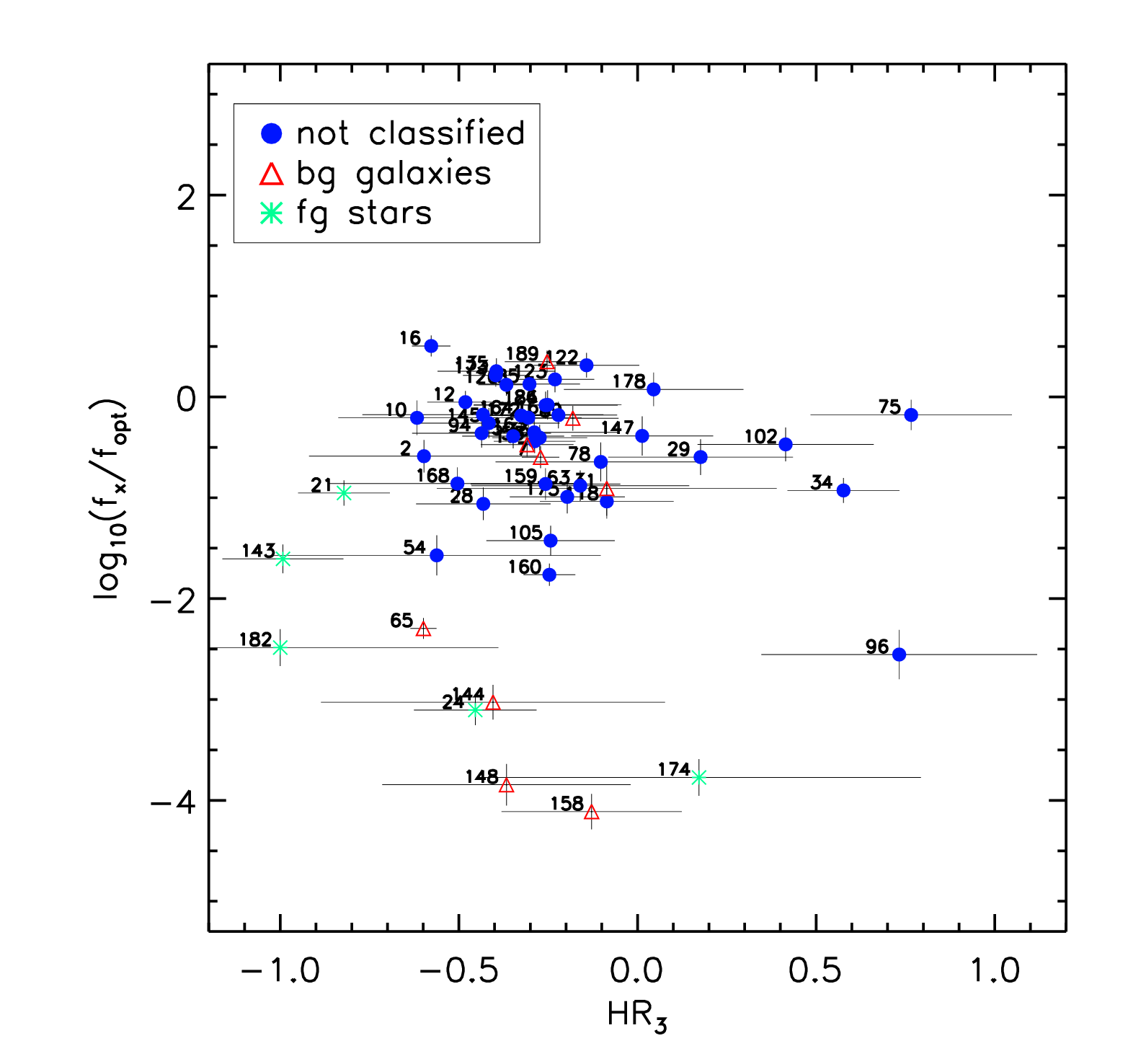}
\end{center}
\caption{Flux ratio $\log (f_{\rm x}/f_{\rm opt})$ over hardness ratios HR$_2$ and HR$_3$.}
\label{fig. log10-fx-fopt}
\end{figure*}

\begin{figure*}
\begin{center}
\includegraphics[bb= 92 360 564 841, clip, width=9cm]{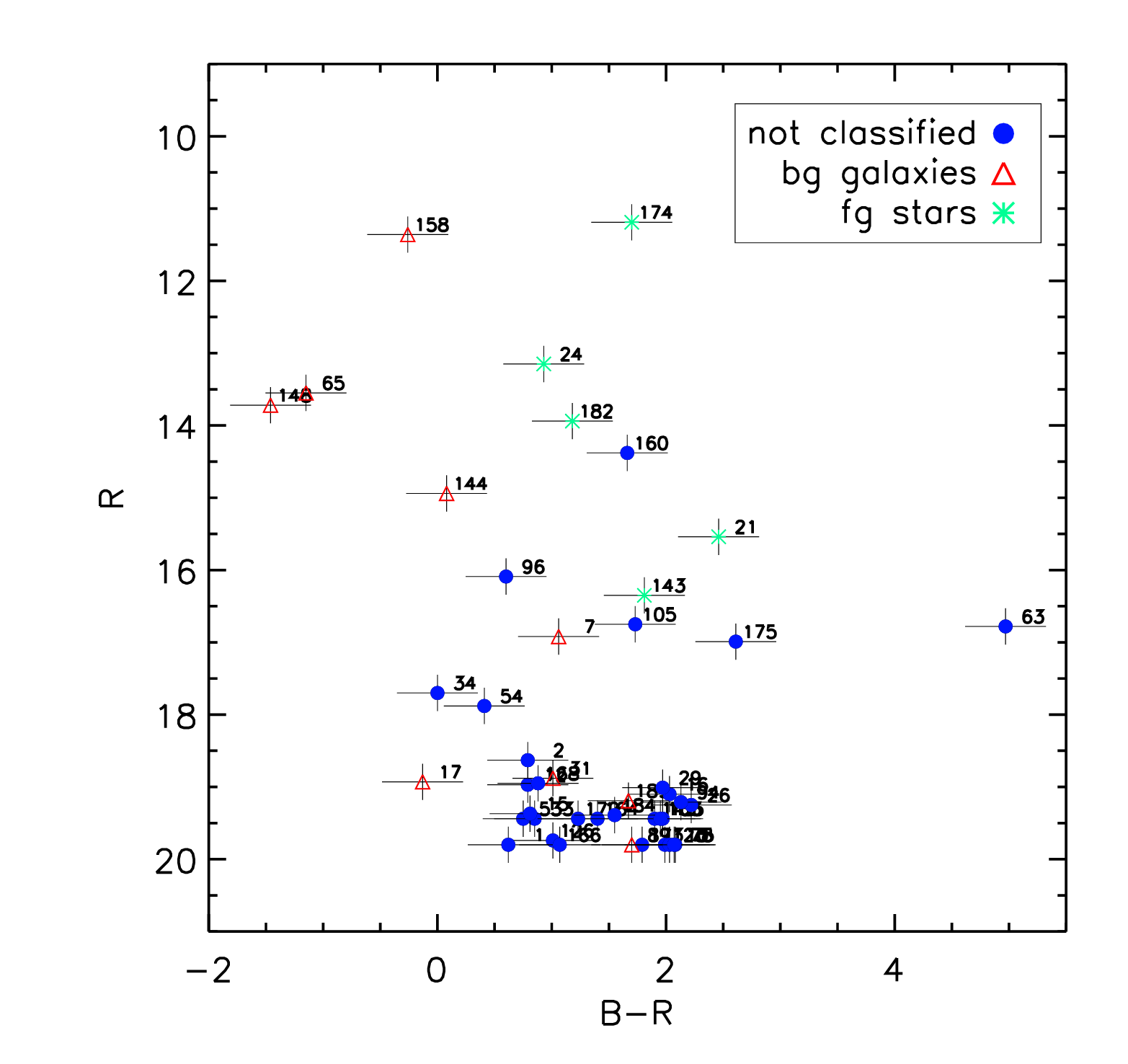}
\includegraphics[bb= 92 360 564 841, clip, width=9cm]{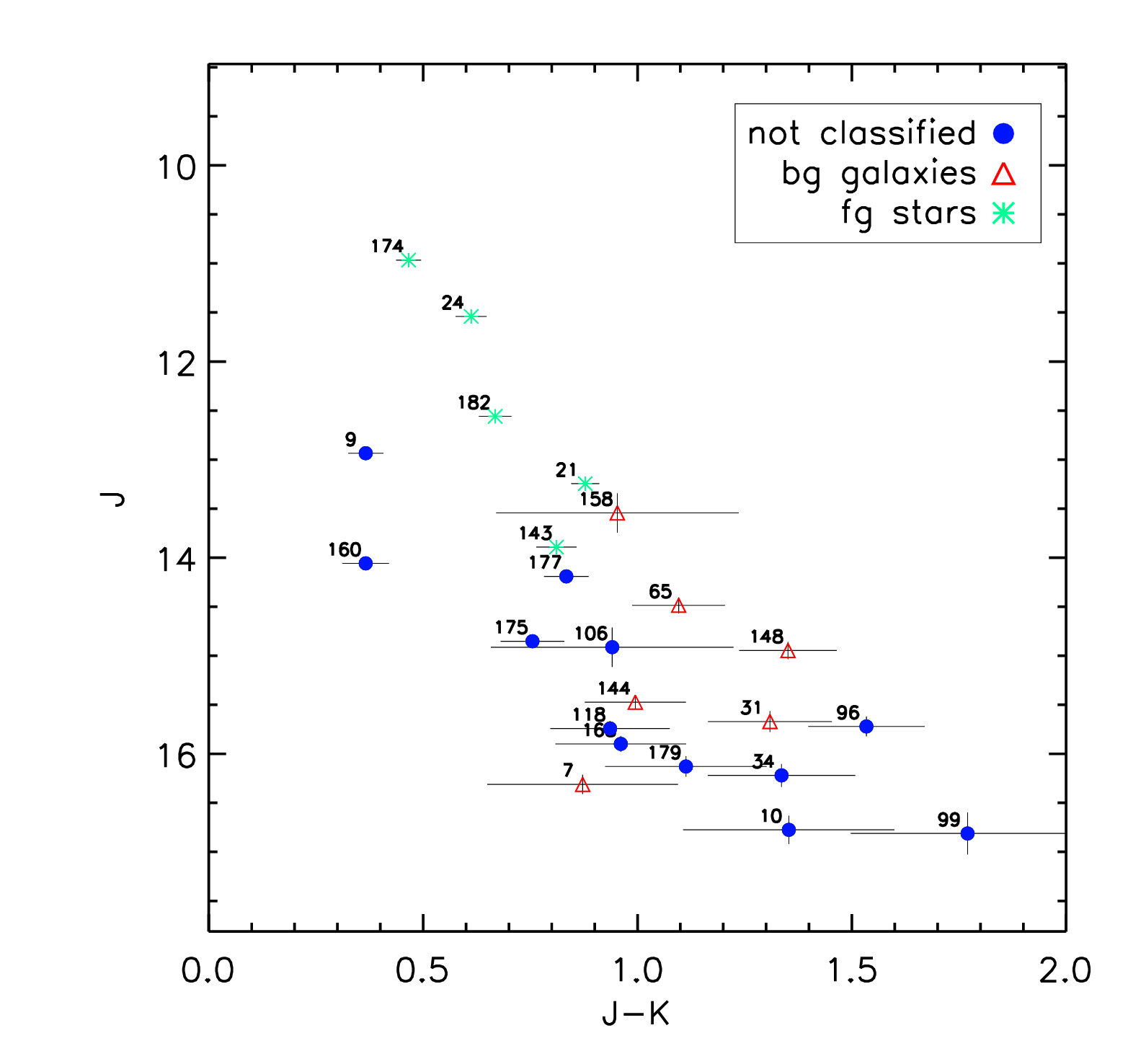}
\end{center}
\caption{Colour-magnitude diagrams of XMM-\emph{Newton} sources correlating with
sources in the USNO-B1 (\emph{left panel}) and 2MASS (\emph{right panel}) catalogues.}
\label{fig. B-R vs R}
\end{figure*}

The X-ray spectra of foreground stars
are relatively soft
and can be described by models
of optically thin plasma in collisional equilibrium
(e.g. \citealt{Raymond77}) with
temperatures ranging from $10^6$ to $10^7$ K.
A common method to distinguish stars from other X-ray sources is comparing the 
X-ray-to-optical flux ratio, as suggested by \citet{Maccacaro88}:
\begin{equation} \label{eq. Maccacaro}
\log_{\rm 10} (f_{\rm x}/f_{\rm opt}) = \log_{\rm 10}(f_{\rm x}) + \frac{m}{2.5} +5.37 \mbox{ ,}
\end{equation}
where $m$ is the visual magnitude $m_{\rm v}$.
In the USNO-B1 catalogue the red and blue magnitudes are given,
thus we assumed $m_{\rm v} \approx (m_{\rm red} + m_{\rm blue})/2$.
We used the blue magnitude $m_{\rm blue}$ as magnitude $m$
when the red magnitude was not available.

For each X-ray source with an optical counterpart,
we distinguished foreground stars from other sources
by plotting X-ray-to-optical flux ratios over
the hardness ratios $HR_2$ 
and $HR_3$ (Fig. \ref{fig. log10-fx-fopt}).
The X-ray-to-optical flux ratios and the hardness ratios 
differ significantly between different classes of sources.

The soft X-ray flux of early-type stars (OB type) scales
with $f_{\rm x} \approx 10^{-7} f_{\rm opt}$ (\citealt{Kudritzki00} and references therein),
while the ratio $f_{\rm x}/f_{\rm opt}$ of late-type stars (F to M)
usually ranges from $10^{-6}$ to $10^{-1}$ (e.g. \citealt{Krautter99}).
In contrast, sources such as SNRs, SSSs, and XRBs
radiate mainly in X-rays.

We also used optical and near-infared magnitudes and colours
to classify foreground stars (Figs. \ref{fig. b-r_j-k.ps}  and \ref{fig. B-R vs R}).
Fig. \ref{fig. b-r_j-k.ps} is the colour-colour diagram for XMM-\emph{Newton} sources 
with optical (USNO-B1) and infrared (2MASS) counterparts. 
Lines show the expected $(B-R)$ and $(J-K)$ colours for main-sequence, 
giant, and supergiant stars belonging to the Milky Way.
We obtained these lines using intrinsic colours calculated by \citet{Johnson66}.
Stars located at the Galactic latitude of M\,83 ($b \approx 32^\circ$)
have on average a colour excess per kiloparsec of $E(B-V) = 0.05 \pm 0.05$ mag kpc$^{-1}$ \citep{Gottlieb69}.
Therefore, the colour excesses $E(J-K)$ and $E(B-R)$
are negligible compared to the optical and infrared magnitude uncertainties \citep{Schild77}.

Figs. \ref{fig. b-r_j-k.ps}  and \ref{fig. B-R vs R} allow to separate 
foreground stars from other classes of sources.
Foreground stars are brighter in $R$ than background objects or members of M\,83,
and sources with $J-K \lesssim 1.0$ and $B-R \lesssim 2.0$ are most likely foreground stars.

From previous considerations, 
we classified foreground stars when these  
conditions were met: 
\begin{itemize}
\item $\log (f_{\rm x}/f_{\rm opt}) \lesssim -1$;
\item $HR_2 \lesssim 0.3$;
\item $HR_3 \lesssim -0.4$;
\item $J-K \lesssim 1.0$; 
\item $B-R \lesssim 2.0$.
\end{itemize}

The five sources classified as foreground star candidates 
are reported in Table \ref{Tab. source-offset list}.
A detailed discussion of the identification and classification
of foreground stars is provided in sections \ref{sect. Discussion foreground stars}
and \ref{sect. Sources which are not foreground stars}.

\begin{table}
\begin{center}
\caption{M83 X-ray sources and their associated candidate sources in our Galaxy.}
\label{Tab. source-offset list}
\resizebox{\columnwidth}{!}{
\begin{tabular}{lccc@{\, \,}c@{\, \,}c}
\hline
\hline
No.      &       RA      &       Dec     &    USNO-B1   & B\,mag. & R\,mag.   \\
         &     (J2000)   &     (J2000)   &              &        &            \\
\hline                                                                           
{ } 21       & 13\,36\,18.73 & -30\,01\,38.1 & 0599-0299962 & 18.0  & 15.5    \\
{ } 24       & 13\,36\,19.95 & -29\,51\,08.3 & 0601-0298625 & 14.1  & 13.2    \\
143      & 13\,37\,27.29 & -29\,55\,45.5 & 0600-0300561 & 18.2  & 16.4    \\    
174      & 13\,37\,44.79 & -30\,07\,49.2 & 0598-0301638 & 12.9  & 11.2    \\
182      & 13\,37\,57.77 & -30\,01\,40.6 & 0599-0300696 & 15.1  & 13.9    \\    
\hline
\end{tabular}
}
\end{center}
\end{table}

\subsection{Background objects}
\label{sect. Background objects}

The identification of AGNs, normal galaxies, and galaxy clusters
is based on SIMBAD and NED correlations,
and is confirmed if there is an optical counterpart
in the \emph{2nd Digitized Sky Survey} (DSS2) image.
New classifications 
are based on the radio counterpart and hardness ratio $HR_2 \geq -0.4$ \citep{Pietsch04}.

We identified nine sources as background galaxies and AGNs 
(sources No. 7, 17, 31, 65, 83, 89, 144, 148, 158, see Table \ref{Tab. list-galaxies}).
We found radio counterparts of the sources No. 20, 37, 189 
and classified them as AGN candidates for the first time
(see section \ref{sect. Discussion Background objects}).
Based on the $\log N - \log S$ calculated by \citet{Cappelluti09} 
(see section \ref{sect. AGN-corrected XLFs}), about $40$ observed sources 
(with a $2-10$ keV flux $F_{\rm x} > 10^{-14}$ erg cm$^{-2}$ s$^{-1}$) 
are expected to be background objects
in each XMM-\emph{Newton} observation of Table \ref{Tab. OBS ID XMM}.
From a comparison with other works (e.g. \citealt{Misanovic06}),
we expect a large difference between the predicted number of background objects 
from background surveys and the number of identified/classified background
objects in an XMM-\emph{Newton} observation.
This difference is due to the difficulty in classifying sources which,
because of their distance, are too faint (and therefore provide little information)
to be classified with the methods at our disposal.

\begin{table}
\begin{center}
\caption{X-ray sources identified and classified as galaxies or AGNs
and their counterparts or previous X-ray classifications.}
\label{Tab. list-galaxies}
\resizebox{\columnwidth}{!}{
\begin{tabular}{lccccc}
\hline
\hline
No.      &     RA     &      Dec      &          Name            \\
         &   (J2000)  &    (J2000)    &        (SIMBAD)          \\
\hline
\multicolumn{4}{c}{Identifications:}\\
\hline                                                                    
{ } { } 7    & 13 36 04.66 & -30 08 30.8  &       QSO B1333$-$298      \\ 
{ } 17   & 13 36 15.42 & -29 57 58.2  &         [I1999] 5$^3$        \\   
{ } 31   & 13 36 28.13 & -29 42 27.9  & 2MASS\,13362821$-$2942266  \\    
{ } 65   & 13 36 45.78 & -29 59 13.0  & 6dFGS gJ133645.8$-$295913  \\     
{ } 83   & 13 36 58.26 & -29 51 04.3  &        [MCK2006] 28$^1$      \\   
{ } 89   & 13 36 59.68 & -30 00 58.8  &        [BRK2009] 7$^2$       \\  
144  & 13 37 27.46 & -30 02 28.3  & 6dFGS gJ133727.5$-$300228  \\     
148  & 13 37 29.36 & -29 50 27.4  & 6dFGS gJ133729.5$-$295028  \\     
158  & 13 37 32.94 & -29 51 01.2  &        ESO 444$-$85        \\     
\hline
\multicolumn{4}{c}{New classifications:}\\
\hline
{ } 20   & 13 36 18.21 & -30 15 00.5  & NVSS\,J133618$-$301459 \\       
{ } 37   & 13 36 30.53 & -30 16 57.0  & NVSS\,J133630$-$301651 \\       
189  & 13 38 05.57 & -29 57 45.4  & NVSS\,J133805$-$295748 \\         
\hline
\end{tabular}
}
\end{center}
Notes:\\
$^1$: \citet{Maddox06};\\
$^2$: \citet{Bresolin09};\\
$^3$: \citet{Immler99}.
\end{table}

\subsection{Nuclear sources}
\label{sect. nuclear sources}

We detected two bright sources in the nuclear region of M\,83
with the source detection procedure: sources No.\,92 and No.\,95.
They are separated by $\sim6.3^{\prime\prime}$
and are the brightest sources detected with XMM-\emph{Newton} in M\,83
($F_{\rm No.\,90} = [1.03 \pm 0.25] \times 10^{-12}$ erg cm$^{-2}$ s$^{-1}$;
$F_{\rm No.\,93} = [2.59 \pm 0.15] \times 10^{-13}$ erg cm$^{-2}$ s$^{-1}$; $0.2-12$ keV,
assuming an absorbed powerlaw spectrum with index 1.8 
and a foreground Galactic absorption of $N_{\rm H}=3.69 \times 10^{20}$ cm$^{-2}$).
The two nuclear sources coincide with $\sim 18$ 
\emph{Chandra} sources and the bright diffuse emission of the starburst nucleus,
not resolved by XMM-\emph{Newton} because of its high PSF, 
which causes source confusion in crowded regions, 
such as the nuclear region of M\,83.

\subsection{X-ray binaries}
\label{sect. X-ray Binaries}

We classified sources as XRBs if the X-ray spectra
or hardness ratios were compatible with the typical
spectra of XRBs and we detected a flux periodicity.

We identified two X-ray binaries (Nos.\,81 and 120), 
previously classified by SW03 using \emph{Chandra} observations
(section \ref{sect. class. X-ray binaries}).

\subsection{Supernova remnants}
\label{sect. supernova remnants}

We assume that the X-ray spectra of SNRs are well described by the 
thermal plasma model APEC \citep{Smith01}, with temperatures ranging from $0.2$ to $1.5$ keV.
At this distance we are unable to resolve an SNR 
or to verify a more detailed spectral model
assuming, e.g., a non-equilibrium ionisation.

We classified an X-ray source as SNR 
if $HR_1 >0.1$, $HR_2 < -0.4$, the source was not a foreground star,
and did not show a significant variability \citep{Pietsch04}.

We identified the source No.\,79
as source [SW03]\,27, classified as a young SNR candidate by SW03
(section \ref{sect. class. Supernova remnants}).

\paragraph{\textbf{SN1957D}}
\citet{Long12} reported the first detection of SN1957D in X-rays with \emph{Chandra}.
The source shows a luminosity of $1.7 \times 10^{37}$ erg cm$^{-2}$ s$^{-1}$
($d=4.61$ Mpc, \citealt{Saha06}; $0.3-8$ keV), and the spectrum is well modelled with
an absorbed powerlaw with an index $\sim 1.4$, a foreground Galactic absorption
of $N_{\rm H}=4 \times 10^{20}$ cm$^{-2}$ and an intrinsic column density of 
$N_{\rm H}=2 \times 10^{22}$ cm$^{-2}$.
We did not detect SN1957D in the XMM-\emph{Newton} observations. 
In observation 1 the source is located near to the centre of the field of view,
and in the other two observations the source is located at the edge of the field of view.
Assuming the spectral parameters found by \citet{Long12}, 
we calculated a $3\sigma$ upper-limit in observation 1 
of $\sim 2.4 \times 10^{-14}$ erg cm$^{-2}$ s$^{-1}$ ($0.2-12$ keV),
corresponding to a luminosity of $\sim 5.8 \times 10^{37}$ erg s$^{-1}$,
well above the luminosity detected by \citet{Long12}.

\subsection{Super-soft sources}
\label{sect. super-soft sources}

Super-soft sources are a class of sources that
are believed to be binary systems containing  a white dwarf.
The white dwarf accretes matter from a Roche-lobe-filling companion
at high rates ($\dot{M}_{\rm acc} \sim 10^{-7}$ M$_\odot$ yr$^{-1}$),
which leads to quasi-steady nuclear burning on its surface 
(see e.g. \citealt{vandenHeuvel92}).
SSSs show soft spectra with blackbody temperatures of $15-150$ eV
and X-ray luminosities ranging from $\sim 10^{35}$ erg s$^{-1}$ to $10^{38}$ erg s$^{-1}$
(\citealt{DiStefano03}; \citealt{Kahabka97}).
An additional harder component, due to interactions 
of the radiation with matter near to the white dwarf
or wind interactions can be observed \citep{DiStefano03}.
Moreover, SSSs are often observed as transient X-ray sources (see \citealt{Greiner00}).
Other classes of sources with soft spectra can be confused with SSSs.
For example, some X-ray pulsars observed outside 
the beam of the pulsed radiation can show a soft ($\sim 30$ eV) component
(\citealt{Hughes94}; \citealt{DiStefano03}).
Moreover, stripped cores of giant stars can be classified as SSSs \citep{DiStefano01}.

As our classification criteria, we assumed blackbody 
temperatures of $kT_{\rm bb} \leq 100$ eV (in agreement with the 
selection procedure proposed by \citealt{DiStefano03})
and hardness ratios that do not overlap with
those of other classes of sources.
These criteria are an $HR_1 \lesssim 0$ and $HR_2 - EHR_2 < -0.9$.
We classified a source as SSS only if both criteria are fulfilled.

We identified source No.\,91
as source M83-50, classified as an SSS candidate by \citet{DiStefano03}
using \emph{Chandra} observations (section \ref{sect. class. Super-Soft Sources}).

\subsection{Hard sources}
\label{sect. main hard sources}

Hard sources show hard X-ray spectra (or hard HRs, see Table 5 in \citealt{Pietsch04}).
Using their spectral properties and hardness ratios,
we classified five hard sources (Nos.\,16, 61, 103, 126, and 153; 
see section \ref{section Newly classified hard sources})
and we identified 11 hard sources
(Nos.\,60, 80, 92, 97, 99, 106, 107, 108, 114, 116, 129;
see section \ref{section identification hard sources}).

\subsection{Ultra-luminous X-ray sources} 
\label{sect. Observation of an ULX} 

ULXs are pointlike non-nuclear sources with X-ray luminosities 
in excess of the Eddington limit ($L_{\rm Edd} \simeq 10^{39}$ erg s$^{-1}$)
for a stellar mass black-hole (see e.g. \citealt{FengSoria11}).
They are usually located in active star-forming environments \citep{Miller04},
and their nature is still unclear;
recent studies indicate that ULXs are a heterogeneous sample of objects (e.g. \citealt{Gladstone11}).

Several models have been proposed to explain the high X-ray luminosity of ULXs,
but there are three models that are often used for this class of sources.
The first model requires that ULXs are intermediate-mass black-hole systems (IMBHs)
with masses $M \sim 10^{2}-10^{4}$ M$_\odot$,
accreting at sub-Eddington rates (e.g. \citealt{Colbert99}).
The other models assume that ULXs are stellar-mass black holes
(with masses $M \lesssim 100$ M$_\odot$) in a super-Eddington accretion regime \citep{Poutanen07}
or with beamed radiation (see e.g. \citealt{King09}).

We identified ULX No.\,133,
discovered by \citet{Trinchieri85} with \emph{Einstein} (source H2),
and previously observed in X-rays with ROSAT by \citet{Ehle98} and \citet{Immler99}
(see section \ref{sect. class. Ultra-Luminous X-ray sources}).

\section{X-ray luminosity functions}
\label{sect. X-ray Luminosity Functions}

For each observation, we calculated the XLFs in the energy range $2-10$ keV
excluding the softer bands
to reduce the effect of incompleteness of the observed source sample
due to absorption. 
Moreover, from an XLF calculated in this energy band, 
it is possible to easily subtract the contribution 
of the $\log N-\log S$ of the AGNs, which was calculated
from several surveys performed by XMM-\emph{Newton} 
and \emph{Chandra} (see section \ref{sect. AGN-corrected XLFs}).

We considered for XLFs only sources with
a detection likelihood greater than 6 in the energy range $2-12$ keV.
For each source, we converted the count rates to the $2-10$ keV fluxes
using the ECFs of Table \ref{Tab. ecfs} for the energy bands R4 and R5.

We excluded the region inside
a circle centred on the nuclear region of M\,83 
with radius $R = 26^{\prime\prime}$ from the XLF calculation,
where the large PSF of EPIC in a crowded region causes source confusion effects
(see section \ref{sect. nuclear sources}).
Since we were interested in obtaining XLFs of XRBs,
we also excluded the sources previously classified 
as SNRs, SSSs, ULXs, and foreground stars 
(section \ref{sect. Source classification}).
For each observation, 
we calculated the XLFs of sources detected 
within two regions of M\,83: the inner disc inside the $D_{25}$ ellipse,
and the outer disc outside the $D_{25}$ ellipse.

\subsection{XLFs corrected for incompleteness}
\label{sect. XLFs corrected for incompleteness}

\begin{figure}
\begin{center}
\includegraphics[width=9cm]{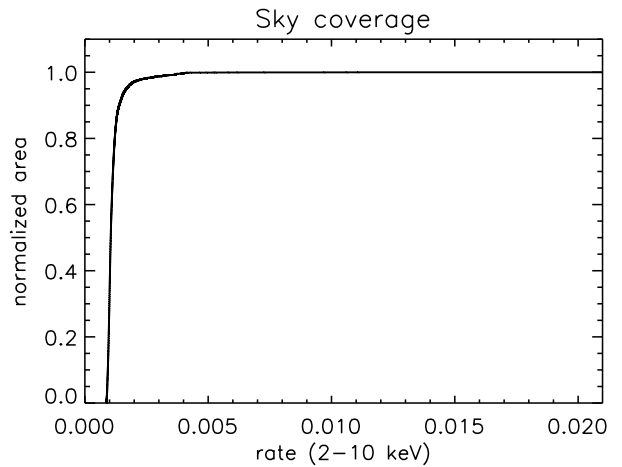}
\end{center}
\caption{Sky coverage as a function of the X-ray flux ($2-10$ keV)
for the region inside the $D_{25}$ ellipse (observation 1),
calculated excluding the region within the circle centered on the nuclear region of M\,83 
with radius $R = 26^{\prime\prime}$.}
\label{fig. sky coverage function}
\end{figure}

The sensitivity of the EPIC instruments depends
on the exposure, background, and PSF, 
which are not uniform across the FOV.
Indeed, the exposure time is relatively high at the centre of the FOV
and decreases with increasing off-axis angle 
(vignetting effect).
The background, modelled by the task {\tt esplinemap}, decreases 
with increasing angular distance from the nuclear region of M\,83 
(due to the diffuse emission in the disc of M\,83),
and the optical properties of the X-ray telescope
introduce a degradation of the PSF with increasing off-axis angle.
Therefore, the sensitivity also varies across the observed area,
allowing the detection of the brightest sources across the entire 
observed area, whereas the effective area for the detection
of faint sources is smaller.
This effect leads to an underestimation of the number 
of sources observed at the faintest flux levels.

We corrected the XLFs  
by taking into account the incompleteness effect described above
by calculating the \emph{sky coverage function},
which is the effective area covered by the observation
as a function of flux.
For each observation, we first created 
the combined sensitivity maps of PN, MOS1, and MOS2
with the SAS task {\tt esensmap}, which requires as input files
the exposure maps, the background images, and the detection masks created 
by the source detection procedure.
We used the sensitivity maps to calculate
the sky coverage function for each observation (Fig. \ref{fig. sky coverage function}).
The cumulative XLF corrected for incompleteness is given by
\begin{equation} \label{eq. xlf}
N(>F_{\rm x}) = A_{\rm tot} \sum_{i=1}^{N_{\rm s}} \frac{1}{\Omega(F_i)} \mbox{ ,} 
\end{equation}
where $N(>F_{\rm x})$ is the number of sources with a flux higher than $F_{\rm x}$,
weighted by the fraction of the surveyed area $\Omega(F_i)/A_{\rm tot}$ 
over which sources with flux $F_i$ can be detected;
$A_{\rm tot}$ is the total area of the sky observed by EPIC,
$\Omega(F_i)$ is the sky coverage
(Fig. \ref{fig. sky coverage function}),
and $N_{\rm s}$ is the total number of the detected sources.
Therefore, with equation (\ref{eq. xlf}), every source is weighted with a factor
correcting for incompleteness at its flux.
The variance of the source number counts is defined as
\begin{equation} \label{eq. var xlf}
\sigma^2 = \sum_{i=1}^{N_{\rm s}} \left( \frac{1}{\Omega_i} \right)^2 \mbox{.}
\end{equation}
%

\subsection{AGN-corrected XLFs}
\label{sect. AGN-corrected XLFs}

The XLFs obtained in section \ref{sect. XLFs corrected for incompleteness} 
consist of sources belonging to M\,83 (XRBs) and AGNs.
We subtracted the AGN contribution using the 
AGN XLF of \citealt{Cappelluti09}, who
derived the XLFs from 
the 2 deg$^2$ of the XMM-COSMOS survey \citep{Scoville07}.
These authors found that the XLF of AGNs in the energy range $2-10$ keV 
is described by a broken powerlaw:
\begin{equation} \label{eq xlf cappelluti09}
\frac{dN}{dF} = 
\left\{
\begin{array}{rl}
AF^{-\alpha_1}  &  F > F_{\rm b} \\
BF^{-\alpha_2}  &  F \leq F_{\rm b} \mbox{ ,}
\end{array}
\right.
\end{equation}
where $A=BF_{\rm b}^{\alpha_1 - \alpha_2}$ is the normalisation,
$\alpha_1=2.46 \pm 0.08$, $\alpha_2=1.55 \pm 0.18$,
$F_{\rm b}=(1.05 \pm 0.16)\times 10^{-14}$ erg cm$^{-2}$ s$^{-1}$,
and $A=413$.

\begin{figure*}
\begin{center}
\includegraphics[width=9cm]{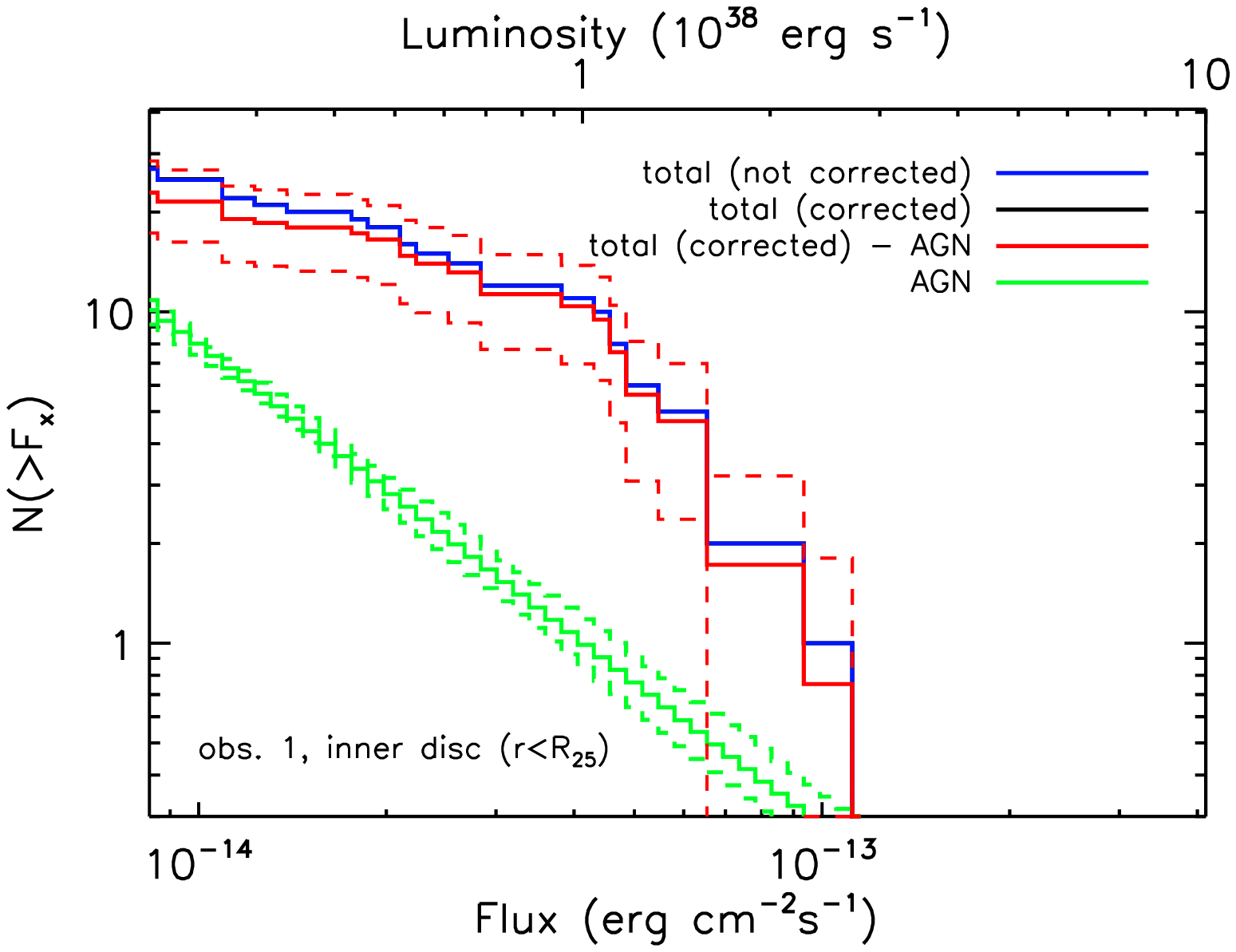}
\includegraphics[width=9cm]{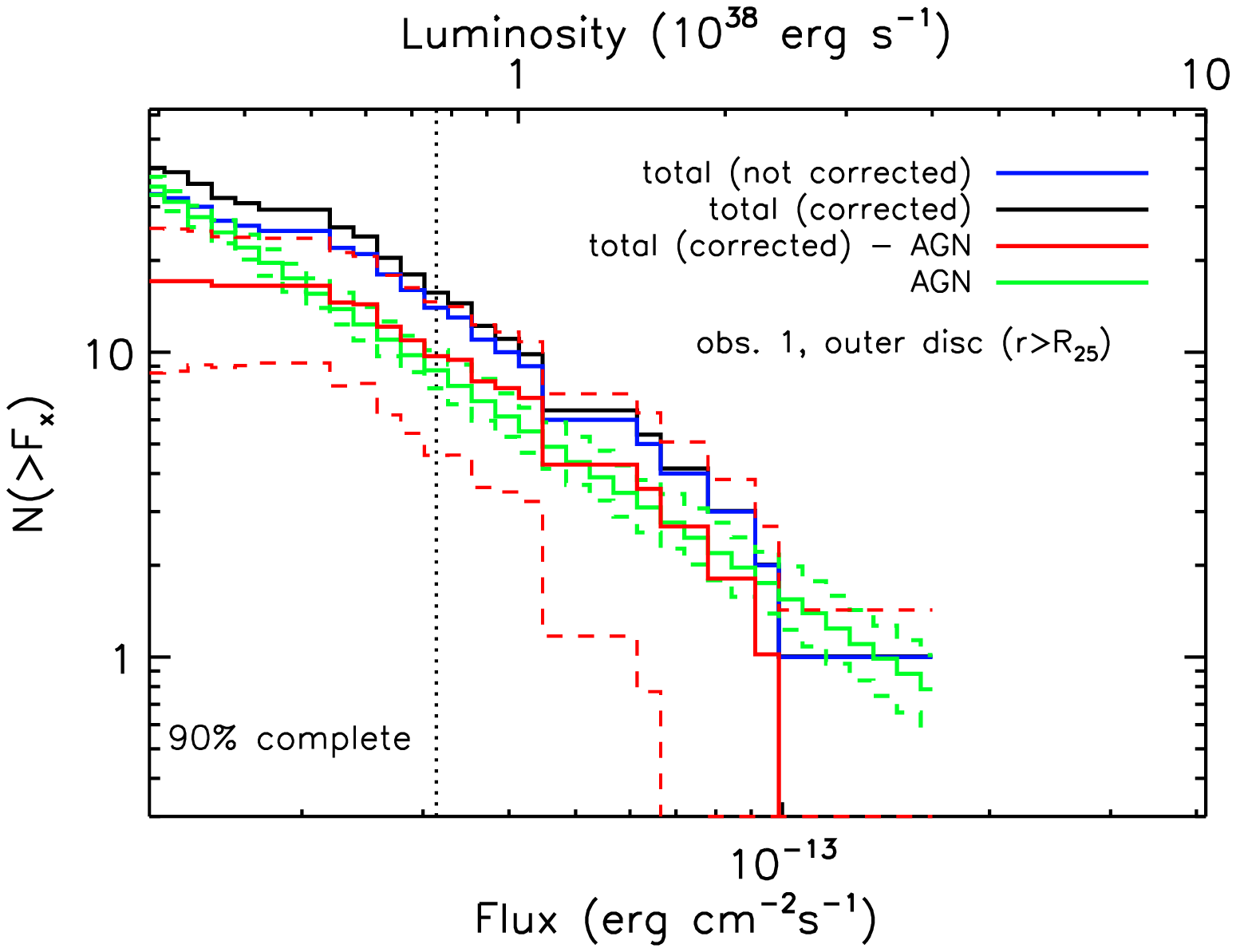}
\includegraphics[width=9cm]{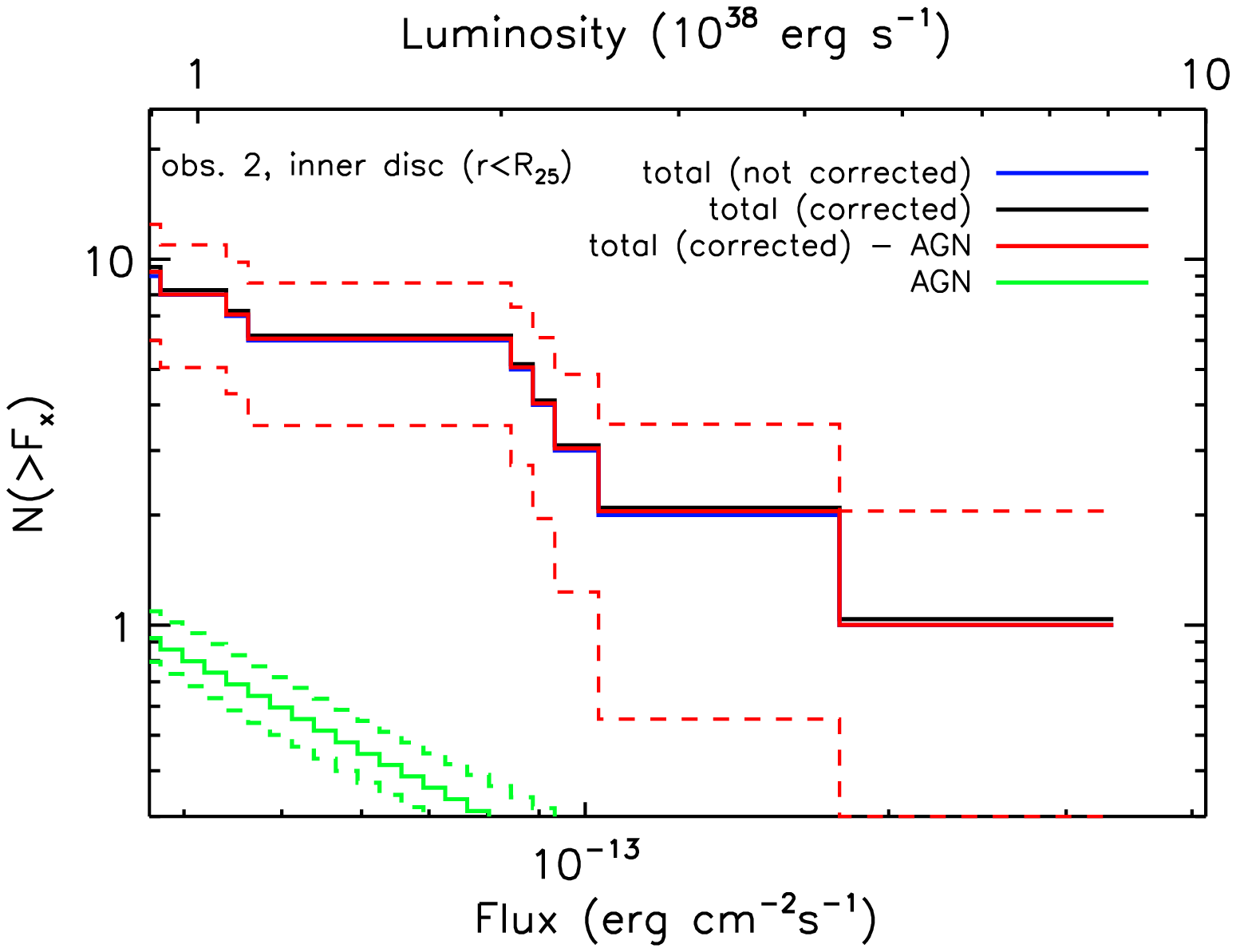}
\includegraphics[width=9cm]{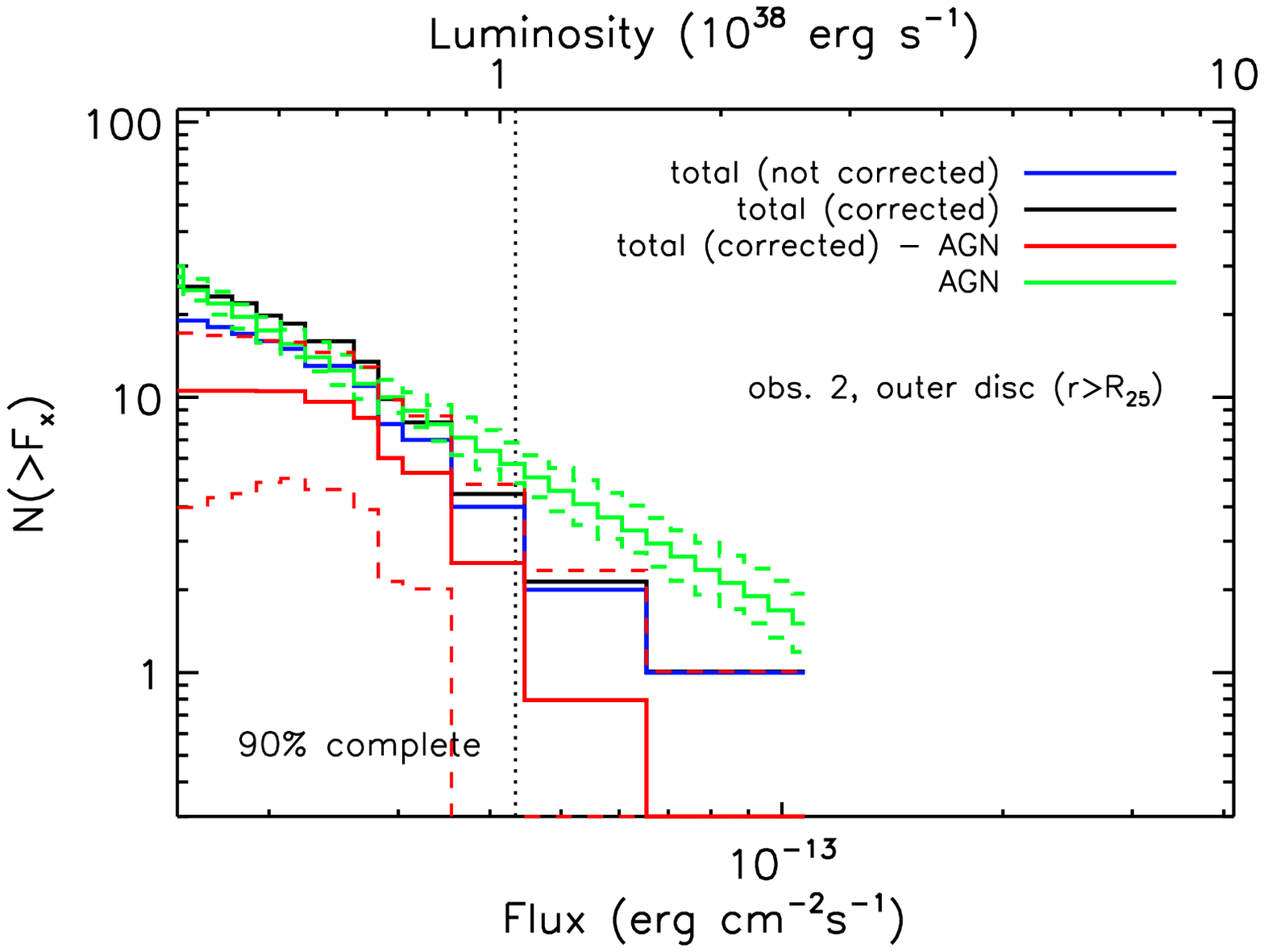}
\includegraphics[width=9cm]{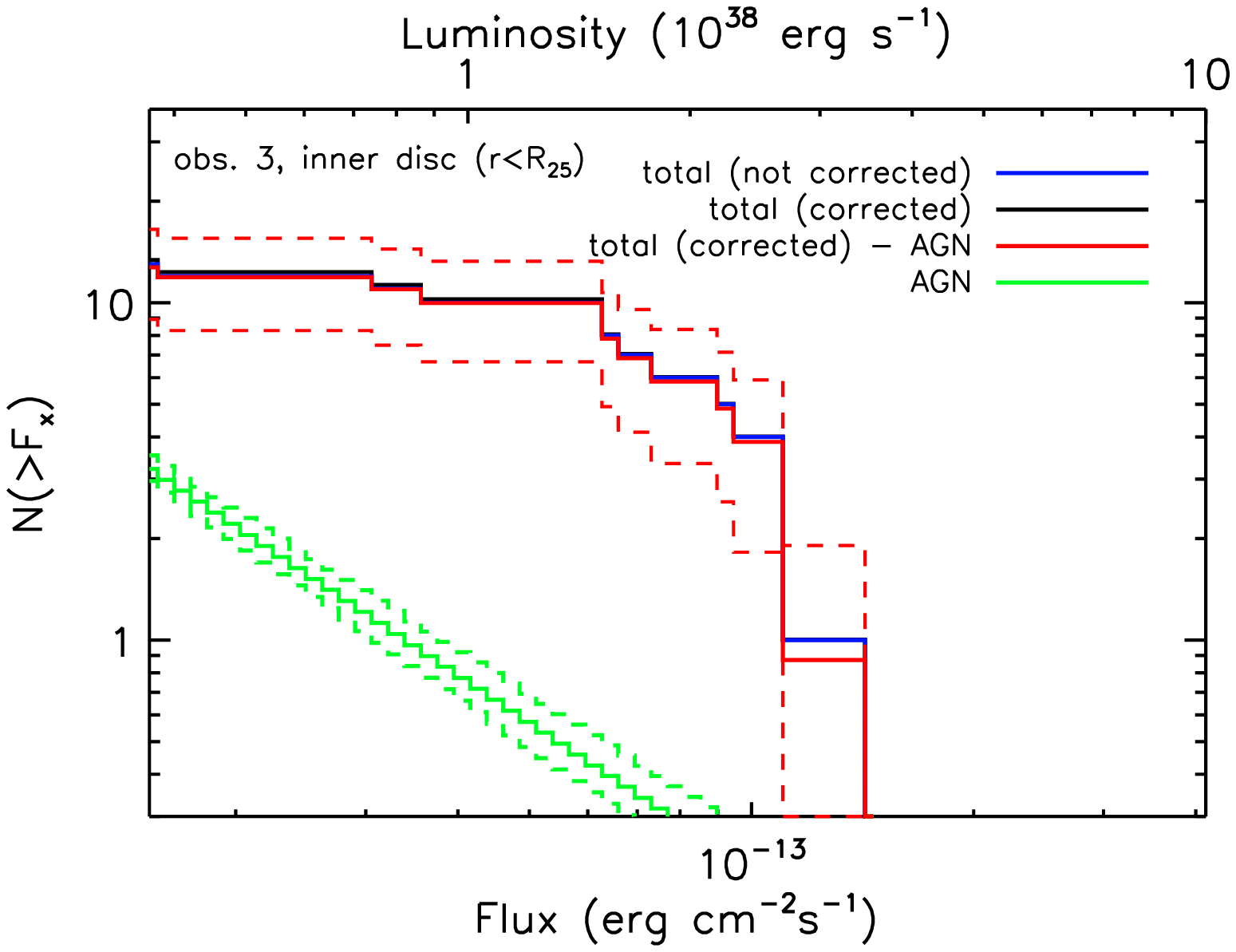}
\includegraphics[width=9cm]{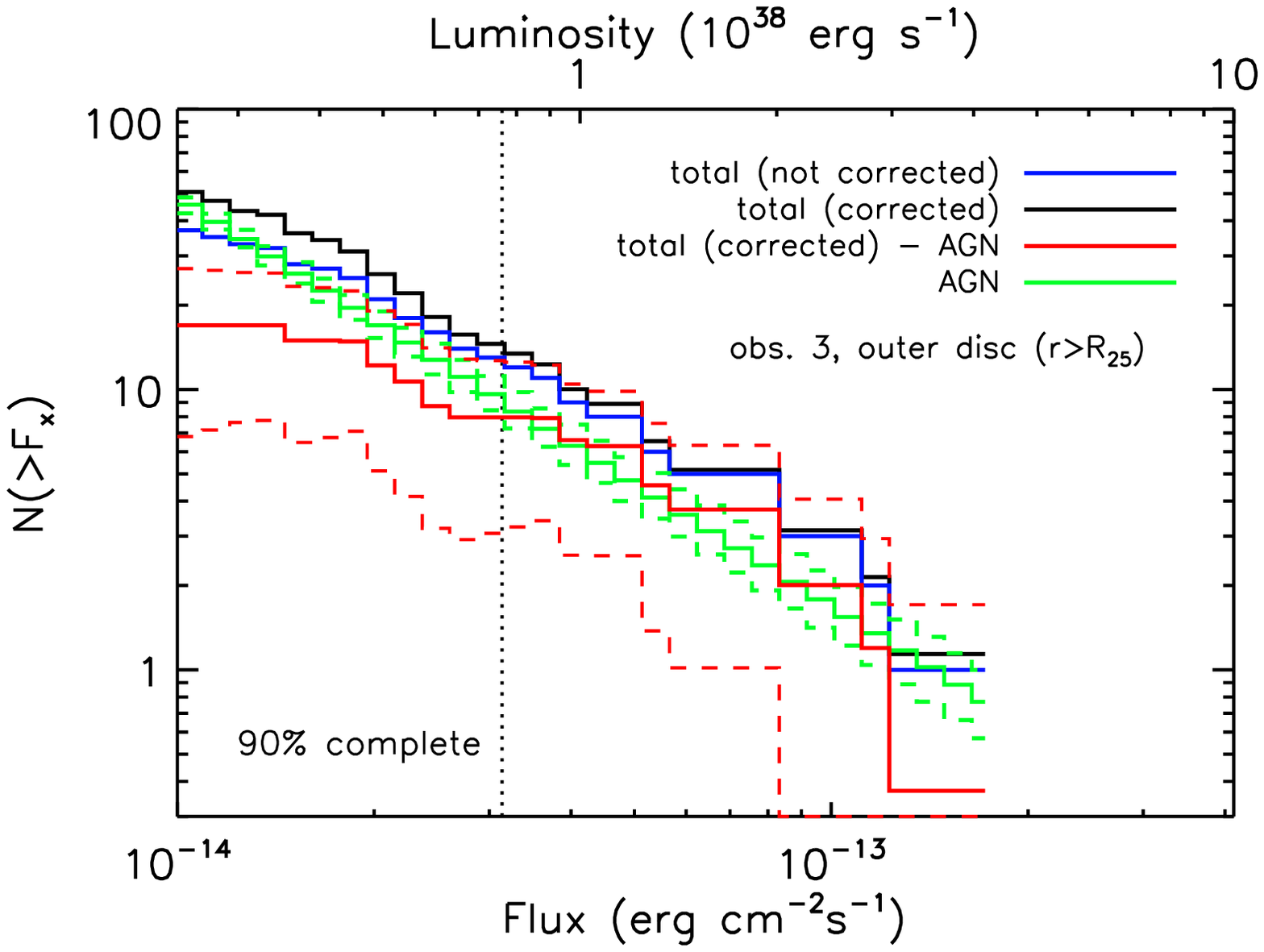}

\end{center}
\caption{Cumulative XLFs in the $2-10$ keV energy band. 
Blue lines correspond to the XLFs without the contribution of SNRs, SSSs, ULX, and foreground stars, not corrected for incompleteness.
Black lines are the XLFs corrected for incompleteness.
Solid green lines are the AGN XLFs of \citet{Cappelluti09},
and dashed green lines are the 90\% confidence errors.
Solid red lines are the XLFs corrected for incompleteness and AGN-subtracted, and the dashed red lines are the resulting uncertainties.}
\label{fig. xlf}
\end{figure*}

Fig. \ref{fig. xlf} shows the XLFs of sources detected
within the $D_{25}$ ellipse and outside, 
calculated for each XMM-\emph{Newton} observation.
Blue lines are the observed XLFs,
and black lines are the XLFs corrected for incompleteness.
Solid green lines are the AGN XLFs of equation (\ref{eq xlf cappelluti09})
with relative uncertainties (dashed green lines).
Solid red lines show the XLFs corrected for incompleteness and AGN-subtracted,
and dashed red lines are the 90\% confidence errors, obtained from 
equation (\ref{eq. var xlf})
and the 90\% confidence errors of the AGN distribution.

Vertical black lines in the right column of Fig. \ref{fig. xlf}
show the level at which the survey is 90\% complete (see section \ref{sect. outer disc}),
defined as the flux at which
$$
\sum_{i=1}^{N_{\rm s}} N(F_i)  /  \sum_{i=1}^{N_{\rm s}} A_{\rm tot}/\Omega(F_i)  = 0.9 \mbox{ .}
$$

\subsection{Fit}
\label{sect. Resulting XLFs}

We fitted the differential XLFs corrected for incompleteness and AGN-subtracted 
with a powerlaw:
\begin{equation} \label{eq pl fit}
A(F) = kF^{\alpha} \mbox{ ,}
\end{equation}
where $k$ is the normalisation and $\alpha$ the powerlaw index.
\begin{table}
\begin{center}
\caption{Best-fitting parameters of the differential XLFs of observations 1, 2, and 3,
calculated for sources within and outside the $D_{25}$ ellipse.
For each observation, the best-fitting parameters were obtained using the total XLF
corrected for incompleteness and AGN-subtracted.}
\label{Tab. all fit xlfs}
\resizebox{\columnwidth}{!}{
\begin{tabular}{lccc}
\hline
\hline
\multicolumn{4}{c}{$<R_{25}$} \\
\hline
                              &      Obs.\,1      &      Obs.\,2     &      Obs.\,3    \\
\hline
\multicolumn{4}{c}{powerlaw$^a$} \\
\hline
$\alpha$                      &      $-1.0 \pm 0.3$      &    $-1.8 \pm 0.4$       &     $-1.6 \pm 1.2$     \\
\smallskip
$k$                           & $10.4{+1.8 \atop -1.6}$ & $11.2{+5.9 \atop -3.9}$ & $21.4{+38.9 \atop -13.8}$\\
\smallskip
$\chi^2$ (d.o.f.)             &        $23.48$ (15)        &        $11.74$ (5)        &        $9.20$ (5)      \\
\hline
\multicolumn{4}{c}{broken powerlaw$^b$} \\
\hline
\smallskip
$\alpha_1$                    &  $-3.0{+0.9 \atop -0.2}$ &   $-2.9{+0.8 \atop 0.2}$ &                       \\
\smallskip
$\alpha_2$                    &  $-1.1{+0.1 \atop -0.5}$ &  $-1.1{+0.1 \atop -0.4}$  &                       \\ 
\smallskip
$k$                           & $20.9{+13.0 \atop -8.0}$& $37.2{+48.0 \atop -20.9}$&                       \\ 
\smallskip
$F_{\rm b}$ ($10^{-14}$ erg cm$^{-2}$ s$^{-1}$)&  $5.6{+1.0 \atop -0.4}$  &  $6.5{+0.8 \atop -0.7}$  &                       \\
\smallskip
$\chi^2$ (d.o.f.)             &        $21.01$ (13)         &          $9.17$ (3)         &                        \\
\hline
\multicolumn{4}{c}{$>R_{25}$} \\
\hline
                    &           Obs.\,1         &           Obs.\,2         &          Obs.\,3          \\
\hline
\multicolumn{4}{c}{powerlaw$^a$} \\
\hline
$\alpha$                      &      $-1.9 \pm 0.5$      &      $-3.3 \pm 1.1$      &      $-1.2 \pm 0.4$      \\
\smallskip
$k$                           & $13.4{+4.7 \atop -3.5}$ & $8.7{+3.5 \atop -2.5}$ & $7.1{+2.0 \atop -1.6}$ \\
\smallskip
$\chi^2$ (d.o.f.)             &        $11.62$ (11)        &         $4.41$ (5)       &         $8.68$ (9)        \\
\hline
\end{tabular}
}
\end{center}
Notes:\\
$^a$: see equation (\ref{eq pl fit});\\
$^b$: see equation (\ref{eq bpo fit});\\
\end{table}
We also fitted the differential XLFs with 
a broken powerlaw:
\begin{equation} \label{eq bpo fit}
A(F) = 
\left\{
\begin{array}{lr}
kF_{\rm b}^{\alpha_2 - \alpha_1} F^{\alpha_1} &  F > F_{\rm b} \\
kF^{\alpha_2}                        &  F \leq F_{\rm b} \mbox{ ,}
\end{array}
\right.
\end{equation}
where $F_{\rm b}$ is the break point. 
The resulting parameters obtained from the fit are reported in Table \ref{Tab. all fit xlfs}.

\subsubsection{Inner disc}
\label{sect. Inner disc}

From \emph{Chandra} observation,
SW03 calculated the XLFs 
of sources located in the inner region 
(distance $<60^{\prime\prime}$ from the nucleus) and outer region
($60^{\prime\prime}<d<R_{25}$) of the optical disc. 
They found that 
the inner region sources have a powerlaw luminosity distribution 
with an differential index of $-1.7$,
while the luminosity distribution of the outer region sources
shows a lack of bright sources above $\sim 10^{38}$ erg s$^{-1}$.
These authors modelled the XLF of these sources with a broken powerlaw 
with a break around $\sim 10^{38}$ erg s$^{-1}$
and differential indices of $-1.6$ and $-2.6$.
They explained the XLF of the inner region sources in terms of current starburst activity,
while the XLF of the outer region may result from an older population of disc sources
mixing with a younger population.

We recall that we cannot study the innermost
region because of poor spatial resolution of
XMM-\emph{Newton} compared to \emph{Chandra}.
We compared the best-fitting parameters of the XLF 
of the outer region sources ($60^{\prime\prime}<d<R_{25}$)
obtained by SW03 with those obtained 
from the XMM-\emph{Newton} analysis (Table \ref{Tab. all fit xlfs}).
In particular, we considered the broken powerlaw fit of sources detected in observation 1.
Only during this observation was the whole optical disc of M\,83 observed.
We found that the indices $\alpha_1$, $\alpha_2$ and the break $F_{\rm b}$ of equation \ref{eq bpo fit}
agree within the uncertainties
with the parameters found by SW03.

\citet{Grimm03} studied the XLFs of a sample of galaxies and 
found the probable existence 
of a universal HMXB XLF (in the luminosity range 
$\sim 4 \times 10^{36}-10^{40}$ erg s$^{-1}$),
described by a powerlaw with differential slope of $-1.6$.
They found that the number of HMXBs with $L_{\rm x} > 2 \times 10^{38}$ erg s$^{-1}$
in a star-forming galaxy
is directly proportional to the SFR,
and proposed that the number and the total X-ray luminosity of HMXBs can be used
to measure the star formation rate of a galaxy.
Based on a much larger sample of galaxies, \citet{Mineo12} 
found that the properties of populations of HMXBs
and their relation with the SFR agree
with those obtained by \citet{Grimm03}.
We estimated the SFR in the optical disc of M\,83 using
the $N_{\rm HMXBs} - $SFR relation of \citet{Mineo12}:
\begin{equation} \label{eq. mineo}
N (> 10^{38} \mbox{erg s}^{-1}) = 3.22 \times \mbox{SFR\,(M}_\odot\mbox{ yr}^{-1}\mbox{)\,.} 
\end{equation}
We assumed that the XLF we used for this calculation
provides a good approximation of the HXMB XLF in M\,83.
The contribution of LMXBs to the XLF is negligible
for a starburst galaxy such as M\,83
when $L_{\rm x} \gtrsim 10^{38}$ erg s$^{-1}$ \citep{Grimm03}.
Moreover, the contribution of LMXBs to the XLF is minimized
by excluding the nuclear region of the galaxy,
from which a strong contribution to the total number of LMXBs is expected.
Using the XLF of sources detected in observation 1 within the $D_{25}$ ellipse,
from equation \ref{eq. mineo} we found an SFR\,$\approx 3.1$ M$_\odot$ yr$^{-1}$, 
in agreement with the SFR estimates obtained from observations in other wavelengths
(see e.g. \citealt{Boissier05}; \citealt{Dong08}; \citealt{Grimm03} and references therein).

\subsubsection{Outer disc}
\label{sect. outer disc}

The XLFs of the outer disc ($d > R_{25}$) show an
excess of sources (with respect to the
expected number of AGNs) in the luminosity 
range $\sim 10^{37}$ to $\sim 2\times 10^{38}$ erg s$^{-1}$ (Fig. \ref{fig. xlf}). 

We are interested in calculating the probability of the luminosity distribution 
of the observed sources to be
consistent with the luminosity distribution of equation (\ref{eq xlf cappelluti09})
which represents the AGN distribution.
Therefore, we compared for each observation 
the luminosity distribution of the sources detected in the outer disc ($d> R_{25}$)
that was not corrected for incompleteness (see section \ref{sect. X-ray Luminosity Functions})
with a distribution of simulated sources over the EPIC FOV
obtained from a uniform spatial distribution of sources with a luminosity distribution
given by equation (\ref{eq xlf cappelluti09}),
filtered to exclude sources with a flux below the detection threshold calculated
at the position of each source in the sensitivity map.
The Kolmogorov-Smirnov test applied to these source samples showed that
the probabilities that the luminosity distributions of the observed sources
are consistent with the luminosity distribution of AGNs (equation \ref{eq xlf cappelluti09})
are almost zero, being $0.04$\% in observation 1, $0.7$\% in observation 2, and $0.6$\% in observation 3.

\begin{figure}
\begin{center}
\includegraphics[width=9cm]{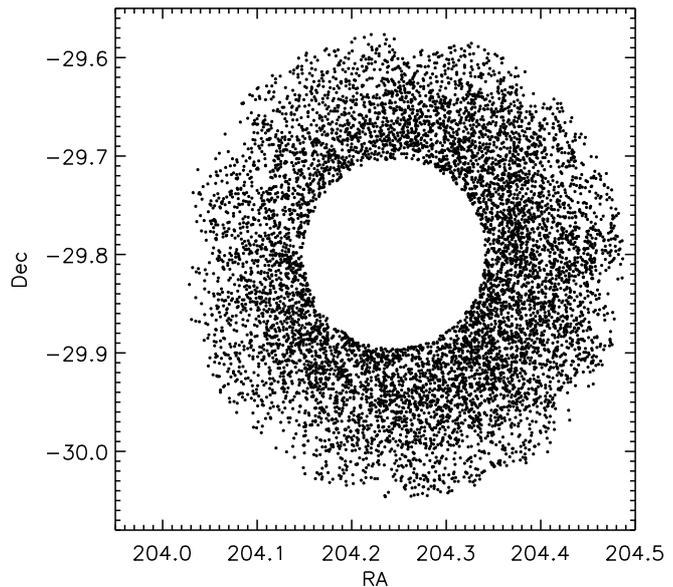}
\end{center}
\caption{Sample of $10^4$ simulated sources, distributed over 
         the EPIC field of view of observation 1 and located outside the $D_{25}$ ellipse.}
\label{fig. agn distrib. simul.}
\end{figure}
To quantify the probability that the set of X-ray sources
located outside the $D_{25}$ ellipse are AGNs
(which are expected to be uniformly distributed across the sky)
or XRBs (whose distribution should not be uniform, 
because the position of XRBs should correlate 
with the arms extending out of the optical disc),
we performed a two-dimensional  
Kolmogorov-Smirnov test (\citealt{Fasano87}; \citealt{Peacock83}).
This test is based on the statistic $\delta$, which in the unidimensional
Kolmogorov-Smirnov test represents the largest 
difference between two cumulative distributions.
We applied this test to two data samples:
\begin{enumerate}
\item all X-ray sources detected in observation 1
      that are located outside the $D_{25}$ ellipse. 
      The number of these sources is $N_1=39$;
\item a distribution of simulated sources in the EPIC FOV of observation 1,
      obtained from a uniform spatial distribution of sources
      (which represents the uniform spatial distribution of AGNs)
      modified to take into account the incompleteness effect described 
      in section \ref{sect. XLFs corrected for incompleteness}. 
      We obtained this spatial distribution of sources as follows.
      We first generated a uniform spatial distribution of sources
      with fluxes given by the XLF of AGNs described in section \ref{sect. AGN-corrected XLFs}.
      Then, we selected sources with flux higher than that 
      corresponding to the position of each source in the sensitivity map.
      We additionally selected sources with luminosity $> 10^{37}$ erg s$^{-1}$ 
      in the energy range $2-10$ keV that are located outside the $D_{25}$ ellipse.
      With this method, we generated a sample of $N_2=10^4$ coordinate pairs (RA, Dec) 
      of sources (see Fig. \ref{fig. agn distrib. simul.}).
\end{enumerate}

From the number of data points $N_1$ and $N_2$ of the two data sets,
the significance level was calculated from the probability distribution
of the quantity
\begin{equation} \label{eq. Zn}
Z_n \equiv \delta \sqrt{n} \mbox{,}
\end{equation}
where $n=N_1N_2/(N_1 + N_2)$.
The analytical formula for calculating of the probability
that the two data samples come from the same distribution
is accurate enough for large data sets with $n>80$ \citep{Fasano87}.
Since in our case $n \approx 39$, we needed to use Monte Carlo simulations.
We generated many synthetic data samples  
simulating the uniformly distributed AGNs 
with the same method previously used to calculate sample 2;
each of the synthetic data samples has the same number 
of sources as the observed data set 1 ($N_1=39$).
For each data set we applied the 2D Kolmogorov-Smirnov test 
by comparing the synthetic data set with the set of $10^4$ sources
distributed across the EPIC FOV previously described,
then we calculated the quantity $Z_n$ using equation (\ref{eq. Zn}).
The probability of the observed $Z_n$ is given by the fraction 
of the times the simulated $Z_n$ are larger than the observed $Z_n$.

Applying this statistical method to our data, 
we found a probability of $99.5$\% that the
observed sample 1 and the simulated homogeneously distributed sample 2 
are significantly different,
which suggests a non-uniform distribution of the observed X-ray sources
and therefore a possible correlation between the positions of these sources
and the extended arms of M\,83.

The incompleteness correction given by equation (\ref{eq. xlf})
is based on the hypothesis that sources are uniformly distributed.
However, we have demonstrated that the X-ray sources located outside the $D_{25}$ ellipse have
a non-uniform distribution, hence the associated XLFs corrected for incompleteness 
of Fig. \ref{fig. xlf} (right column) are not reliable at low luminosities.
Therefore we only considered the part of the XLFs with luminosities higher than 
the level at which the survey is 90\% complete 
(to the right of the vertical black lines in Fig. \ref{fig. xlf}).
We found that the 90\% complete XLFs of observations 1 and 3 
(for which we have enough data points to find a good fit)
are well fitted with a powerlaw with differential slopes $\alpha = -2.2 \pm 0.5$ (observation 1),
and $\alpha = -1.7 \pm 0.4$ (observation 3),
which are consistent with each other within errors.
These are also consistent with the AGN slope of \citet{Cappelluti09}.

Assuming that the spatial distribution of AGNs and their number density
are not subject to strong fluctuations on small angular scales
corresponding to different directions in the M\,83 field,
the observed excess of sources
(with respect to the AGN distribution) in the luminosity 
range $\sim 10^{37}$ to $\sim 2\times 10^{38}$ erg s$^{-1}$ (Fig. \ref{fig. xlf}) 
can probably be ascribed
to a population of XRBs located in the outer disc of M\,83.
The recent star-forming activity discovered by GALEX in this region
indicates that a large portion of the observed X-ray sources
are HMXBs. However, the observed XLF slope is steeper than
the slope of the universal HMXB XLF inferred by \citet{Grimm03}.
A possible explanation for the difference between the two slopes
could be that the observed XLFs are the result of a mix
of XRB populations formed after starbursts of different ages.
An alternative explanation is that 
the mass distribution of the population of stars
in the low-density regions of the outer disc of M\,83
is described by a truncated initial mass function (IMF),
whose existence was proposed to explain the production of fewer high-mass stars 
(compared to the standard IMF) in low-density environments
(see e.g. \citealt{Krumholz08}; \citealt{Meurer09}).
The universality of the IMF is still a matter of debate \citep{Bastian10};
in this context, a recent Subaru H$\alpha$ observation of the outer disc of M\,83
revealed O stars even in small clusters ($M \lesssim 10^3$ M$_\odot$),
which supports the hypothesis that the IMF is not truncated 
in low-density environments \citep{Koda12}.

\section{Summary}
\label{sect. conclusions}

We presented an analysis of three XMM-\emph{Newton} observations of M\,83.
We performed the source detection procedure 
separately for images of each observation,
and we obtained a catalogue containing 189 sources.

Based on cross-correlations with other catalogues
we identified counterparts for 103 sources,
12 of which were identified or classified as background objects
and 5 as foreground stars (one as candidate CV).
We performed spectral analysis of the sources with the largest
number of counts, as well as studies of the X-ray variability
and the hardness ratio diagrams.
The spectral analysis of ULX No.\,133 in observations 2 and 3
showed good fits with the standard IMBHs model as well as with 
accreting stellar-mass black-hole model, in agreement with
the results obtained by \citet{Stobbart06} from observation 1.

In Sect. \ref{sect. X-ray Luminosity Functions} 
we presented the XLFs of sources in the $2-10$ keV energy band,
within and outside the $D_{25}$ ellipse.
We corrected the XLFs for incompleteness and subtracted the
contribution of background AGNs from the total XLF
to obtain the XLFs of XRBs.
The XLF of the optical disc is well fitted with a powerlaw
or a broken powerlaw, while the XLF of the outer disc 
is well fitted with a simple powerlaw.
The broken powerlaw fit parameters  
agree (within the uncertainties)
with the parameters found by SW03 with \emph{Chandra}.
From the XMM-\emph{Newton} XLF, we obtained an
SFR\,$\approx 3.1$ M$_\odot$ yr$^{-1}$ in the optical disc of M\,83,
which agree with 
previous estimates obtained in other wavelengths.

The XLFs of these sources show an excess of sources
(compared to the AGNs distribution) in the luminosity range 
$\sim 10^{37}$ to $\sim 2\times 10^{38}$ erg s$^{-1}$.
The application of the Kolmogorov-Smirnov test to the X-ray sources
detected outside the $D_{25}$ ellipse allowed us to find that 
this population of sources is significantly different 
from the population of background AGNs,
which is supposed to have a homogeneous distribution.
These results led us to suggest 
that a part of the X-ray sources observed outside the $D_{25}$
ellipse belongs to the outer disc of M\,83.
The 90\% complete XLFs of the outer disc are well fitted with a simple powerlaw with
differential slope $\alpha = -2.2 \pm 0.5$ (observation 1),
and $\alpha = -1.7 \pm 0.4$ (observation 3)
steeper than the universal HMXB XLF
discovered by \citet{Grimm03}.
We proposed as a possible origin for the steep slope of the observed XLF
that the observed XLFs are the result of a mix
of XRB populations of different ages,
or, as an alternative explanation, 
that the IMF in the low-density regions of the outer disc of M\,83
is truncated, as previously suggested by e.g. \citet{Krumholz08}
and \citet{Meurer09} to explain the low production of high-mass
stars in low-density environments.
Additional X-rays and UV observations of the outer disc of M\,83,
analysed with  
most effective methods such as the one used by \citet{Bodaghee12}
to measure the spatial cross-correlation of HMXBs
and OB star-forming complexes in the Milky-Way,
will be fundamental to confirm our hypothesis.

\begin{acknowledgements}
We thank the referee Eric M. Schlegel for constructive comments, 
which helped to improve the manuscript.
This research is funded by the Deutsche Forschungsgemeinschaft
through the Emmy Noether Research Grant SA 2131/1.
This research has made use of the SIMBAD database,
operated at CDS, Strasbourg, France, and
of the NASA/IPAC Extragalactic Database (NED), 
which is operated by the Jet Propulsion Laboratory, 
California Institute of Technology, under contract 
with the National Aeronautics and Space Administration. 
This publication has made use of data products from the 
Two Micron All Sky Survey, which is a joint project of the 
University of Massachusetts and the Infrared Processing and Analysis Center, 
funded by the National Aeronautics and 
Space Administration and the National Science Foundation.
This research has made use of SAOImage DS9, 
developed by Smithsonian Astrophysical Observatory.
\end{acknowledgements}

\bibliographystyle{aa} 
\bibliography{lducci_m83}

\Online

\begin{appendix}

\section{Classification and identification of the XMM-\emph{Newton} sources}
\label{sect. Discussion of classification and identification of the XMM-Newton sources}

\subsection{Foreground stars}
\label{sect. Discussion foreground stars}

\paragraph{\textbf{Sources No.\,21, 143, 182, and 174}}
Using the criteria in Sect. \ref{sect. Foreground stars},
we classified sources No.\,21, 143, 182, and 174 as foreground stars
according to their optical and infrared properties
(Figs. \ref{fig. b-r_j-k.ps}, \ref{fig. B-R vs R}), 
and their optical-to-X-ray ratios as a function of the hardness ratios (Fig. \ref{fig. log10-fx-fopt}).
Although the hardness ratio criterion $HR_3 \lesssim -0.4$ of source No.\,174 is not fulfilled,
we classified this source as a foreground star because of the large uncertainty of the
hardness ratio (see Fig. \ref{fig. log10-fx-fopt}).

\paragraph{\textbf{Source No.\,24}} has optical and infrared counterparts
and $\log_{10}(f_{\rm x}/f_{\rm opt})<-1$,
but violates the hardness ratio $HR_2$ criterion (see Fig. \ref{fig. log10-fx-fopt}).
The optical counterpart is bright ($m_{\rm B,No.\,24}=14.1$),
and the $B-R$ and $J-K$ colours are consistent with those of foreground stars 
(Figs. \ref{fig. b-r_j-k.ps} and \ref{fig. B-R vs R}),
thus this source most likely belongs to the Milky Way.
It has been detected in observations 1 and 2 in all three EPIC cameras. 
In all cases, source No.\,24 shows hard HR$_2$ 
(Fig. \ref{fig. log10-fx-fopt}, left panel), 
inconsistent with the expected X-ray spectra of foreground stars.
The properties of the optical companion and the hard X-ray spectra
may indicate a cataclysmic-variable nature for this source.
This class of sources can show short- and long-term time variability,
therefore we produced the X-ray lightcurve in the energy range $0.5-4.5$ keV
to give more evidence for this identification.
However, the resulting X-ray lightcurve (with a bin-time of 2000 s)
shows neither short- nor long-term variability.

\subsection{Sources that are not foreground stars}
\label{sect. Sources which are not foreground stars}

\paragraph{\textbf{Sources No.\,12, 137, 164, and 189}}
coincide with ROSAT sources H2, H31, H34 and H36. 
They were classified by \citet{Immler99} as foreground stars
based on positional coincidences with optical sources 
of the \emph{APM Northern Sky Catalogue} \citep{Irwin94}.
We found possible optical counterparts in the USNO-B1 catalogue
for source No.\,164 (USNO$-$B1\,$0601-0299090$) 
and source No.\,12 (USNO$-$B1\,$0602-0301227$).
However, their X-ray-to-optical flux ratios (equation \ref{eq. Maccacaro})
are $\log(f_{\rm x}/f_{\rm opt}) \approx 0.10$ and $0.11$ respectively 
($f_{\rm opt}$ of both sources was calculated using visual magnitude),
hence the foreground star classification for these sources is ruled out.
The refined positions of sources Nos.\,137 and 189
obtained with XMM-\emph{Newton}, allowed us to exclude
their association with the optical counterparts proposed by \citet{Immler99}.
Source No.\,189 can be associated with a new optical counterpart, USNO$-$B1\,$0600-0300832$, which is
$\sim 3$ orders of magnitude fainter than the previous one (USNO$-$B1\,$0600-0300831$).
However, the new X-ray-to-optical flux ratio is $\log(f_{\rm x}/f_{\rm opt}) \approx 0.68$
($f_{\rm opt}$ was calculated using visual magnitude),
too high for a foreground star (see Sect. \ref{sect. Discussion Background objects}).
Hardness ratios of sources No.\,164 and 137 are consistent with a powerlaw or disk-blackbody spectrum.
Therefore, the spectra of these sources are too hard 
to be classified as foreground stars.

\subsection{Background objects}
\label{sect. Discussion Background objects}

We found radio counterparts of the sources No.\,20, 37, and 189 
and classified them as AGN candidates for the first time.

\paragraph{\textbf{Source No.\,20}} is located outside the $D_{25}$ ellipse
($D_{25}=11.5^\prime$; \citealt{Tully88})
at $\sim 0.41^\circ$ from the centre of the galaxy.
It coincides with the radio source NVSS\,J133618$-$301459. 
We detected this source with XMM-\emph{Newton} 
in observations 2 and 3 in the outer disc of M\,83.
Source No.\,20  shows a significant long-term variability (Table \ref{Tab. variability}),
and the hardness ratios are roughly consistent with
a spectrum described by an APEC model
with a temperature of $kT_{\rm apec} \sim 0.5$ keV ($HR_2=-0.2 \pm 0.1$; $HR_3=-0.81 \pm 0.11$).
Therefore, source No.\,20 can be classified as an AGN candidate 
(with a soft spectral component) or an SNR candidate.
The distance of this source from the nuclear region of M\,83 of $\sim32$ kpc rather indicates
that source No.\,20 does not belong to the galaxy, 
therefore it is more likely an AGN than an SNR candidate.

\paragraph{\textbf{Sources No.\,37 and 189}} coincide with the radio sources 
NVSS\,J133630$-$301651 and NVSS\,J133805$-$295748, respectively. 
Source No.\,189 was previously classified as a foreground star 
by \citet{Immler99} (see Sect. \ref{sect. Sources which are not foreground stars}).
We detected these sources with XMM-\emph{Newton} in observation 3.
Their hardness ratios are consistent with a spectrum described with a powerlaw or
disc-blackbody model
(No.\,37: $HR_2=0.62 \pm 0.12$; $HR_3=-0.37 \pm 0.13$;
No.\,189: $HR_2=0.07 \pm 0.10$; $HR_3=-0.25 \pm 0.12$).
Therefore, they can be classified as AGN candidates.

\subsection{X-ray binaries}
\label{sect. class. X-ray binaries}

\paragraph{\textbf{Source No.\,81}}
coincides with the \emph{Chandra} source [SW03]\,33,
classified as an accreting X-ray pulsar, with a hard spectrum ($\Gamma \approx 1.7$)
and a spin period of 174.9 s.

We observed source No.\,81 in all XMM-\emph{Newton} observations.
The hardness ratios are consistent with an absorbed powerlaw spectrum,
and this source shows a significant long-term X-ray variability 
($V_{\rm f}=2.5$, $S=3.0$, Table \ref{Tab. variability}).
We applied a Fourier transform periodicity search and a Z$_n^2$ analysis 
(section \ref{sect. Variability of the sources}),
which did not reveal any significant periodicity.
We calculated the upper-limit on the pulsed fraction
(defined as the semi-amplitude of the sinusoidal modulation
divided by the mean count rate) using the procedure
described by \citet{Vaughan94}.
The upper limit on the pulsed fraction 
obtained from the combined PN and MOS events of observation 1
is 16\% at the 99\% confidence level.
This upper limit is marginally compatible
with the pulsed fraction of $(50 \pm 15)$\%
of source [SW03]\,33.

\paragraph{\textbf{Source No.\,120}}
corresponds to the X-ray source [SW03]\,113.
Using the spectral properties and the 201.5 s periodicity detected with \emph{Chandra},
SW03 classified source [SW03]\,113 as an XRB in a soft state.

We observed source No.\,120 with XMM-\emph{Newton} in observations 1 and 3.
The hardness ratios of this source are consistent with an 
absorbed powerlaw spectrum with $N_{\rm H} \sim 5 \times 10^ {21}$ cm$^{-2}$
and $\Gamma \sim 1.5$.
Similarly to source No.\,81, a Fourier transform
periodicity search and a $Z_n^2$ analysis did not reveal
any significant periodicity.
At the 99\% confidence level, the upper limit
on the pulsed-fraction of source No.\,120
derived from the MOS events is 49\%.
This upper limit is compatible with the $(50 \pm 19)$\%
pulsed fraction of [SW03]\,113.

\subsection{Supernova remnant candidates}
\label{sect. class. Supernova remnants}

\paragraph{\textbf{Source No.\,79}}
The position of this source corresponds to the position 
of the ROSAT source H15 \citep{Immler99} and
the \emph{Chandra} source [SW03]\,27.
The \emph{Chandra} spectrum shows emission lines, suggesting the possibility of emission
from optically thin thermal plasma, and has been fitted by 
SW03 with an absorbed powerlaw with $\Gamma \sim 1.4$ and $N_{\rm H}\sim 7 \times 10^{20}$ cm$^{-2}$.
SW03 classified this source as a young SNR candidate. 
Another possible explanation for the hard powerlaw spectrum with 
superposition of emission lines of [SW03]\,27 is that the source is an XRB 
surrounded by a photoionised nebula (SW03). 
However, XRBs showing these spectral properties usually
have a higher absorbing column density than that of [SW03]\,27 (see e.g. \citealt{Sako99}). 

The XMM-\emph{Newton} hardness ratios of source No.\,79  below 2 keV 
are consistent with an APEC model with temperature 
$kT_{\rm apec} \gtrsim 1.5$ keV, 
while at higher energies the hardness ratios 
are consistent with a powerlaw with photon index $\sim 2$.
The spectral shape of source No.\,79 derived from XMM-\emph{Newton} 
hardness-ratio diagrams agrees with the X-ray spectrum
of [SW03]\,27 presented by SW03 (see Figure 6 in SW03)
and can be interpreted as an SNR exhibiting 
both a thin-thermal emission (below $\sim 2$ keV)
and an additional hard component, 
which dominates at energies above $\sim 2$ keV.
Also, source No.\,79 does not show any significant 
long-term variability (see Table \ref{Tab. variability}).

\subsection{Super-soft source candidates}
\label{sect. class. Super-Soft Sources}

\begin{figure}
\begin{center}
\includegraphics[bb=75 371 560 825,clip,width=8cm]{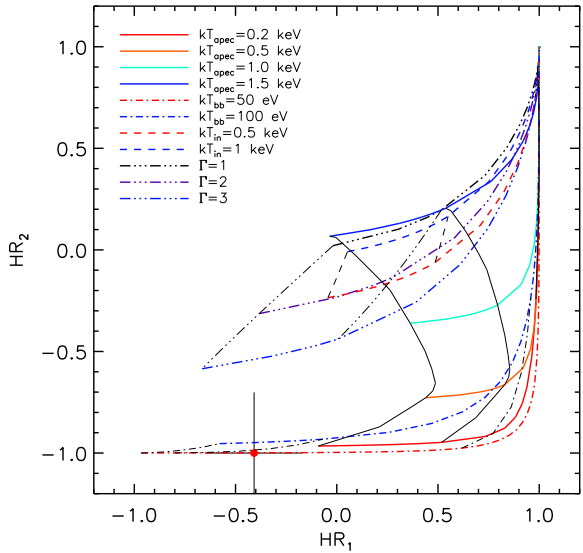}
\end{center}
\caption{Hardness-ratio diagram of source No.\,91 observed with XMM-\emph{Newton}.
Thick lines are different spectral models as function of the $N_{\rm H}$,
thin lines are different column densities $N_{\rm H}$ 
(from left to right: $10^{20}$, $10^{21}$, $10^{22}$ cm$^{-2}$)
as a function of the spectral parameters.}
\label{fig. col-col source 93}
\end{figure}

\paragraph{\textbf{Source No.\,91}}
coincides with \emph{Einstein} source 3 \citep{Trinchieri85}
and \emph{Chandra} source [SW03]\,55 
classified by \citet{DiStefano03} as an SSS candidate (source M83-50 in \citealt{DiStefano03}).
\citet{DiStefano03} fitted the X-ray spectrum of M83-50 with an absorbed blackbody
with a temperature of $kT_{\rm bb}= 66{+13 \atop -24}$ eV,
a column density of $N_{\rm H}=2.4{+7.4 \atop -2.4} \times 10^{20}$ cm$^{-2}$,
and a luminosity of $L_{\rm x}=2.8 \times 10^{37}$ erg s$^{-1}$ ($0.3-7$ keV, $d=4.5$ Mpc).

We detected source No.\,91 in observation 1, where the hardness ratios are consistent with
a blackbody spectrum (with column density in the range $\approx 10^{20}-10^{21}$ cm$^{-2}$)
and marginally compatible with an APEC spectrum with temperature in the range $\approx 0.2-0.5$ keV 
(Fig. \ref{fig. col-col source 93}).
Source No.\,91 has a $0.2-4.5$ keV luminosity of $L_{\rm x}=(2.2 \pm 0.2)\times 10^{37}$ erg s$^{-1}$
and does not show any significant variability compared to the \emph{Chandra} observation.

\subsection{Ultra-luminous X-ray sources}
\label{sect. class. Ultra-Luminous X-ray sources}

\begin{figure*}
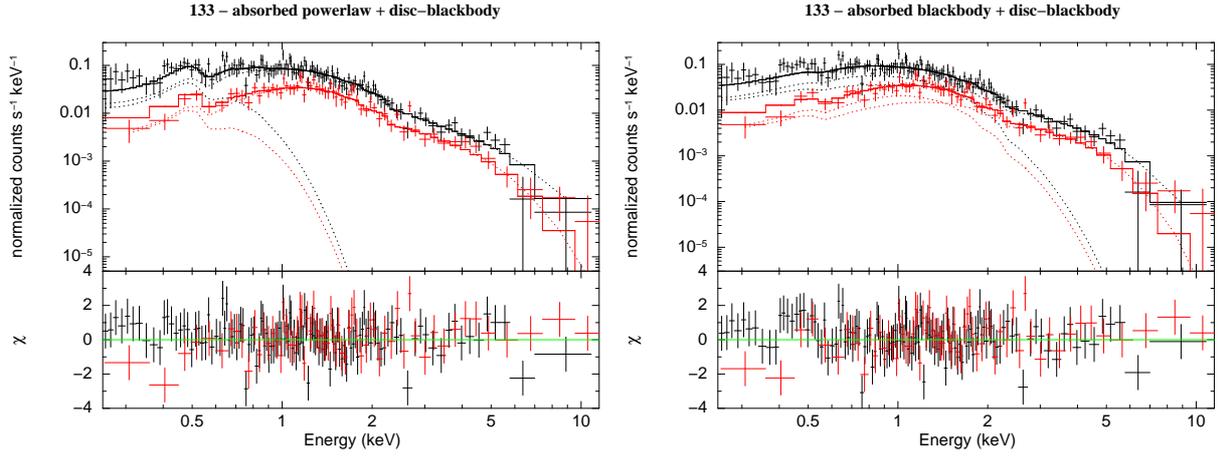

\begin{center}
\includegraphics[bb=25 -21 583 720, clip, angle=-90,width=8cm]{s3_obs2_pnm2_phabs_phabs_po_diskbb.ps}
\includegraphics[bb=25 -21 583 720, clip, angle=-90,width=8cm]{s3_obs2_pnm2_phabs_phabs_bb_diskbb.ps}
\end{center}
\caption{EPIC counts spectra, together with residuals in units of standard deviations
for source No.\,133 detected in the observation 2.
\emph{Left panel} shows the fit with an absorbed cool disc-blackbody plus hard powerlaw,
while the \emph{right panel} shows the fit with an absorbed cool blackbody plus a warm disc-blackbody (see Table \ref{Tab. ulx spectral parameters}).}
\label{fig. ulx-obs2}
\end{figure*}

\begin{table*}
\begin{center}
\caption{Best-fitting parameters of the X-ray spectra of source No.\,133 (errors at 90\% confidence level).}
\label{Tab. ulx spectral parameters}
\begin{tabular}{lccc}
\hline
\hline
                                                 &                          \multicolumn{2}{c}{obs. 2}                             &         obs. 3                \\
\hline
model                                           &         {\sc powerlaw + diskbb}        &        {\sc bbody + diskbb}        &         {\sc powerlaw}              \\
\hline
\smallskip
$N_{\rm H}$ ($10^{22}$ cm$^{-2}$)                 &$0.30{+0.12 \atop -0.09}$               &               $\leq 0.03$          &          $0.12 \pm 0.05$     \\
\smallskip
$\Gamma$ or $kT_{\rm bb}$ (keV)                   &$2.64{+0.19 \atop -0.18}$              &         $0.30{+0.02 \atop -0.04}$   &        $2.6 {+0.3 \atop -0.2}$      \\
\smallskip
norm.                                           &$4.4{+1.0 \atop -0.4}\times 10^{-4}$    &$3.6{+0.8 \atop -0.9}\times 10^{-6}$  & $1.4{+0.3 \atop -0.3}\times 10^{-4}$ \\
\smallskip
$kT_{\rm in}$ (keV)                               &$0.09{+0.02 \atop -0.03}$              &        $1.4{+0.3 \atop -0.2}$       &                                     \\
\smallskip
norm.                                           &$4.8{+47.4 \atop -4.4}\times 10^3$     &$7.8{+9.7 \atop -3.1}\times 10^{-3}$  &                                     \\
\smallskip
$\hat{\chi}^2_\nu$ (d.o.f.)                      &          0.946 (184)                  &              0.994 (184)            &            0.918 (62)              \\
\smallskip
$F_{\rm x}$ ($0.2-12$ keV, erg cm$^{-2}$ s$^{-1}$)&$8.9{+57.5 \atop -3.7}\times 10^{-13}$&$8.5{+24.5 \atop -5.2}\times 10^{-13}$&$3.7{+2.2 \atop -1.4}\times 10^{-13}$ \\
\smallskip
$L_{\rm x}$ ($d=4.5$ Mpc, erg s$^{-1}$)         &$1.3{+20.4 \atop -0.6}\times 10^{40}$ &$2.2{+6.3 \atop -1.3}\times 10^{39}$    & $2.2{+0.9 \atop -0.5}\times 10^{39}$ \\
\hline
\end{tabular}
\end{center}
\end{table*}

Two ULXs have been discovered in M\,83: 
H2 \citep{Trinchieri85}, and a transient ULX discovered with \emph{Chandra}
on 23 December 2010 with a luminosity of $L_{\rm x} \sim 4 \times 10^{39}$ erg s$^{-1}$ ($0.3-10$ keV)
by \citet{Soria10}, and classified as an accretion-powered black hole 
with mass $M_{\rm BH} \approx 40-100$ M$_\odot$ \citep{Soria12}.
This ULX has not been detected in the XMM-\emph{Newton} data.
\citet{Soria12} measured 
an upper limit to the X-ray luminosity of $\sim 10^{37}$ erg s$^{-1}$ ($0.3-10$ keV)
from the three XMM-\emph{Newton} observations.

\paragraph{\textbf{Source No.\,133}}
We observed the ULX as source No.\,133 in all XMM-\emph{Newton} observations.
\citet{Ehle98} and \citet{Immler99} found 
a faint extended optical source within the error circle of the ROSAT source position.
\citet{Roberts08} used HST images in three Advanced Camera for Survey (ACS) filters
to find the counterparts to six ULXs in different galaxies.
For the ULX in M\,83, they compared the optical position 
with the X-ray position from a \emph{Chandra} High Resolution Camera for Imaging (HRC-I) observation.
They detected a counterpart to the ULX with magnitudes $B=25.66\pm 0.13$, $V=25.36\pm 0.17$.
They also noticed that the ULX is located at $\sim 5^{\prime\prime}$
from the centre of a background galaxy, and although the latter is outside the error circle,
\citet{Roberts08} did not completely rule out 
a possible association between the ULX and the background galaxy.

\citet{Stobbart06} reported the XMM-\emph{Newton} spectral analysis 
of source No.\,133 during observation 1.
They found that the X-ray spectrum is well fitted with 
a cool disc-blackbody ($kT_{\rm in} \sim 0.2$ keV) plus a powerlaw ($\Gamma \sim 2.5$),
or with a cool blackbody ($kT_{\rm bb} \sim 0.2$ keV) plus a warm disc-blackbody ($kT_{\rm in} \sim 1.1$ keV).
The first spectral model is the standard IMBH model,
where the low disc temperature is due to a black hole with mass of $\sim 1000$ M$_\odot$,
while the origin of the powerlaw component is still not clear (see \citealt{Roberts05}).
Instead, the spectral parameters obtained with the second spectral model suggest that No.\,133
is a stellar-mass black hole accreting close to the Eddington limit.
In this model, the cool blackbody component represents the optically thick wind
from the stellar-mass black-hole accreting at or above the Eddington limit,
while the high temperature of the disc follows the standard trend $L_{\rm x} \propto T^4$ 
shown by the Galactic stellar-mass black-hole binaries.
 
We analysed all XMM-\emph{Newton} observations of the ULX No.\,133
and fitted the PN, MOS1 and MOS2 spectra simultaneously
with a model assuming an IMBH ({\sc phabs*[diskbb + powerlaw]} in XSPEC),
and a model assuming a stellar-mass BH ({\sc phabs*[bbody + diskbb]}).  
We used two absorption components:
the Galactic absorption column density ($N_{\rm H}=3.69\times 10^{20}$ cm$^{-2}$)
and the absorption within M\,83 plus the intrinsic column density of the ULX.
In all fits we obtained a good fit with both spectral models
with the resulting spectral parameters in agreement with those obtained by 
\citet{Stobbart06} from observation 1.
However, the spectral parameters in observation 3 are only poorly constrained
due to the poor statistics (only MOS1 and MOS2 data were available for this observation).
Therefore, we fitted the spectrum of observation 3 with a single component model
and found that an absorbed powerlaw can adequately fit the data 
(Fig. \ref{fig. ulx-obs2}, Table \ref{Tab. ulx spectral parameters}).

\subsection{Hard sources}
\label{section Hard sources}

\subsubsection{New classifications}
\label{section Newly classified hard sources}

\begin{table*}
\begin{center}
\caption{Best-fitting parameters of sources 
No.\,16, 61, 103, 126, 153.
We fitted the spectra with an absorbed powerlaw.
$\Gamma$ is the powerlaw photon-index, $F_{\rm x}$ is the absorbed flux in the energy range $0.2-12$ keV,
$L_{\rm x}$ is the X-ray luminosity in the same energy range of $F_{\rm x}$  (errors at 90\% confidence level).}
\label{Tab. spectral parameters xrbs}
\begin{tabular}{lcccccccc}
\hline
\hline
Source & \multicolumn{6}{c}{Parameters} & \multicolumn{2}{c}{Analysed data} \\
\hline
       &         $N_{\rm H}$        &        $\Gamma$       &                norm.               &$\hat{\chi}^2_\nu$ (d.o.f.)&             $F_{\rm x}$               &                $L_{\rm x}$           & obs. &instrument    \\
       &    ($10^{21}$ cm$^{-2}$)  &                       &                                    &                          &      (erg cm$^{-2}$ s$^{-1}$)        &             (erg s$^{-1}$)         &      &              \\  
\hline
\smallskip
{ } 16     & $0.8{+0.4 \atop -0.3}$   & $2.6{+0.3 \atop -0.3}$ & $4.1{+0.9 \atop -0.7}\times 10^{-5}$&      $0.928$ $(38)$     & $1.4{+0.8 \atop -0.5}\times 10^{-13}$ & $6.3{+2.7 \atop -1.3}\times 10^{38}$ &  1   & PN,MOS1,MOS2\\
\smallskip
{ } 61     & $2.01{+0.75 \atop -0.65}$& $2.4{+0.3 \atop -0.3}$ & $2.9{+0.9 \atop -0.6}\times 10^{-5}$ &     $1.04$ $(36)$       & $8.0{+6.8 \atop -3.7}\times 10^{-14}$ & $4.3{+1.4 \atop -0.6}\times 10^{38}$ &  1   & PN,MOS1,MOS2 \\
\smallskip
103    & $0.7{+1.0 \atop -0.7}$   & $1.8{+0.4 \atop -0.4}$ & $3.7{+1.8 \atop -1.2}\times 10^{-5}$ &     $0.876$ $(19)$      & $2.2{+3.3 \atop -1.3}\times 10^{-13}$ & $6.6{+6.7 \atop -2.7}\times 10^{38}$ &  1   &   PN,MOS2    \\
\smallskip
126    & $0.01{+0.56 \atop -0.01}$& $1.8{+0.4 \atop -0.2}$ & $8.8{+2.7 \atop -1.2}\times 10^{-6}$ &     $0.743$ $(17)$      & $6.3{+3.1 \atop -2.9}\times 10^{-14}$ & $1.5{+0.8 \atop -0.4}\times 10^{38}$ &  2   & PN,MOS1,MOS2 \\ 
\smallskip
153    & $0 < N_{\rm H} \leq 1.5$  & $1.4{+0.7 \atop -0.3}$ & $7.6{+5.8 \atop -1.6}\times 10^{-6}$ &      $1.027$ $(15)$      & $8.8{+13.7 \atop -6.2}\times 10^{-14}$& $2.1{+3.3 \atop -1.2}\times 10^{38}$ &  1   & PN,MOS1,MOS2 \\
\hline
\end{tabular}
\end{center}
\end{table*}

\paragraph{\textbf{Source No.\,16}}
coincides with the ROSAT source H3 discovered by \citet{Immler99}.
This source is located outside the optical disc of M\,83,
and its position overlaps with the outer disc of M\,83
observed by GALEX (e.g. \citealt{Thilker05}).

We detected source No.\,16 in all XMM-\emph{Newton} observations,
but only in observation 1 was it bright enough to allow spectral analysis.
The spectrum can be well fitted with an absorbed powerlaw with $\Gamma = 2.6{+0.3 \atop -0.3}$,
compatible with that of an XRB or an AGN (see Table \ref{Tab. spectral parameters xrbs}). 
Source No.\,16 shows a significant long-term variability ($S=9.5$)
with a variability factor of $V_{\rm f}=10.6 \pm 0.3$ (Table \ref{Tab. variability}).
It also shows a significant variability within observation 1,
with a variability factor of $V_{\rm f}=6.6 \pm 4.5$ and significance $S=4.0$.

\paragraph{\textbf{Source No.\,61}}
is in the field of view of XMM-\emph{Newton} during observation 1,
where it shows an X-ray luminosity of $L_{\rm x} \approx 4\times 10^{38}$ erg s$^{-1}$ 
(see Table \ref{Tab. spectral parameters xrbs}).
It has not been previously detected in X-ray, optical, radio, infrared, or UV.
The X-ray spectrum is well fitted with an absorbed powerlaw
with $\Gamma = 2.4{+0.3 \atop -0.3}$
or a disc-blackbody model with temperature $kT_{\rm in}=0.82{+0.13 \atop -0.11}$ keV 
(Table \ref{Tab. spectral parameters xrbs}).
Source No.\,61 shows a significant long-term variability ($S=8.6$)
with a variability factor of $V_{\rm f}=4.3 \pm 0.1$ (Table \ref{Tab. variability}).

\paragraph{\textbf{Source No.\,103}} is located at a distance of $\sim 6^{\prime\prime}$ from a radio source 
(6 in \citealt{Cowan94}, 36 in \citealt{Maddox06}),  
and at $1.6^{\prime\prime}$ from the \emph{Chandra} source [SW03]\,84,
which shows hardness ratios compatible with a powerlaw or a disc-blackbody spectrum.

We detected source No.\,103 only in the XMM-\emph{Newton} 
observation 2, with a flux of $(2.23{+3.26 \atop -1.34})\times 10^{-13}$ erg cm$^{-2}$ s$^{-1}$ ($0.2-12$ keV).
The X-ray spectrum is well fitted with an absorbed powerlaw with $\Gamma =1.8{+0.4 \atop -0.4}$
(Table \ref{Tab. spectral parameters xrbs}).
We did not detect source No.\,103 in observations 1 and 3,
thus we calculated the flux upper-limits and we found a significant ($S = 7.2$) long-term variability,
with a variability factor of $V_{\rm f}=12.78\pm0.12$ 
(Table \ref{Tab. variability}, Fig. \ref{fig. variab}).

\paragraph{\textbf{Source No.\,126}}
coincides with X-ray source 30 \citep{Ehle98} discovered with ROSAT.
Source No.\,126 also cross-correlates with the optical counterpart USNO-B1\,$0599-0300335$,
but the ratio $\log_{10}(F_{\rm x}/F_{\rm opt})$ 
does not match the criteria previously specified to classify foreground stars.
Source No.\,126 is located outside the optical disc of M\,83, and its position overlaps
with an extended arm of the galaxy. 

We observed source No.\,126 in all XMM-\emph{Newton} observations.
The X-ray spectra extracted from each observation can be well fitted
with an absorbed powerlaw with $\Gamma \approx 1.8$ and the flux 
is consistent with that measured by \citet{Ehle98}
(Table \ref{Tab. spectral parameters xrbs}).

\paragraph{\textbf{Source No.\,153}}
is detected in all XMM-\emph{Newton} observations,
and has not been previously detected in X-rays, optical,
radio, infrared, or UV bands.
It is located in the extended arms observed by \emph{GALEX},
$\approx 10^\prime$ away from the nuclear region of M\,83.

The spectra extracted from each observation can be well fitted 
with an absorbed powerlaw with $\Gamma \approx 1.5$,
suggesting an XRB nature for this source
(see Table \ref{Tab. spectral parameters xrbs}).
%
%

\subsubsection{Identifications}
\label{section identification hard sources}

\begin{table*}
\begin{center}
\caption{Best-fitting parameters of sources 
No.\,97, 106, 107, 108, 114, and 129.
We fitted the spectra with an absorbed powerlaw.
$\Gamma$ is the powerlaw photon-index, $F_{\rm x}$ is the absorbed flux in the energy range $0.2-12$ keV,
$L_{\rm x}$ is the X-ray luminosity in the same energy range of $F_{\rm x}$ (errors at 90\% confidence level).}
\label{Tab. spectral parameters xrbs appendix}
\begin{tabular}{lcccccccc}
\hline
\hline
Source & \multicolumn{6}{c}{Parameters} & \multicolumn{2}{c}{Analysed data} \\
\hline
       &        $N_{\rm H}$       &        $\Gamma$       &                norm.               &$\hat{\chi}^2_\nu$ (d.o.f.)  &             $F_{\rm x}$              &            $L_{\rm x}$                 & obs. &instrument    \\
       &   ($10^{21}$ cm$^{-2}$) &                       &                                    &                            &      (erg cm$^{-2}$ s$^{-1}$)      &          (erg s$^{-1}$)               &      &              \\  
\hline
\smallskip
{ } 97     & $4.4{+1.1 \atop -0.9}$ & $2.4{+0.3 \atop -0.3}$ & $7.4{+2.3 \atop -1.7}\times 10^{-5}$&        $1.03$ $(45)$      & $1.7{+1.5 \atop -0.8}\times 10^{-13}$ & $1.1{+0.4 \atop -0.2}\times 10^{39}$ &  1   &  PN,MOS2  \\
\smallskip
106    & $0.3{+0.5 \atop -0.3}$ & $1.8{+0.4 \atop -0.3}$ & $8.3{+2.5 \atop -1.9}\times 10^{-6}$&        $1.38$ $(27)$      & $5.4{+5.0 \atop -2.7}\times 10^{-14}$ & $1.5{+1.0 \atop -0.5}\times 10^{38}$ &  1   & PN,MOS1,MOS2\\
\smallskip
107    & $3.3{+1.0 \atop -0.8}$ & $2.8{+0.4 \atop -0.3}$ & $3.9{+1.4 \atop -1.0}\times 10^{-5}$&        $0.821$ $(43)$     & $6.5{+6.5 \atop -3.3}\times 10^{-14}$ & $6.6{+5.1 \atop -2.3}\times 10^{38}$ &  1   & PN,MOS1,MOS2\\
\smallskip
108    & $3.3{+0.7 \atop -0.6}$ & $2.7{+0.3 \atop -0.2}$ & $8.0{+1.9 \atop -1.5}\times 10^{-5}$&        $0.91$ $(68)$      & $1.5{+0.9 \atop -0.6}\times 10^{-13}$ & $1.3{+0.5 \atop -0.3}\times 10^{39}$ &  1   & PN,MOS1  \\
\smallskip
114    & $4.5{+3.8 \atop -3.1}$ & $1.7{+0.4 \atop -0.4}$ & $1.8{+1.3 \atop -0.7}\times 10^{-5}$&        $0.952$ $(20)$     & $9.6{+24.0 \atop -6.9}\times 10^{-14}$& $3.5{+5.68 \atop -1.8}\times 10^{38}$ &  1   & PN,MOS1,MOS2\\ 
\smallskip
129    & $5.4{+4.3 \atop -3.0}$ & $2.0{+0.6 \atop -0.5}$ & $2.0{+1.9 \atop -0.9}\times 10^{-5}$&        $0.867$ $(18)$     & $6.7{+20.9 \atop -5.1}\times 10^{-14}$& $3.1{+5.4 \atop -1.4}\times 10^{38}$ &  1   &  PN,MOS1  \\
\hline
\end{tabular}
\end{center}
\end{table*}

\paragraph{\textbf{Source No.\,60}}
correlates with the \emph{Chandra} source [SW03]\,5 
SW03 suggested that this source is an XRB candidate.

We observed source No.\,60 in all XMM-\emph{Newton} observations.
The source shows a significant long-term variability ($V_{\rm f}=2.4$, $S=3.8$, Table \ref{Tab. variability}) 
with respect to the \emph{Chandra} observation.
X-ray colours of No.\,60 are consistent with a powerlaw  or disc-blackbody spectrum,
in agreement with the spectral analysis of SW03.

\paragraph{\textbf{Source No.\,80}}
correlates with the \emph{Chandra} source [SW03]\,31. 
From the spectral properties, SW03 suggested that [SW03]\,31 is an XRB candidate.

We observed source No.\,80 with XMM-\emph{Newton} in observation 1.
The hardness ratios are consistent with a powerlaw  or disc-blackbody spectrum with column density 
of $\sim 10^{21}$ cm$^{-2}$ 

\paragraph{\textbf{Source No.\,92}}
coincides with the \emph{Chandra} source [SW03]\,60.
SW03 suggested that No.\,92 is a XRB candidate because of its hard spectrum ($\Gamma \sim 1.6$).

We observed source No.\,92 with XMM-\emph{Newton} in observation 1.
The hardness ratios are consistent with a spectrum described by an absorbed powerlaw model with $\Gamma \sim 2$.
Source No.\,92 also shows a high long-term variability by a factor of $V_{\rm f}=2.7$, 
with a variability significance of $S=4.3$ 
(see Table \ref{Tab. variability}).

\paragraph{\textbf{Source No.\,97}}
coincides with the \emph{Chandra} source [SW03]\,72 
and with a ROSAT source (source 7 in \citealt{Ehle98} and source H20 in \citealt{Immler99}).

We observed source No.\,97 in all XMM-\emph{Newton} observations.
The spectra extracted from each observation can be well fitted
with an absorbed powerlaw or a disc-blackbody model 
(Table \ref{Tab. spectral parameters xrbs appendix}),
with spectral parameters in agreement with the spectral analysis of SW03.
Source No.\,97 shows a significant long-term variability between XMM-\emph{Newton} 
and \emph{Chandra} observations ($V_{\rm f}=2.8 \pm 0.1$, $S=6.6$; Table \ref{Tab. variability}).
Within observation 1 we found a variability of $V_{\rm f}=6.4 \pm 2.7$
with a significance of $S=4.8$.

\paragraph{\textbf{Source No.\,99}}
coincides with the \emph{Chandra} source [SW03]\,73,
and it is associated with the radio source MCK\,34 (\citealt{Maddox06}). 
located in a HII region (RK\,137, \citealt{Rumstay83}).
From a spectral study, 
SW03 proposed that [SW03]\,73 is more likely an XRB than a young SNR.

We observed source No.\,99 with XMM-\emph{Newton} in observations 2 and 3.
The source shows a significant variability ($S=5.0$),
with a variability factor of $V_{\rm f}=15.3$ (Table \ref{Tab. variability}).
and the hardness ratios are consistent with an absorbed powerlaw or disk-blackbody spectrum.

\paragraph{\textbf{Source No.\,106}}
corresponds to the X-ray source H25 observed by  \citet{Immler99} in a ROSAT observation
and the \emph{Chandra} source [SW03]\,85. 

We observed source No.\,106 in all the observations.
During observation 1 the source was bright enough to allow spectral analysis.
The spectrum can be well fitted with an absorbed powerlaw
(see Table \ref{Tab. spectral parameters xrbs appendix}),
with spectral parameters in agreement with those 
previously obtained by SW03.

\paragraph{\textbf{Source No.\,107}}
was detected by \citet{Ehle98} (source 9) and \citet{Immler99} (source H26) 
in ROSAT (PSPC and HRI) observations. 
\citet{Immler99} found that H26 coincides with a compact radio source (source 8 in \citealt{Cowan94}),
and with a giant HII region \citep{Rumstay83}.
Hence, they classified this source as an SNR candidate.
Moreover, also the observation of H$\alpha$ and H$\beta$ emission anti-coincident with HI emission \citep{Tilanus93} supports
the SNR hypothesis.
Source No.\,107 was also observed in 2000 April 29 by \emph{Chandra} (source [SW03]\,86). 
From a spectral analysis, SW03 proposed that No.\,110 is more likely an XRB (BH candidate) 
than an SNR.

We detected source No.\,107 in all XMM-\emph{Newton} observations 
with a luminosity of $\sim 7 \times 10^{38}$ erg s$^{-1}$.
In observations 1 and 3 the source was bright enough to allow spectral analysis.
The spectra can be well fitted with an absorbed powerlaw or a disc-blackbody
(see Table \ref{Tab. spectral parameters xrbs appendix}).
The obtained spectral parameters are consistent with those 
previously found by SW03 with \emph{Chandra}.
Source No.\,107 shows a significant long-term variability
between XMM-\emph{Newton} observations ($V_{\rm f}=1.97 \pm 0.12$, $S=5.1$).

\paragraph{\textbf{Source No.\,108}}
was first detected in X-rays by \citet{Trinchieri85} 
(source 4) with the \emph{Einstein} satellite
and by \citet{Ehle98} (source 8) and \citet{Immler99} (source H27) with ROSAT.
It also coincides with the \emph{Chandra} source [SW03]\,88.

We observed source No.\,108 in all XMM-\emph{Newton} observations.
During observation 1 source No.\,108 was in the centre of the field of view,
providing enough statistic to allow spectral analysis.
We extracted the PN and MOS1 spectra (the position of source No.\,108 was in a gap of MOS2)
and we found that an absorbed powerlaw or an absorbed disc-blackbody provide acceptable fits
(Table \ref{Tab. spectral parameters xrbs appendix}),
with spectral parameters consistents with those obtained by SW03.
Source No.\,108 shows a significant long-term X-ray variability 
($V_{\rm f}=1.4 \pm 0.1$, $S=3.3$ Table \ref{Tab. variability}), 
and during observation 1 we found a variability of $V_{\rm f} = 4.2 \pm 1.5$, 
with a significance of $S=4.4$.

\paragraph{\textbf{Source No.\,114}}
coincides with the \emph{Chandra} source [SW03]\,104. 

We observed source No.\,114 in all XMM-\emph{Newton} observations.
During observation 1 source No.\,114 was in the centre of the field of view,
providing enough statistics to allow a spectral analysis.
The spectrum is well fitted with an absorbed powerlaw 
with spectral parameters consistent with those found by SW03 with \emph{Chandra}
(see Table \ref{Tab. spectral parameters xrbs appendix}).
Source No.\,114 also shows a significant long-term variability 
($V_{\rm f} = 2.4 \pm 0.2$, $S=4.1$).

\paragraph{\textbf{Source No.\,116}}
coincides with [SW03]\,105. 
\citet{DiStefano03} observed [SW03]\,105 with \emph{Chandra} (source M83-88 in \citealt{DiStefano03})
and classified it as an SSS candidate.
\citet{Blair04} compared the list of \emph{Chandra} sources of SW03 
with a list of optical SNR candidates
and associated [SW03]\,105 with the optical SNR candidate BL53.

\begin{figure}
\begin{center}
\includegraphics[bb=167 238 457 549,clip,width=4.4cm]{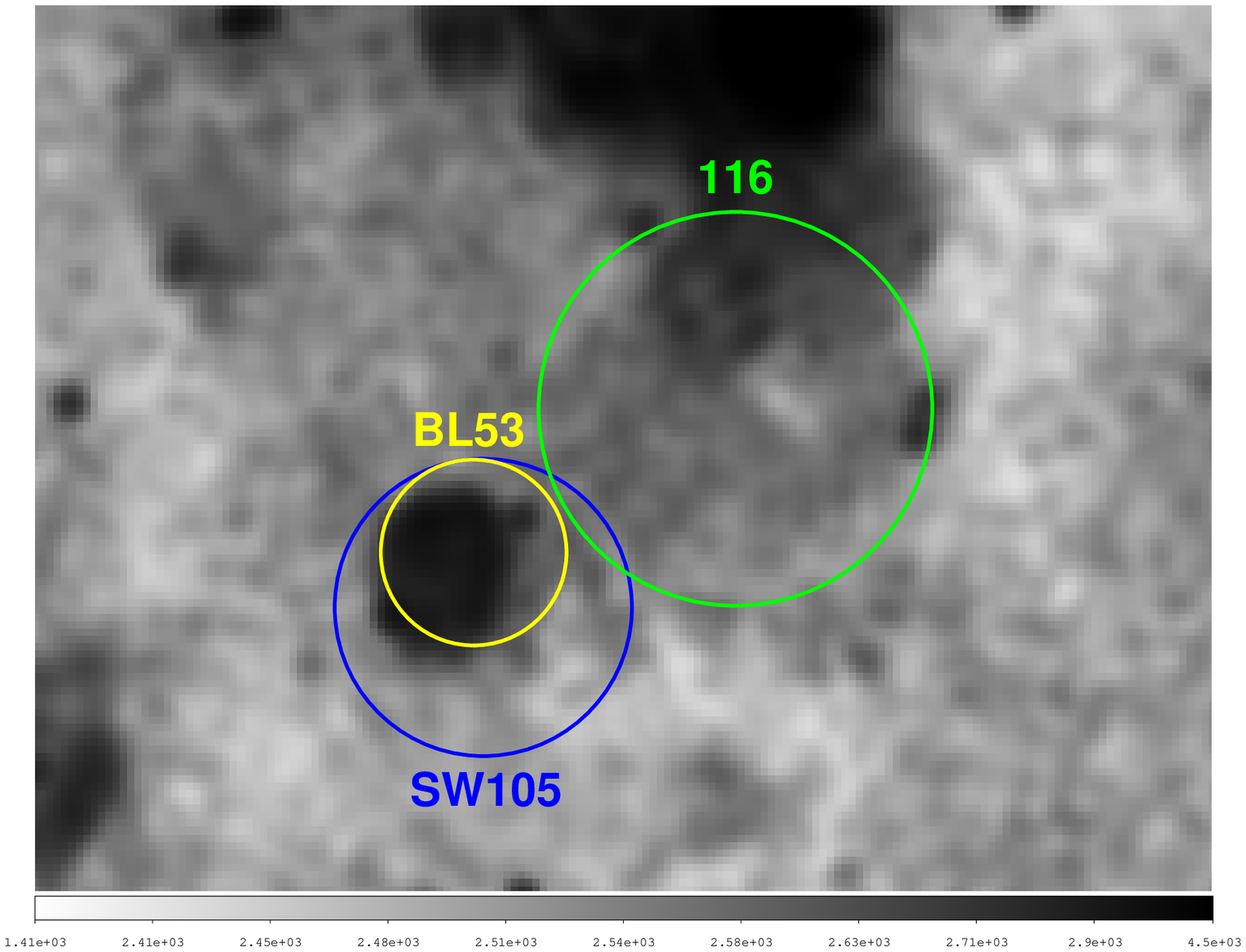}
\includegraphics[bb=154 205 481 558,clip,width=4.4cm]{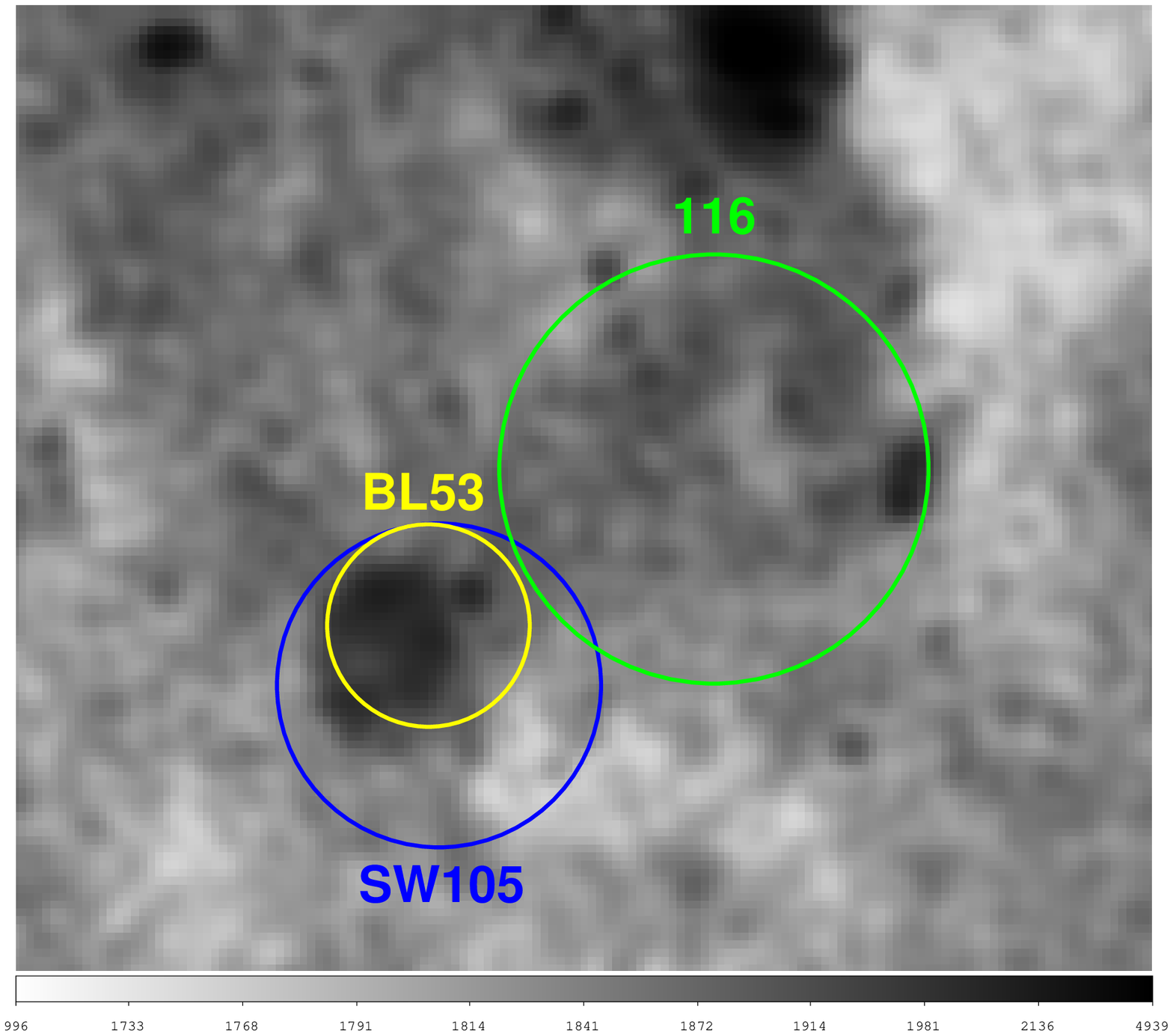}
\end{center}
\caption{Emission line \emph{HST}/WFC3 images of the region surrounding source No.\,116. 
\emph{Left panel:} WFC3 image with the narrowband filter F657N, corresponding to H$\alpha$ line emission.
\emph{Right panel:} WFC3 image with the narrowband filter F673N, corresponding to SII line emission.
The radii of the circles of \emph{Chandra} (SW105) and XMM-\emph{Newton} (116) sources give the 3$\sigma$ accuracy of the position of the sources.
The circle labelled BL53 gives the position of the opitcal SNR candidate.}
\label{fig. HST/WFC3 source 119}
\end{figure}

We observed source No.\,116 with XMM-\emph{Newton} in observation 1, 
where it shows a significant X-ray variability ($V_{\rm f}=20.8$, $S=6.2$, Table \ref{Tab. variability}) 
compared to the \emph{Chandra} observation, 
and the X-ray hardness ratios are consistent with a hard spectrum.
These properties indicate that source No.\,116 is most likely an XRB.
We overlaid the $3\sigma$ error circles of source No.\,116, [SW03]\,105, and BL53
on the emission line images H$\alpha$ and SII
obtained from the public 
\emph{Wide Field Camera 3 (WFC3)} observation of 2009-08-20 (Fig. \ref{fig. HST/WFC3 source 119}).
H$\alpha$ and SII images are used in extragalactic searches of SNRs 
because their optical spectra show high [SII]:H$\alpha$ ratios
compared to the spectra of normal HII regions (see e.g. \citealt{Blair04}).
Fig. \ref{fig. HST/WFC3 source 119} shows that the shell of the optical SNR is located
only in the error circles of BL53 and [SW03]\,105, 
indicating that  source No.\,116 and [SW03]\,105 cannot be the same source.

Therefore, source No.\,116 is more likely a transient source not associated
with BL53.

\paragraph{\textbf{Source No.\,129}}
coincides with the \emph{Chandra} source [SW03]\,121 
and with a ROSAT source (source 12 in the catalogue of \citealt{Ehle98},
source H29 in the catalogue of \citealt{Immler99}).

We observed source No.\,129 during observation 1, where it was in the
XMM-\emph{Newton} field of view.
The spectrum is well fitted with an absorbed powerlaw with
spectral parameters consistent with those found by SW03 with \emph{Chandra}
(see Table \ref{Tab. spectral parameters xrbs appendix}).
We did not detect source No.\,129 in observations 2 and 3,
thus we calculated the flux upper limits and we found a significant 
($S\gtrsim 6.4$) long-term variability
with a variability factor of $V_{\rm f}=3.94\pm0.11$ 
(Table \ref{Tab. variability}).

\section{X-ray source catalogue of the XMM-\emph{Newton} EPIC M\,83 observation}
\label{sect. catalogue-table}

\ 

\onecolumn

\begin{landscape}
\scriptsize
\begin{longtable}{lccccccccccccc}
\caption{\emph{(Only in the electronic version)} M\,83 sources detected by XMM-\emph{Newton}.}
\label{Tab. source list}\\
\hline
\hline

\hline
No. &   RA(2000)  &  DEC(2000)   &          pos. err.      &        rate (obs1)    &     rate (obs2)       &      rate (obs3)      &    \multicolumn{3}{c}{likelihood}        &                    HR1                 &                     HR2               &                     HR3                &                   HR4                   \\
    &             &              &                         &                       &                       &                       &   (obs1)     &    (obs2)    &    (obs3)   &                                        &                                       &                                        &                                        \\
\hline
\endfirsthead

\multicolumn{14}{r}{Continued from previous page}\\
\hline
No. &   RA(2000)  &  DEC(2000)   &          pos. err.      &        rate (obs1)    &     rate (obs2)       &      rate (obs3)      &    \multicolumn{3}{c}{likelihood}        &                    HR1                 &                     HR2               &                     HR3                &                   HR4                   \\
    &             &              &                         &                       &                       &                       &   (obs1)     &    (obs2)    &    (obs3)   &                                        &                                       &                                        &                                        \\
\hline

\endhead

\hline \multicolumn{14}{r}{{Continued on next page}} \\
\endfoot

\hline
\endlastfoot

\hline
\smallskip 
{ } { }   1 & 13 35 48.83 & -29 56 39.5 & $ 1.57^{\prime\prime}$ & $0.012 \pm   0.002$ &                     &                       & $  33.2$ &          &          & $ -0.07 \pm   0.21$ & $ -0.06 \pm   0.23$ & $ -0.43 \pm   0.34$ & $ -0.58 \pm   1.11$  \\ 
\smallskip
{ } { }   2 & 13 35 49.82 & -30 00 38.4 & $ 2.94^{\prime\prime}$ &                     & $0.016 \pm   0.004$ &                       &          & $  21.2$ &          & $  0.06 \pm   0.31$ & $  0.27 \pm   0.25$ & $ -0.60 \pm   0.32$ & $  0.50 \pm   0.44$  \\ 
\smallskip
{ } { }   3 & 13 35 52.30 & -30 02 59.8 & $ 1.72^{\prime\prime}$ &                     & $0.019 \pm   0.004$ & $0.0048 \pm   0.0008$ &          & $  28.2$ & $  57.0$ & $  1.00 \pm   0.08$ & $  0.06 \pm   0.17$ & $ -0.64 \pm   0.20$ & $ -1.00 \pm   0.76$  \\ 
\smallskip
{ } { }   4 & 13 35 55.46 & -30 13 23.3 & $ 1.96^{\prime\prime}$ &                     & $0.013 \pm   0.004$ &                       &          & $  16.2$ &          & $  0.22 \pm   0.28$ & $ -0.79 \pm   0.31$ & $  0.64 \pm   0.50$ & $  0.41 \pm   0.40$  \\
\hline
\end{longtable}
\end{landscape}

\begin{landscape}
\scriptsize
\begin{longtable}{lcccccc}
\caption{\emph{(Only in the electronic version)} M\,83 sources detected by XMM-\emph{Newton} cross-correlated with 
optical and radio counterparts and X-ray sources
detected with ROSAT (\citealt{Ehle98}: [E1998]; \citealt{Immler99}: [I1999]),
\emph{Einstein} (\citealt{Trinchieri85}: [T1985]), and \emph{Chandra} [SW2003].
For each source the identification proposed by the respective authors and our classification are given.
Uncertain classifications are given in brackets. 
For each counterpart, we assigned a ``grade'' A, B, or C if the positions of the X-ray source 
and its counterpart are closer than 1$\times$, 2$\times$, 3$\times$ combined statistical errors, respectively.}
\label{Tab. source list classification}\\
\hline
\hline
\multicolumn{1}{l}{No.} &
\multicolumn{1}{c}{USNO B1} &
\multicolumn{1}{c}{2MASS} &
\multicolumn{1}{c}{optical} &
\multicolumn{1}{c}{radio} &
\multicolumn{1}{c}{X-ray} &
\multicolumn{1}{c}{class.}\\
\hline
\endfirsthead

\multicolumn{7}{r}{Continued from previous page}\\
\hline
\multicolumn{1}{l}{No.} &
\multicolumn{1}{c}{USNO B1} &
\multicolumn{1}{c}{2MASS} &
\multicolumn{1}{c}{optical} &
\multicolumn{1}{c}{radio} &
\multicolumn{1}{c}{X-ray} &
\multicolumn{1}{c}{class.}\\
\hline

\endhead

\hline \multicolumn{7}{r}{{Continued on next page}} \\
\endfoot

\hline
\endlastfoot

{ } { } 1   & 0600-0299949 (A) &                      &                        &                            &                                                  & \\
\hline
{ } { } 2   & 0599-0299809 (B) &                      &                        &                            &                                                  & \\
\hline
{ } { } 3   &                  &                      &                        &                            &                                                  & \\
\hline
{ } { } 4   &                  &                      &                        &                            &                                                  & \\
\hline
{ } { } 5   &                  &                      &                        &                            &                                                  & \\
\hline
{ } { } 6   &                  &                      &                        &                            &                                                  & \\
\hline
{ } { } 7   & 0598-0300930 (C) & 13360480-3008319 (C) &  QSO\,1333-298 (C)     &                            &                                                  & AGN \\
\hline
{ } { } 8   &                  &                      &                        &                            &                                                  & \\
\hline
\end{longtable}
\end{landscape}
\twocolumn

\end{appendix}

\end{document}